\documentclass[useAMS,usenatbib]{mn2e}
\usepackage{epsfig,graphicx,latexsym,amsmath,amssymb}
\usepackage{natbib}
\usepackage{mathrsfs}
\usepackage{wasysym}
\usepackage{url}
\usepackage{lastpage}

\usepackage{latexsym}
\usepackage{graphicx}
\usepackage{eepic}
\usepackage{epic}
\usepackage{subfigure}
\usepackage{alltt}

\bibpunct[,]{(}{)}{;}{a}{}{,}

\title[Hybrid methods: A new composite algorithm]
       {Hybrid methods in planetesimal dynamics (I) :\\
        Description of a new composite algorithm}

\author[P. Glaschke, P. Amaro-Seoane \& R. Spurzem]
{
P. Glaschke$^{1}$
,
P. Amaro-Seoane\thanks{E-mail: Glaschke@ari.uni-heidelberg.de (PG);
                Pau.Amaro-Seoane@aei.mpg.de (PAS, corresponding author);
                Spurzem@bao.ac.cn (RS)}$^{2,\,3,\,4}$ \&
R. Spurzem$^{3,\,1,\,5}$\\
$^{1}$Astronomisches Rechen-Institut, M{\"o}nchhofstra{\ss}e 12-14,
      Zentrum f\"ur Astronomie, Universit\"at Heidelberg, Germany\\
$^{2}$Max Planck Institut f\"ur Gravitationsphysik
      (Albert-Einstein-Institut), D-14476 Potsdam, Germany\\
$^{3}$National Astronomical Observatories of China, Chinese Academy of
       Sciences, 20A Datun Lu, Chaoyang District, 100012, Beijing, China\\
$^{4}$Institut de Ci{\`e}ncies de l'Espai (CSIC-IEEC), Campus UAB,
      Torre C-5, parells, $2^{\rm na}$ planta, ES-08193, Bellaterra,
      Barcelona, Spain\\
$^{5}$Kavli Institute for Astronomy and Astrophysics, Peking
      University, China
}

\begin{document}

\date{draft \today}

\pagerange{\pageref{firstpage}--\pageref{lastpage}} \pubyear{2011}

\maketitle

\label{firstpage}

\begin{abstract}
The formation and evolution of protoplanetary systems, the breeding grounds of
planet formation, is a complex dynamical problem that involves many orders of
magnitudes. 
To serve this purpose, we present a new hybrid algorithm that combines a Fokker-Planck approach with the
advantages of a pure direct-summation $N-$body scheme, with a very accurate
integration of close encounters for the orbital evolution of the larger bodies
with a statistical model, envisaged to simulate the very large number of
smaller planetesimals in the disc.  Direct-summation techniques have been
historically developped for the study of dense stellar systems such as open and
globular clusters and, within some limits imposed by the number of stars, of
galactic nuclei.
The number of modifications to adapt direct-summation $N-$body techniques to
planetary dynamics is not undemanding and requires modifications. These include
the way close encounters are treated, as well as the selection process for the
``neighbour radius'' of the particles and the extended Hermite scheme, used
for the very first time in this work, as well as the
implementation of a central potential, drag forces and the adjustment of the
regularisation treatment.
For the statistical description of the planetesimal disc we employ a
Fokker-Planck approach. We include dynamical friction, high- and low-speed
encounters, the role of distant encounters as well as gas and collisional
damping and then generalise the model to inhomogenous discs.  We then describe
the combination of the two techniques to address the whole problem of
planetesimal dynamics in a realistic way via a transition mass to integrate the
evolution of the particles according to their masses.
\end{abstract}


\section{Introduction \label{PlanetForm}}

The formation of a planetary system is closely related to the formation of the
host star itself.  Cool molecular clouds collapse and fragment into smaller
substructures which are the seeds for subsequent star formation. Angular
momentum conservation in the forming clumps forces the infalling matter into a
disc-like structure. The subsequent viscous evolution of the disc leads to a
transport of angular momentum which channels gas to the protostar in the
centre. These protoplanetary discs are the
birth place of planets \citep[for a detailed review see][]{Armitage2010}.
Embedded dust grains are the seed for the enormous growth to bodies of
planetary size.  The first hint to the structure of protoplanetary discs has been provided
by our own solar system. Through ``smearing out'' all planets and adding the
missing fraction of volatile elements, one can estimate the structure and mass
of the protoplanetary disc. Since the efficiency of planet formation is
unknown, this yields only a lower limit -- the {minimum mass solar nebula}
\citep{Hayashi1981}.  The inferred surface density decreases with the distance
from the sun as $\propto r^{-3/2}$.

Further insight has been obtained by the detection of an {infrared excess}
of some stars, i.\,e. additional infrared radiation that does not originate
from the star but an unresolved disc. Advancements in observation led in the
mid-90s to the direct imaging of nearby star forming regions which opened a new
flourishing field in astronomy \citep[see][for an example with Hubble Space
Telescope images)]{Odell1993}. Since then a big amount of observations of
protoplanetary discs across the electromagnetic range has been obtained and
interpreted, both space-based (ISO, Spitzer and now Herschel) and in the ground
(VLT, VLT-I, Subaru, Keck). We refer the reader to ``The Star Formation Newsletter''
URL\footnote{\url{http://www.ifa.hawaii.edu/users/reipurth/newsletter.htm}} 
for an overview of the recent papers in star- and planet formation
and the book review ``Protostars and Planets V'' \citep{ProtostarsPlanetsV}.

Protoplanetary discs masses cover a range from $10^{-3}$ to $0.1 M_{\odot}$
with a peak around $0.01 M_{\odot}$ \citep[data from
Taurus/Ophiuchus][]{Beckwith1996}, in accordance with mass estimates deduced
from the minimum mass solar nebula.  Since the disc lifetime can not be
measured directly, it is derived indirectly from the age of young, naked
(i.\,e. discless) stars which sets an upper limit.  \mbox{Pre-main} sequence
evolutionary tracks are used to gauge the stellar ages, providing lifetimes of
a few $10^6$~years.  The subsequent evolution of the disc proceeds in several
stages.

Two different scenarios have been proposed to explain the formation of kilometre-sized planetesimals.
\begin{enumerate}
\item
One process is the gravitational instability of the dust component in a protoplanetary disc
that leads to the direct formation of larger bodies. \cite{Goldreich1973} propose that an initial growth phase of dust grains leads to a thin dust disc that undergoes a
gravitational collapse. As the dense dust layer decouples from the gas, it rotates with the local Keplerian
velocity, whereas the gas component rotates slower as it is partially pressure supported. This gives
rise to a velocity shear at the boundary, which may excite turbulence through the
{Kelvin-Helmholtz-instability}. Since the motion of small dust grains in the boundary layer is coupled
to the gas, the turbulent velocity field could  suppress the formation of a stratified dust layer,
which is a necessary prerequisite for the gravitational instability \citep{Weidenschilling77}.
\item
The collisional agglomeration of dust particles is an opposed formation mechanism.
Relative velocities are dominated by the Brownian motion in the early phases of the
growth process. This mechanism becomes increasingly inefficient with
growing mass, but successively the particles
decouple from the gas and settle
to the midplane -- a process that yields even larger velocities with increasing mass.
The sedimentation initiates a growth mode that is
similar to the processes in rain clouds: Larger grains drop faster, thus accreting
 smaller grains on their way to the midplane.
Turbulence may modify this basic growth scenario by forcing the dust grains in a
convection-like motion. Dust grains still grow during the settling process,
but the turbulent velocity field could mix up dust from the midplane, and a
new cycle begins. Each cycle adds a new layer to the dust grains -- a mechanism
that also operates in hail clouds -- until the grains are large enough to decouple
from the turbulent motion.
Again, turbulence plays an important role in determining the growth mode and the relative velocities.
While the relative velocities are high enough to allow for a fast growth, it is not
clear {\em a priory} that collisions are sticky enough to allow for a net growth.
High speed encounters lead to fragmentation, which counteracts agglomeration \citep[e.g.][and references therein]{BlumWurm2000}.
An important bottleneck in the agglomeration process is the fast
orbital decay of 1~m~sized boulders. Their orbital lifetime is as short
as 100 years, and a quick increase in size -- at least over one order of magnitude -- is needed
to reduce the radial drift significantly.
\end{enumerate}
To overcome the difficulties associated with each of these scenarios,
modifications have been proposed.  {M}agneto-{h}ydro{d}ynamic simulations
include electro-magnetic interactions in hydrodynamical calculations. See the
reviews of \citealt{Balbus1998} and \citealt{Balbus2003}. The MHD simulations
by \cite{Johansen2006} show that trapping of larger particles in turbulent
vortices helps in increasing the orbital lifetime, but could also trigger local
instabilities that may lead to the direct formation of planetesimals
\citep{Ina2005}. \cite{JohansenEtAl07} describe a gravoturbulence mechanism as 
a feasible pathway to planetesimal formation in accreting circumstellar discs.

The details of agglomeration have drawn a lot of attention and are still under
question \citep[see][]{KempfPfalznerHenning99,PaszunDominik09,WadaEtAl09}, but
the successive agglomeration of planetesimals is commonly accepted.

\subsection{Formation of protoplanets}

The further growth of planetesimals proceeds through mutual collisions,
where the initial phase involves a large number of particles and is 
well described by a coagulation equation \citep{Safronov1969}.
While earlier works \citep[e.g.][]{Naka1983} focussed
mainly on the evolution of the size distribution, subsequent refinements
of the evolution of the random velocities showed that it is important to
evolve the size distribution and the velocity dispersion in a consistent way.
A fixed velocity dispersion is an oversimplification, which changes the growth mode
and increases the growth timescale as well \citep{Wetherill1989}.

The initial growth is quite democratic. All planetesimals grow roughly at the
same rate and the maximum of the size distribution is shifted gently towards
larger sizes. As soon as {gravitational focusing} and {dynamical
friction} become important, the growth mode changes to a qualitatively
different mechanism.  Efficient gravitational focusing leads to a {growth
timescale} (which we denote as $M/\dot M$) that
decreases with mass. Hence larger bodies grow faster than the smaller
planetesimals, a trend that is further supported by energy equipartition due to
planetesimal--planetesimal encounters. This  {dynamical friction} keeps the
largest bodies on nearly circular orbit, thus the relative velocities are small
and gravitational focusing remains efficient. Smaller planetesimals are stirred
up into eccentricity orbits, which slows down their growth rate compared to the
largest bodies.  This accelerated growth, denoted as {runaway growth} of
planetesimals \citep[see e.g][]{Greenberg1978,Wetherill1989,Wetherill1993},
shortens the growth timescale to a few $10^5$~years.  The term {runaway
growth} stresses that the {growth timescale} of a particle decreases with
mass, hence the largest body ``runs away'' to the high mass end of the
distribution \citep[see][]{Koku1995}.

While energy equipartition increases the velocity dispersion with decreasing mass
of the planetesimals, additional damping due to the gaseous disc leads to a turn-over at smaller
sizes. However, higher relative velocities may lead nevertheless to destructive encounters, but these
{fragmentation events} could even speed up the growth \citep{Chambers2006,BromleyKenyon06,BromleyKenyon10}. Since smaller
bodies are more subjected to gas drag, their velocity dispersion is smaller which allows
an efficient accretion by the runaway bodies.
Moreover, smaller particles damp the velocity dispersion of the largest bodies more efficiently.

As runaway proceeds, the system becomes more and more dominated by few big bodies -- the {protoplanets }
. Due to the dominance of few, very large bodies one can not use statistical
methods anymore to simulate the problem.

\cite{KobayashiEtAl10} derived analytical expressions for the final masses of
these planetary ``embryos'', as they call them, including the role of
planetesimal depletion due to collisional disruption.  They conclude that the
final mass in the minimum-mass solar nebula at several AU can achieve $\sim
0.1$ Earth mass within $10^7$ years.

\subsection{Oligarchic growth}

The runaway growth of large bodies (i.\,e. protoplanets) ceases to be efficient
as soon as the protoplanets start to control the velocity dispersion of the
remaining planetesimals in their vicinity.  Gravitational focusing becomes less
effective, therefore the  growth timescale increases with size and the growth
mode changes to {oligarchic growth}. The protoplanets still grow faster
than the field planetesimals\footnote{The term ``field planetesimals'' denotes
in the following the smooth component of smaller planetesimals.}, but the
masses of the protoplanets remain comparable.

A combination of dynamical friction due to the field planetesimals and perturbations
from the neighbouring protoplanets conserves a separation of five to ten Hill radii
between neighbouring bodies. Therefore only planetesimals from a limited area,
the {feeding zone}, are accreted by a given protoplanet. If this zone is
emptied, they have reached their final {isolation mass} \citep{Koku1998,KokuboIda2000,KokuboIda2002}.

As the damping of the remaining field planetesimals is weak enough, further growth is
dominated by mutual perturbations among the protoplanets, which leads
to giant impacts.
Protoplanets beyond the {``snow line''} (or ice condensation point) can grow larger than 5--15 earth masses and
initiate the formation of giant planets. If the protoplanetary disc is very
massive in the inner planetary system, this may lead to an in-situ formation
of hot Jupiters \citep[see][]{Bodenheimer2000}.

\cite{KokuboKominamiIda2006} used the oligarchic growth model of protoplanets
to statistically quantify the giant impact stage (i.e. collisions of
protoplanets to form planets) with $N-$body simulations. They found that for
steeper surface density profiles, large planets usually form closer to the
star.

\subsection{Migration}

The proposed three-stage scenario of planet formation covers the dominant growth processes,
but a major mechanism is still missing -- the {migration} of bodies in the
system. {Migration} is a generic term that summarises a set of different
mechanisms that lead to secular radial drift of bodies \citep[see e.g. the review of][]{Papalo2006,Armitage2010}:
\begin{enumerate}
\item
  The dissipation due to the remaining gas disc leads to an orbital decay of the
  planetesimals. While this poses a severe problem for 1 m--sized objects, larger
  bodies drift very slowly inward. One denotes this process as {type 0} migration.
\item
  Planets which are embedded in a gaseous disc launch spiral density waves at the inner and outer {Lindblad resonances},
  which leads to an exchange of angular momentum with the resonant excited waves.
  This {type I} migration leads to a robust inward migration independent of the
  density profile.  The perturbation from the planet is small, hence
  linear perturbation theory is in principle applicable \citep{Ward1986}, but
  recent work proves that non-linear and radiative
  effects can be quite important \citep{PaardekooperEtAl11}.
\item
  If the protoplanet is massive enough, it opens a gap in the gaseous disc and excites waves through tidal interaction with
  the gaseous disc. An imbalance of the exchange of angular momentum with the inner and outer part
  of the disc leads to {type II} migration. The strong interaction between planet and disc leaves
  the linear regime and requires a numerical solution of the hydrodynamic equations \citep{Lin1979}.
\item
 Even with an opened gap, the planet still channels gas between the inner and outer part
 of the disc. While this {corotational flow} is  to some extent already present during type II
 migration, it dominates the angular momentum balance in the case of {type III} migration due
 to an asymmetry in the leading and trailing part of the flow.
 An imbalance allows for an efficient exchange of angular momentum with
 this corotational gas stream, which gives rise to a remarkably fast migration \citep{MassetPapaloizou03}.
 Type III migration is subjected to a positive feedback: A faster migration increases the asymmetry in the corotational
 flow, which speeds up the migration. Both inward and outward migration are possible.
\end{enumerate}
The four  migration types modify the three-stage scenario in different ways.\\
{\em Type 0} migration is most efficient for small planetesimals (i.\,e. 1 km or smaller).
It poses a severe problem during the early stages of planet formation, as it may
induce a significant loss of solid material, accompanied by a global change in the
initial surface density \citep{Kornet2001}.
The importance of this process  diminishes as planetesimal growth proceeds,
but it still leads to the loss of collisional fragments during the final disc clearing.\\
{\em Type I} is most important for protoplanets (i.\,e. 0.1 earth masses or larger). It leads
to an orbital decay of protoplanets, but this does not only imply a loss of protoplanets, but also breaks the conditions
for isolation. Migrating bodies can accrete along their way through the disc and are thus
not constrained to a fixed feeding zone, which may increase the isolation mass.\\
{\em Type II} and {\em type III} mainly influences giant planets (larger than 10 earth masses)
causing an inward or outward migration, depending on the angular momentum exchange balance. It can explain
the large number of giant planets close to their host star ({hot Jupiters}) found in extrasolar systems. An important
issue is the timescale of the migration process. If the migration is too fast, virtually
all planets spiral inward and leave an empty system behind.

Migration is a powerful process with the capability to reshape an entire planetary
system, at least for gas giants, since terrestrial planets are likely not affected by
migration in gaseous discs \citep{BromleyKenyon11}. However, it also requires some ``parking mechanism'' which terminates migration
before all planets (or protoplanets) are lost. Inhomogeneities in the gaseous disc
may change the crucial momentum balance of the inner and outer part of the disc, thus stalling
or even reversing the drift of a planet.

The migration processes end after the dissipation of the gaseous disc due to
photoevaporation or star-star encounters (i.\,e. after a few $10^6$ to $10^7$
years).

The formation of a planetary system is a vital process that is driven by the
interplay between the different growth phases and the migration of planets and
protoplanetary cores (i.\,e. the precursors of giant planets). While the
preceding sections only summarised the main evolutionary processes, even more
processes could influence the formation of planetary systems. A fast accretion
of giant planets in the outer parts of a planetary system could introduce
further perturbations on the inner part and may even trigger the formation of
terrestrial planets. Moreover, the stellar environment in dense star clusters
and multiple stellar systems also perturbs planet formation, which would
require an even broader view on the problem.
Last but not least, the
excitation due to gravitational coupling to gas surface density fluctuations
plays an important role \citep[see e.g.][]{IdaGuillotMorbidelli08}.

Any approach to planet formation can hardly include this wealth of different
phenomena, thus it is important to focus on a well-defined subproblem.
In this work we focus on the formation of protoplanets for the following reasons:
\begin{enumerate}
\item
 The size, growth timescale and spacing of the protoplanets is a key element
 in the planet formation process.
\item
 The protoplanet growth is well-defined by different growth  modes.
 It starts with the already formed planetesimals ($\approx 1$--$5$ km) and ends,
 when orbital crossing of the protoplanets initiates the final growth phase.
\item
 The planetesimals are large enough to treat the remaining gaseous disc
 as a small perturbation.
\item
 The protoplanets are small enough to neglect tidal interaction with the disc
 in the inner planetary system. Collective  planetesimal--protoplanet
 interaction are also negligible.
\end{enumerate}
Though the protoplanet formation is a well-posed subproblem, our approach
has to incorporate various mechanisms and techniques to cover the full
size range of the problem \citep[see the review of][for a extensive review of the problem and
references therein]{Armitage2010}. However, it is still accessible
to theoretical calculations to some extent which provide a safe ground for the analysis of
the results.

In next section \ref{Theory} we briefly summarise the analytical model that we
use for our initial models. Then we introduce the problem that we will address
later, as well as the timescales and quantities of interest for the general
problem of planetesimal formation: In sections \ref{KeplerOrb}, \ref{HillProb},
\ref{ProtoGrowth} we give a description for a test particle moving around
a body with a central mass $M$, the Hill's problem and the protoplanet growth
from a theoretical point of view, respectively.  

Subsequently, in section \ref{Integrator} and its corresponding subsections, we
ascertain the direct-summation part of the general algorithm, as well as the
required modifications and additions to tailor it to the specific planetesimal
problem. In section \ref{CollModel} we introduce our collisional treatment,
including fragmentation, collisional cascades, migration and coagulation. The
statistical part of the algorithm is described in detail in section
\ref{StatModel}. Finally, in section \ref{HybridCode} we explain how to bring
together both schemes into a single numerical tool and in section
\ref{discussion} we give our conclusions and compare our scheme to other
numerical approaches.

\section{Initial models}
\label{Theory}

The basis for all planet formation models is the structure of the protoplanetary disc.
We summarise the pioneering work of \cite{Hayashi1981} to have a robust initial model at hand.
Subsequent evolution of the disc may change this simple approach, but it is still a
valuable guideline.

A basic estimate of the minimum surface density of solid material in the disc can be deduced
from the mass and location of the present planets in the solar system:
\begin{align}
 \Sigma_{\mathrm{solid} }(r) &=  \left\{
  \begin{array}{rcl}
   7.1 \left( r/\mathrm{1 AU}\right)^{-3/2} \mathrm{g}/\mathrm{cm}^2 & 0.35 \leq r \leq 2.7 \nonumber \\
  30.0 \left( r/\mathrm{1 AU}\right)^{-3/2} \mathrm{g}/\mathrm{cm}^2 & 2.70 \leq r \leq 36.0
        \end{array} \right.
\end{align}
The discontinuity at 2.7 AU stems from the location of the ice condensation point (or {snow line}) that
allows the formation of icy grains in the outer solar system. Furthermore, the total surface density is estimated through
the chemical composition of the disc, which gives the ratio of gas to solids. A fiducial value is 1:0.017 \citep[see][]{Cameron1973}.
The surface density of the gas component is therefore:
\begin{equation}
\Sigma_{\mathrm{gas} }(r)  =    1700 \left( \frac{r}{\mathrm{1 AU}}\right)^{-3/2} \frac{\mathrm{g}}{\mathrm{cm}^2}
\end{equation}
Since the dust content is rather low, the gaseous component is transparent to the visible solar radiation.
Thus the gas temperature follows from the radiation balance:
\begin{equation}
T  =  T_0 \left(\frac{r}{\mathrm{1 AU}} \right)^{-1/2} \left( \frac{L}{L_{\odot}} \right)^{1/4}  \qquad T_0 =280\mbox{ K}
\end{equation}
$L$ is the solar luminosity during the early stages, normalised by the present value $L_{\odot}$.
The three-dimensional density structure is given by an isothermal profile
\begin{equation}
 \rho_{\mathrm{gas}}(r,z) = \frac{\Sigma_{\mathrm{gas}}}{\sqrt{2\pi}h} \exp\left( -\frac{z^2}{2h^2} \right)
\end{equation}
with a radially changing {scale height} $h$:
\begin{align}
 h &=  \frac{c_s}{\Omega} \qquad  c_s = \left(\frac{k_B T}{\mu m_{\mathrm{H}}} \right)^{1/2} \qquad \mu=2.34 \nonumber \\
   &=  \frac{c^{(0)}_s}{\Omega_0} \left(\frac{r}{\mathrm{1 AU}} \right)^{5/4} \left( \frac{L}{L_{\odot}} \right)^{1/8}
   \qquad c^{(0)}_s=  993.56 \frac{\mathrm{m}}{\mathrm{s}}
\end{align}
$c_s$ is the sound velocity of an ideal gas with a mean molecular weight $\mu$ in units of the hydrogen mass $m_{\mathrm{H}}$.
\noindent
Since the density profile is related to a radially varying pressure, the gas velocity  deviates
from the local Keplerian velocity. The balance of forces relates the angular velocity $\Omega_g$
to the pressure gradient:
\begin{equation}
        r\Omega_g^2  =  r\Omega^2+\frac{1}{\rho}\frac{dP}{dr}
\end{equation}
Thus the angular velocity $\Omega_g$ of the gas is \citep[see e.g.][]{Adachi1976}:
\begin{align}
     \Omega_g &=  \Omega \sqrt{1-2\eta_g(r)} \nonumber \\
     \eta_g &=  -\frac{1}{2}\frac{d\ln(\rho_{\mathrm{gas}} c_s^2)}{d\ln(r)}\left(\frac{c_s}{v_K} \right)^2
\end{align}
It is more appropriate to formulate the rotation of the gaseous disc in terms
of a velocity lag $\Delta v_g$ normalised to the local Keplerian velocity $v_K$:
\begin{align}
     \Delta v_g &=  r(\Omega_g - \Omega) \nonumber \\
                &\approx   -\eta_g v_K
\end{align}
A typical value of $\Delta v_g$ for the minimum mass solar nebula at 1 AU is $\Delta v_g = -60$ m$/$s.

This simple model provides a brief description of the initial disc. However, the subsequent
evolution  further modifies the structure of the protoplanetary disc. Since embedded dust grains
are coupled to the gas, it is likely that a global migration of solid material changes the surface
density. Moreover, the dust grains are chemically processed, depending on the local temperature and composition
which introduces additional spatial inhomogeneities. When the growing particles pass the critical
size of $\sim 1$ metre, the strong onset of radial migration may lead to a final reshaping of
the distribution of solid material.
While these restrictions weaken the validity of this approach as the ``true'' initial model, it is still a robust guideline to choose reasonable
surface densities for the solid and the gaseous component after the formation of planetesimals.

\section{Kepler Orbits}
\label{KeplerOrb}

Planetesimals in a protoplanetary disc are subjected to various perturbations: Close encounters
change their orbits, a small but steady gas drag gives rise to a radial drift and accretion
changes the mass of the planetesimals. While all theses processes drive the disc evolution
on a timescale of at least a few thousand years, each planetesimal moves most of the time
on an orbit close to an unperturbed Kepler ellipse. Though the protoplanetary disc introduces
additional perturbations, the central potential dominates for typical disc masses around
$0.01\,M_{\odot}$. Therefore the classical orbital elements still provide a proper framework
to study planetesimal dynamics.

The orbital elements of a test particle moving around a mass $M$ are:

\begin{align}
  E       &=  \frac{v^2}{2} -\frac{GM}{r} \nonumber \\
  a       &=  -\frac{GM}{2E}              \nonumber \\
  e       &=  \sqrt{1-\frac{L^2}{GMa}}    \nonumber \\
  \cos(i) &=  \frac{L_z}{L}
\end{align}

\begin{align}
 {\bf L} &=  {\bf r} \times {\bf v} \nonumber \\
  e\cos(\phi_E) &=  \frac{rv^2}{GM}-1 \nonumber \\
  e\sin(\phi_E) &=  \frac{{\bf r} \cdot {\bf v}}{\sqrt{GMa}}
\end{align}

\noindent
$a$ is the semimajor axis, $e$ is the eccentricity and $i$ is the inclination of the orbit.
As long as no dominant body is structuring the protoplanetary disc, it is justified
to assume axisymmetry. Hence the argument of the perihelion $\omega$, the longitude of the
ascending node $\Omega$ and the eccentric anomaly $\phi_E$  are omitted in the
statistical description.

The deviation of planetesimal orbits from a circle is quite small. Thus it is appropriate to expand
the above set of equations. A planetesimal at a distance $r_0$, in a distance $z$ above the midplane
and with a velocity  $(v_r,v_{\phi},v_z)$ with respect to the local circular velocity $v_K$
has orbital elements (leading order only):
\begin{align}
 a   &\approx   r_0 +2\frac{r_0v_{\phi}}{v_K}          \label{Approx_a} \\
 e^2 &\approx   \frac{v_r^2+4v_{\phi}^2}{v_K^2}        \label{Approx_e} \\
 i^2 &\approx   \frac{z^2}{r_0^2} +\frac{v_z^2}{v_K^2} \label{Approx_i}
\end{align}
These expressions allow us later a convenient transformation between the statistical representation
through orbital elements and the utilisation of a velocity distribution function.

\section{Hill's Problem}
\label{HillProb}

When two planetesimals pass close by each other, they exchange energy and angular momentum
and separate with modified orbital elements. Successive encounters transfer energy between
planetesimals with different masses, driving an evolution of the overall velocity
distribution.

It seems that an encounter is a two-body problem, as there are only two planetesimals
involved, but the central mass has also a major influence turning the problem
into a {three-body encounter}\,\footnote{The term {three-body encounter} does not imply a close passage
of all involved bodies, but emphasises the strong influence of a third one.}
The complexity of the problem is considerably simplified by reducing it to Hill's problem
\citep{Hill1878}.

Consider two masses $m_1$ and $m_2$ that orbit a much larger mass
$M_c$, where both masses are small compared to the central mass $M_c$.
The mass ratio $m_1:m_2$ could be arbitrary. This special type of
a three body problem is denoted as Hill's problem, originally
devised to calculate the orbit of the moon. It provides a
convenient framework to examine planetesimal encounters in the
potential of a star. The equations of motion\footnote{The 
following derivation is quite common to the literature 
\citep[see e.g.][]{Ida1990,Henon1986}. The later work uses a slightly different scaling.} 
of the two
planetesimals including the central potential and their mutual
interaction are:
\begin{align}
  {\bf\ddot r}_1 &=  -{\bf r}_1 \frac{GM_c}{r_1^3}-({\bf r}_1-{\bf r}_2)\frac{Gm_2}{r_{12}^3} \nonumber \\
  {\bf\ddot r}_2 &=  -{\bf r}_2 \frac{GM_c}{r_2^3}-({\bf r}_2-{\bf r}_1)\frac{Gm_1}{r_{12}^3}
\end{align}
We now introduce the relative vector
${\bf r}$ and the centre-of-mass ${\bf R}$:
\begin{equation}
  {\bf r} = {\bf r}_2 - {\bf r}_1
  \qquad {\bf R} = \frac{ m_2 {\bf r}_2 + m_1{\bf r}_1 }{m_1+m_2}
\end{equation}
Furthermore, the equations of motion are transformed to a
corotating set of coordinates which are scaled by the {mutual
Hill radius} $ r_{\mathrm{Hill}}$  and the local Kepler frequency
$\Omega$
\begin{align}
  x &=  \frac{r'-a_0}{ r_{\mathrm{Hill}}} \nonumber \\
  y &=  \frac{a_0(\phi-\Omega t)}{ r_{\mathrm{Hill}}} \nonumber \\
  z &=  \frac{z'}{ r_{\mathrm{Hill}}} \nonumber \\
  r_{\mathrm{Hill}} &=  a_0 \sqrt[3]{\frac{m_1+m_2}{3M_c}}
\end{align}
where $(r',\phi,z')$ are heliocentric cylindrical coordinates.
$a_0$ is the radius of a properly chosen reference orbit, while
the {mutual Hill radius} $r_{\mathrm{Hill}}$ is an
intrinsic length scale of the problem : Since the two orbiting planetesimals are small
compared to the central star, the Hill radius is much smaller than
the size $a_0$ of the reference orbit. Hence it is possible to
expand the central potential about the reference orbit. This
yields approximate equations for the relative motion and the
centre-of-mass:

\begin{align}
\ddot x &=   2\dot y +3x -3x/r^3  \nonumber \\
\ddot y &=  -2\dot x    -3y/r^3 \nonumber \\
\ddot z &=   -z    -3z/r^3  \label{EqX1}
\end{align}

\begin{align}
  \ddot X &=   2\dot Y +3X  \nonumber \\
  \ddot Y &=  -2\dot X     \nonumber \\
  \ddot Z &=   -Z  \label{EqX2}
\end{align}
 
The equations~\ref{EqX1} and \ref{EqX2}  have some interesting
properties: Firstly, the centre-of-mass motion  separates from the
interaction of the two bodies. Secondly, the scaled relative
motion is independent of the masses  $m_1$ and $m_2$, implying a
fundamental similarity of planetesimal encounters\footnote{section~\ref{StatModel}
makes extensive use of this property.}. As
Eq.~\ref{EqX2} is a simple linear differential equation, one
readily obtains the solution of the centre-of-mass motion
\begin{align}
  X &=  b-e\cos(t-\tau) \nonumber \\
  Y &=  -\frac{3}{2}b t+\psi +2e\sin(t-\tau) \nonumber \\
  Z &=  i\sin(t-\omega)    \label{Eq_CM_S}
\end{align}
which is equivalent to a first-order expansion of a Kepler ellipse.
$\omega$ and $\tau$ are the longitudes of the ascending node and
the pericentre, while $e$ and $i$ are the eccentricity and
inclination scaled by the {reduced} (i.\,e. dimensionless) {\em
Hill radius}  $r_{\mathrm{Hill}}/a_0 $. The value of $b$ depends on
the choice of the reference orbit, but it is natural to set $b=0$
which implies that the centre-of-mass defines the reference orbit.
\begin{figure}
\begin{center}
\includegraphics[scale=0.4]{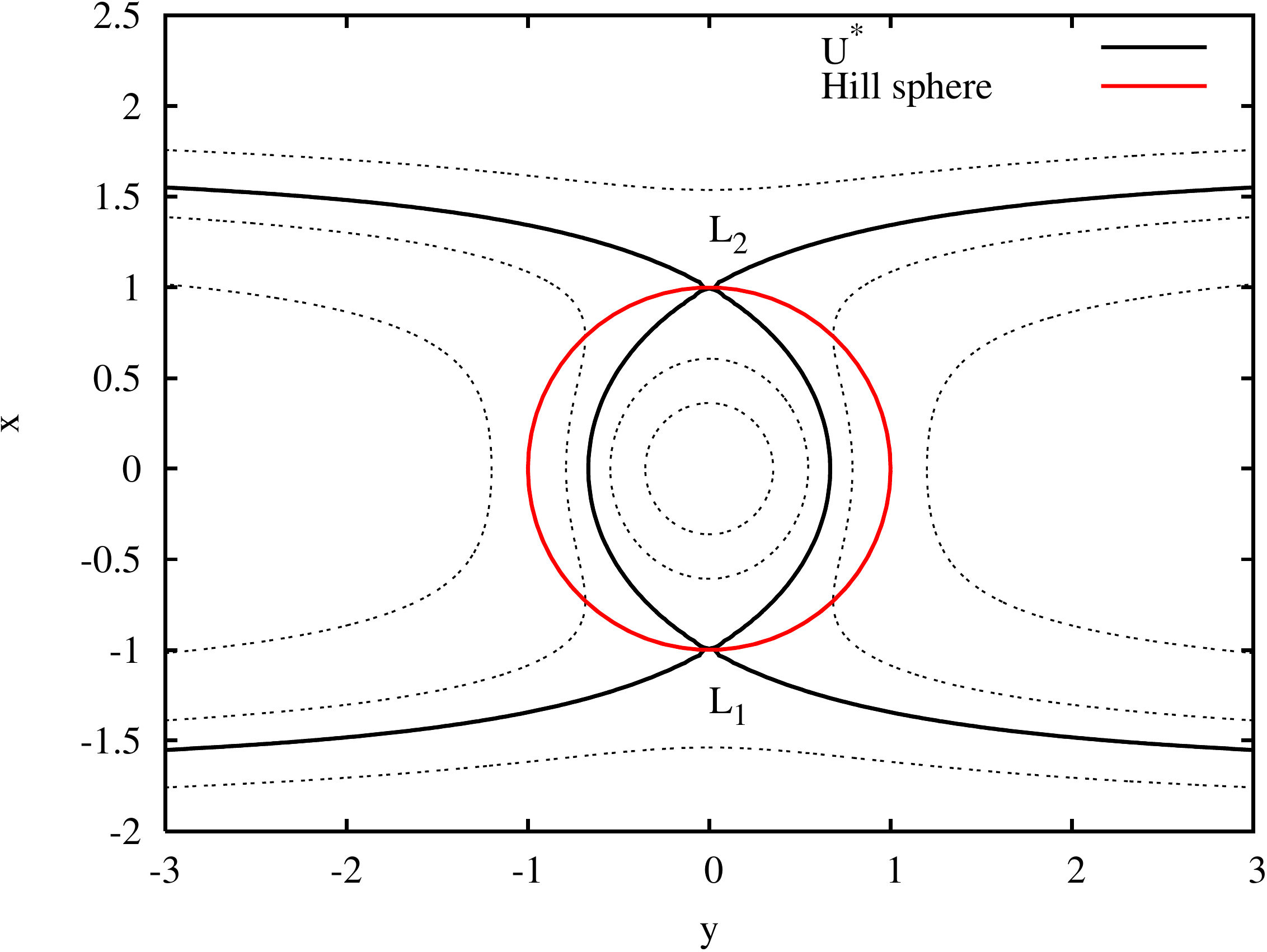}
\end{center}
\caption[Hill sphere]{Equipotential lines for the effective potential U
at $z=0$ (see Eq.~\ref{U_eff}). $U=U^*$ refers to the largest
allowed volume, which is enclosed by the Hill sphere and the two
Lagrange points $L_1$ and $L_2$.\label{Hill_Plot}}
\end{figure}

While the nonlinear nature of the relative motion (see Eq.~\ref{EqX1})
prevents any general analytical solution, Eq.~\ref{Eq_CM_S}
provides at least an asymptotic solution for a large
separation of the planetesimals, where $b$ can be interpreted as
an {impact parameter}. Nevertheless, small $b$ do not necessarily imply
close encounters, as opposed to the standard definition of the
impact parameter. However, the expression $b=a_2-a_1$ provides a
measure of the distance of the two colliding bodies without
invoking the complicated encounter geometry. A special solution to Eq.~\ref{EqX1} are
the unstable equilibrium points  $L_1$ and $L_2$ at $(x,y,z)=(\pm 1,0,0)$, denoted as
{Lagrange points}\,\footnote{The additional Lagrange points $L_3$--$L_5$ are missing due
to the Hill approximation.}. In
addition, an inspection of Eq.~\ref{EqX1} reveals that the {\em
Jacobi energy} $E_J$ is conserved:
\begin{equation}
  E_J  =  \frac{1}{2}\left( \dot x^2 +\dot y^2 +\dot z^2 + z^2 -3x^2 \right)-\frac{3}{r}
\end{equation}
Since the kinetic energy is always a positive quantity, the
following inequality holds:
\begin{equation}
   E_J   \geq U = \frac{1}{2}\left(z^2 - 3x^2  \right)-\frac{3}{r} \label{U_eff}
\end{equation}
Thus the allowed domain of the particle motion is enclosed by the
equipotential surfaces of the effective potential $U$. A subset of
these equipotential surfaces restricts the allowed domain to the
vicinity of the origin (see Fig.~\ref{Hill_Plot}). The largest of
these surfaces passes through the Lagrange points $L_1$ and
$L_2$. Hence we identify the Hill radius  (or Hill sphere) as the
maximum separation which allows the bound motion of two
planetesimals\footnote{The same argument applies to  the tidal
boundary in cluster dynamics or the Roche lobe in stellar
dynamics, which are equivalent to the Hill sphere.}.

Beside the numerical solution of the equations of motion, it is
useful to define {osculating} (or {instantaneous})
orbital elements

\begin{align}
  b &=  4x+2\dot y \nonumber \\
  i^2 &=  z^2+{\dot z}^2 \nonumber \\
  e^2 &= {\dot x}^2+ (3x+2\dot y)^2
\end{align}

\begin{align}
 \psi &=  - 2\dot x +y+\frac{3}{2}b t \nonumber \\
  \omega &=  t-\arctan( z ,\dot z )\nonumber \\
  \tau   &=  t-\arctan( \dot x ,3x+2\dot y )
\end{align}

\begin{equation}
E_J  =  \frac{1}{2}(e^2+i^2)-\frac{3}{8}b^2-\frac{3}{r}
\end{equation}

\noindent which provide a convenient description of the initial
relative orbit and the modified orbit after the encounter.

\section{Protoplanet Growth}
\label{ProtoGrowth}

Our work follows the evolution of a planetesimal disc into few protoplanets, including the
full set of interaction processes. Hence we summarise the main aspects
of protoplanet growth first to provide a robust framework.

Although the sizes of the planetesimal cover a wide range, they virtually form two main groups: The smaller field
planetesimals and the embedded protoplanets (or their precursors). This {two-group approximation} \citep[e.g.][]{Wetherill1989,Ida1993b}
allows one to have a clearer insight into the growth process.

During the initial phase all planetesimals share the same velocity dispersion independently
of their mass. The initial random velocities are low enough for an efficient gravitational focusing.
Hence, the growth rate of a protoplanet with mass $M$ radius $R$ can be estimated as \citep[e.g.][]{Ida1993}:
\begin{equation}
 \dot M  \approx  v_{\mathrm{rel}}\frac{\Sigma}{H}\pi R^2\left( 1+\frac{2GM}{Rv_{\mathrm{rel}}^2} \right) \label{AccM}
\end{equation}
Eq.~\ref{AccM} is the {two-body accretion rate}, which should be modified due to the gravity of the
central star. Nevertheless this approximation gives an appropriate description to discuss the basic properties of
the growth mode. The scale height $H$ (Eq.~\ref{ScaleHeight}) and the relative velocity $ v_{\mathrm{rel}}$ are related
to the mean eccentricity $e_m =\sqrt{\langle e^2\rangle }$
of the field planetesimals:
\begin{equation}
  H  \approx   v_{\mathrm{rel}}/\Omega \qquad v_{\mathrm{rel}}\approx e_m a\Omega
\end{equation}
Thus the accretion rate in the limit of strong gravitational focusing ($2GM/R \gg v_{\mathrm{rel}}^2$) is:
\begin{align}
 \dot M &\approx  2\pi  R \frac{\Sigma GM}{a^2\Omega e_m^2} \label{RunAcc} \\
        &\propto   M^{4/3}
\end{align}
If the protoplanets are massive enough, they start to control the velocity dispersion
of the planetesimals in their vicinity. The width $\Delta a$ of this sphere of influence, the {heating zone},
is related to the Hill radius $R_{\mathrm{Hill}}$ of the protoplanet \citep[][]{Ida1993b}:
\begin{align}
\Delta a &=  \Delta \tilde a R_{\mathrm{Hill}} = 4R_{\mathrm{Hill}} \sqrt{\frac{4}{3}(\tilde e_m^2+\tilde i_m^2)+12} \nonumber \\
       h &=  \sqrt[3]{\frac{M}{3M_c}}
\end{align}
$\tilde e_m$ and $\tilde i$ are eccentricity and inclination of the field planetesimals, scaled by the {reduced
Hill radius} $h$ of the protoplanet. $M_c$ is the mass of the central star. The condition that the protoplanet controls the
velocity dispersion of the field planetesimals reads \citep[][]{Ida1993}:
\begin{equation}
\frac{2M^2}{2\pi a\Delta a }  >  \Sigma m
\end{equation}
This condition is equivalent to a lower limit of the protoplanetary mass:
\begin{equation}
 \frac{M}{m}  >  \left(\frac{\pi\Delta\tilde a}{\sqrt[3]{3}} \right)^{3/5} \left(\frac{\Sigma a^2}{M_c} \right)^{3/5}
                 \left(\frac{m}{M_c} \right)^{3/5} \label{Mheat}
\end{equation}
$M/m$ depends on several parameters, but reasonable values yield $M/m \approx 50$--$100 $. The velocity
dispersion in  the heated region is roughly
\begin{equation}
  v  \approx   R_{\mathrm{Hill}} \Omega \label{v_Heat}
\end{equation}
which gives an interesting relation to the condition that leads to {gap formation}.
A protoplanet can open a gap in the planetesimal component if it is larger than a
critical mass $M_{\mathrm{gap}}$ \citep{Rafikov2001}

\begin{eqnarray}
 \frac{M_{\mathrm{gap}}}{M_c} & \approx &
 \left\{
 \begin{array}{lll}
 \frac{\Sigma a^2}{M_c} \left(\frac{m}{M_c}\right)^{{1}/{3}} & \mbox{if} & v \lesssim \Omega r_{\mathrm{Hill}} \\
 \frac{\Sigma a^2}{M_c} \left(\frac{m}{M_c}\right)^{1/3}\left(\frac{\Omega r_{\mathrm{Hill}}}{v}\right)^2 & \mbox{if} & v \gg \Omega r_{\mathrm{Hill}}
 \end{array}     \right. \label{Mgap}
\end{eqnarray}

\noindent
where $r_{\mathrm{Hill}}$ is the Hill radius of the field planetesimals.
If the velocity dispersion $v$ is controlled by the protoplanet, Eq.~\ref{Mgap} together with Eq.~\ref{v_Heat} demonstrate
that the condition for gap formation is equivalent to Eq.~\ref{Mheat}, i.\,e. the efficient heating of  the field planetesimals
implies gap formation and vice versa.
The higher velocity dispersion of the field planetesimals (see Eq.~\ref{v_Heat}) reduces the growth rate given by equation~\ref{RunAcc} to

\begin{align}
\dot M & \approx   6\pi\Sigma \Omega \frac{RR_{\mathrm{Hill}}}{\tilde e_m^2} \label{MdotOli} \\
       &\propto   M^{2/3}
\end{align}

\noindent
Different mass accretion rates imply different growth mode. If two protoplanets
have different masses $M_1$ and $M_2$, their mass ratio evolves as:
\begin{equation}
 \frac{d}{dt} \frac{M_2}{M_1}   =   \frac{M_2}{M_1}\left(\frac{\dot  M_2}{M_2}-\frac{\dot M_1}{M_1} \right)
\end{equation}
When the growth timescale $M/\dot M$ decreases with mass, a small mass difference increases
with time. This is the case for Eq.~\ref{RunAcc}, which gives rise to {runaway accretion}. As soon
as the protoplanets control the velocity dispersion of the field planetesimals, the growth
timescale increases with mass and therefore the protoplanet masses become more similar as they grow {oligarchically}.

The field planetesimals also damp the excitation due to protoplanet--protoplanet interactions
and keep them on nearly circular orbits. The balance between these scatterings and the dynamical
friction due to smaller bodies establishes a roughly constant orbital separation $b$
\citep{Koku1997}:
\begin{align}
  b &=   R_{\mathrm{Hill}} \sqrt[5]{\frac{7{\tilde e_m}^2M }{2\pi\Sigma a R}  } \label{b_sep} \\
  \tilde b &=  b/R_{\mathrm{Hill}}
\end{align}
$R$ is the radius of the protoplanet, $M$ is its mass and $\tilde e$ is the reduced eccentricity of the field
planetesimals. The stabilised spacing prevents collisions between protoplanets, but it also
restricts the {feeding zone} -- the area from which a protoplanet accretes. If all matter
in the feeding zone is accreted by the protoplanet, it reaches its final {isolation mass} \citep{KokuboIda2000}:
\begin{equation}
  M_{\mathrm{iso}}  =  2\pi b a \Sigma
\end{equation}
Inserting Eq.~\ref{b_sep} yields the isolation mass in units of the mass of the host star $M_c$:
\begin{align}
  M_{\mathrm{iso}}/M_c & = (112\pi^4)^{3/8}\left(\frac{1}{3}\right)^{5/8}\left(\frac{4\pi}{3}\right)^{1/8}
                       \left({\tilde e_m}^2\right)^{3/8} \nonumber \\
                       & \left( \frac{a^2\Sigma}{M_c}\right)^{3/2}\left( \frac{a^3\rho}{M_c}\right)^{1/8}
          \approx  19.67 \times  \left({\tilde e_m}^2\right)^{3/8} \nonumber \\
                       & \left( \frac{a^2\Sigma}{M_c}\right)^{3/2}\left( \frac{a^3\rho}{M_c}\right)^{1/8}
 \end{align}

There is a weak dependence on the density $\rho$ of the protoplanet, but the most important
parameter is the surface density $\Sigma$. If we take the minimum mass solar nebula
as an example, the radial dependence of the surface density implies an isolation mass that
grows with increasing distance to the host start. Hence protoplanets beyond some critical radial distance
are massive enough \citep[larger than $\approx 15\, M_{\oplus}$][]{Boden1986}
to initiate gas accretion from the protoplanetary disc.
\begin{table}
\begin{center}
\begin{tabular}{|c|c|c|} \hline
$\Sigma[$g$/$cm$^2]$ & $M_{\mathrm{iso}}/M_{\odot}$ & $M_{\mathrm{iso}}/M_{\oplus}$ \\ \hline
 2     & $3.91\times 10^{-8}$   & 0.013 \\ \hline
 10    & $4.33\times 10^{-7}$   & 0.144 \\ \hline
 100   & $1.37\times 10^{-5}$   & 4.548 \\ \hline
\end{tabular}
\end{center}
\caption[Isolation masses]{Isolation mass for different surface densities at $r=$ 1 AU and $M_c = 1\, M_{\odot}$.}
\end{table}

As the protoplanets approach the isolation mass, interactions with the gaseous disc
and neighbouring protoplanets become increasingly important. We estimate the onset
of orbit crossing by a comparison of the perturbation timescale $\tau_{\mathrm{pert}}$
of protoplanet--protoplanet interactions with the damping timescale $\tau_{\mathrm{damp}}$
due to planetesimal--protoplanet scatterings. Since the protoplanets are
 \noindent
well separated ($\tilde b \approx 5\dots 10$),
it is possible to apply perturbation theory \citep[see e.g.]{Petit1986}:
\begin{equation}
 \tau_{\mathrm{pert}}   \approx   \frac{\tilde b^5}{7 h \Omega}
\end{equation}
We anticipate section~\ref{StatModel} (see Eq.~\ref{TriDynF}) to derive the damping timescale
\begin{equation}
\tau_{\mathrm{damp}}  \approx  \frac{1}{2}  \frac{T_r^{3/2}}{ \sqrt{2\pi} G^2 \ln(\Lambda) (M+m)n_0 m }
\end{equation}
where $T_r$ and $T_z$ are the radial and vertical velocity dispersion of the field planetesimals.
Hence the criterion for the onset of orbital crossing is:
\begin{equation}
  \tau_{\mathrm{pert}}   <   \tau_{\mathrm{damp}}
\end{equation}
As the protoplanets control the velocity dispersion of the field planetesimals (see Eq.~\ref{v_Heat}),
this condition reduces to:
\begin{align}
 \Sigma_M &>  \Sigma_m \ln(\Lambda) \frac{72}{7\pi} \left(\frac{\tilde b}{\tilde e}\right)^4 \nonumber \\
          &>  \Sigma_m \times f \label{ProtoCross}
\end{align}
Thus orbital crossing sets in when the mean surface density $\Sigma_M$ of the protoplanets
exceeds some fraction $f$ of the field planetesimal density $\Sigma_m$. While the factor
$f$ depends strongly on the separation $\tilde b$ of the protoplanets, a fiducial value
is $f \approx 1$, in agreement with the estimates of \cite{Goldreich2004}.

The onset of migration and the resonant interaction of protoplanets with the disc
and other protoplanets terminates the local nature of the protoplanet accretion
process and requires a global evolution of the planetary system. While the final stage
deserves a careful analysis, further research is beyond the scope of this work.

\section{Direct-summation techniques from the standpoint of planetary dynamics: First steps}
\label{Integrator}

The protoplanet formation is essentially an $N$--body problem. Although we seek
for a more elaborated solution to this problem which benefits from statistical
methods, the pure $N$--body approach is a logical starting point. Direct
calculations with a few thousand bodies have provided us with a valuable
insight into the growth mode \citep[see e.g.]{Ida1992,Koku1996}, but they are
also powerful guidelines that help developing other techniques.  Statistical
calculations rely on a number of approximations and ``exact'' $N$--body
calculations provide the necessary, unbiased validation of the derived formula.

The choice of the integrator is a key element in the numerical solution of the
equations of motion. Our requirements are the stable long-term integration of a
few ten thousand planetesimals with the capability of treating close
encounters, collisions and the perspective to evolve it into an improved hybrid
code.  Approximative methods like the Fast Multipole Method or Tree codes
have a scaling of the computational time close to $N$, but the accuracy in
this regime is too poor to guarantee the stable integration of Keplerian orbits
\citep[compare the discussion in][]{HernquistHutMakino93,Spurzem1999}.

The class of exact methods (all scale asymptotically with $N^2$) roughly divides in two parts:

\begin{enumerate}

\item Symplectic methods (see e.g. \citealt{Wisdom1991} or the {\sc Symba} code, \citealt{Duncan1998}) rely on a careful expansion
of the Hamiltonian which guarantees that the numerical integration follows a perturbed Hamiltonian.
While there is still an integration error, all properties of a Hamiltonian system like conservation
of phase-space volume are conserved by the numerical integration. The drawback of these very
elegant methods is that the symplecticity is immediately broken by adaptive time steps, collisions
or complicated external forces if no special precaution is taken. Even the numerical truncation
error breaks the symplecticity to some extend \citep[][]{Skeel1999}.
\cite{MooreQuillen2010} recently developed a parallel integrator for graphics processing unit (GPU)
that uses symplectic and Hermite algorithms according to the resolution needed. The algorithm is
similar to {\sc Symba}, but less accurate.

\item The second group represents the ``classical'' methods that are based on Taylor expansions of
the solution. They come in different flavours like implicit methods, predictor--corrector
integrators or iterated schemes. Time symmetric methods stand out among these different
approaches, as they show no secular drift in the energy error. These integrators are so
well-behaved that one may even call them ``nearly symplectic''.

\end{enumerate}

Taking all the requirements into account we have chosen {\sc
Nbody6++}\footnote{\cite{Aarseth1999} gives a nice review on the remarkable
history of the {\sc Nbody}-codes, more details are given in
\cite{Aarseth2003}.}, an integrator which is the most recent descendant from
the famous $N$--body code family from S. Aarseths' ``factory''.  This version
was parallelised by \cite{Spurzem1999}, which opened the use of current
supercomputers.

This parallel version, named {\sc Nbody6++}, offers many versatile features that were
included over the past years and more and more refined as time passed by. While
all these elaborated tools deserve attention, we restrict ourselves for
brevity to the components which contribute to the planetesimal problem.  The
main components of the code are:

\begin{enumerate}
 \item
   Individual time steps and a block time step scheme.
 \item
   Ahmad--Cohen neighbour scheme.
 \item
   Hermite scheme.
 \item
   KS--Regularisation for close two-body encounters.
\end{enumerate}

\noindent
We will explain each of these features and highlight their advantages for the
main goal of this work. This is important to understand the immediate first
level of modifications required to adapt the numerical scheme to planetary
dynamics. We present later, in section \ref{ChapOpt}, the next level of
complexity in the modifications to be carried out to mold the numerical tool to
our purposes.

\subsection{Individual Time Steps}

The choice of the time step controls the accuracy as well as the efficiency of any given
integrator. Too small time steps slow down the integration without necessity, whereas
too large values increase the error. An efficient solution to this dilemma is an {adaptive
time step} that is adjusted after each integration step according to a specified accuracy limit.
While the idea is quite clear, there is no unique receipt how to choose the proper time step.
A common approach for $N$--body systems is to use the parameters at hand (like particle velocity,
force, etc.) to derive a timescale of the particle motion. This procedure leaves enough space for a wealth
of different time step criteria. {\sc Nbody6++} uses the standard Aarseth expression \citep{Aarseth1985}
\begin{equation}
 \Delta t  =  \sqrt{\eta \frac{F F^{(2)}+(F^{(1)})^2 }{ F^{(1)} F^{(3)}+ ( F^{(2)})^2}  } \label{AarsethDt}
\end{equation}
which makes use of the force and the time derivatives up to third order.

The time step choice is not unique in multi-particle systems. One solution is
to take the minimum of all these values as a {shared (or global) time step}, but this
is not recommended unless all individual steps are of the same size.

The second option is to evolve each particle track with its own, {individual time step}.
This method abandons the convenient synchronisation provided by a global step, therefore
each integration of a particle demands a synchronisation of all particles through predictions.
Since the prediction of all particles is $\mathcal{O}(N)$, it is counter-balanced by the saving
of force computations. Nevertheless the overhead is still significant, so a further optimisation
might be desired. The basic idea of the {block time step} method is to force particles
in groups that are integrated together, which reduces the number of necessary predictions
by a factor comparable to the mean group size. These groups are enforced through two
constraints. The first condition is a discretisation of the steps in powers of two:
\begin{equation}
  \Delta t  =  2^{-k} \qquad k \geq 0 \label{dTdiscret}
\end{equation}
This condition increases the chances that two different particles share the same timelevel,
but it also reduces roundoff errors since the time steps are now exactly representable numbers.
The second condition locks the ``phases'' of particles with the same time step
\begin{equation}
  T_i  \equiv   0 \mbox{ mod } \Delta t_i  \label{Phases}
\end{equation}
i.\,e. the particle time $T_i$ is an integer multiple of the actual time step $\Delta t_i$
so that all particles with the same step share the same block. Note that a step can not be increased
at any time $T_i$, but only when the second condition (Eq.~\ref{Phases}) for the larger step is met.

\subsection{Ahmad--Cohen Neighbour Scheme}

\begin{figure}
\resizebox{\hsize}{!}
          {\includegraphics[scale=1,clip]{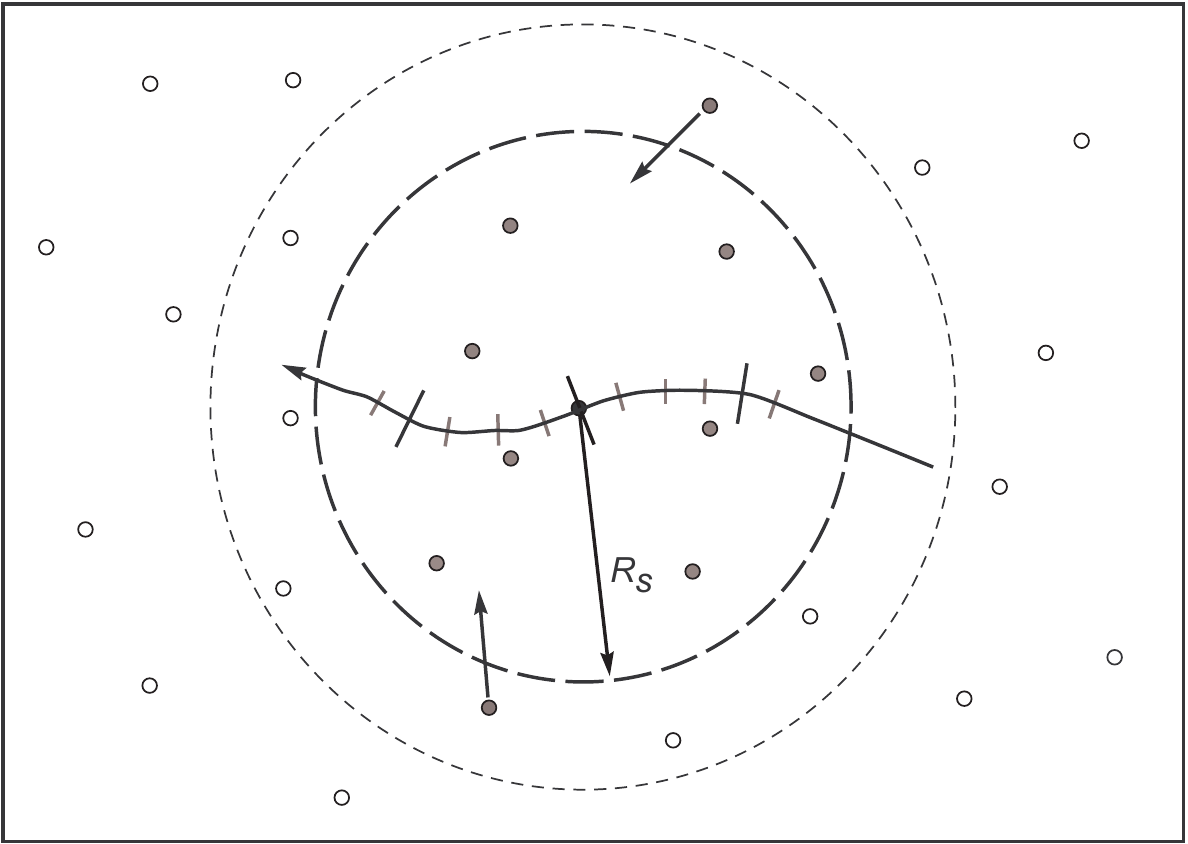}}
\caption
   {  
 All particles inside the neighbour sphere $R_s$ are selected
as neighbours (filled circles). As the neighbour list is fixed during regular steps
(marked by black lines), a second shell includes possible intruders.  
   }
\label{ScetchNeighb}
\end{figure}

The first $N$--body codes calculated always the full force (i.\,e. summation over
all particles) to integrate a particle. But not all particles contribute with the same
weight to the total force. Distant particles provide a smooth, slowly varying {regular force},
whereas the neighbouring particles form a rapidly changing environment which gives
rise to an {irregular force}. The Ahmad-Cohen neighbour scheme \citep{AhmadCohen73} takes advantage
of this spatial hierarchy by dividing the surrounding particles in the already mentioned
two groups according to the {neighbour sphere radius} $R_s$. Both partial
forces fluctuate on  different timescales, which are calculated according to Eq.~\ref{AarsethDt}.
The key to the efficiency of the method is the inequality
\begin{equation}
 \Delta t_{\mathrm{irr}}  \ll  \Delta t_{\mathrm{reg}}
\end{equation}
Regular forces are extrapolated between two full force calculations
\begin{equation}
{\bf F}_{\mathrm{reg}}  =  {\bf F}_{\mathrm{reg}}^{(0)}+\Delta t \, {\bf F}_{\mathrm{reg}}^{(1)}+
    \frac{1}{2}(\Delta t)^2 {\bf F}_{\mathrm{reg}}^{(2)}
   +\frac{1}{6}(\Delta t)^3 {\bf F}_{\mathrm{reg}}^{(3)}
\end{equation}
with a third order accurate expression, whereas the plain {\sc Nbody6++} uses only linear
extrapolation (see the next sections for a detailed discussion).

\subsection{Hermite Scheme}

The {Hermite scheme} is a fourth-order accurate integrator that was applied first
by \cite{Makino1992} to the integration of a planetesimal system. This scheme is used
to integrate single particles and CM-bodies\footnote{CM denotes centre--of--mass, see the section on KS-Regularisation for more details.}
in the main part of {\sc Nbody6++}. It is accomplished by a predictor and a corrector step.
The prediction is second order accurate:
\begin{align}
  {\bf x}_p &=  {\bf x}_0 + {\bf v}_0\Delta t + \frac{1}{2}{\bf a}_0(\Delta t)^2+\frac{1}{6}{\bf\dot a}_0(\Delta t)^3 \nonumber \\
  {\bf v}_p &=  {\bf v}_0 + {\bf a}_0\Delta t + \frac{1}{2}{\bf\dot a}_0(\Delta t)^2
\end{align}
Now the acceleration is evaluated at the predicted position to derive the second and third order
time derivatives of the force:
\begin{align}
{\bf a}^{(2)}_0 &=  \frac{-4 {\bf\dot a}_0 -2{\bf\dot a}_p }{\Delta t}
                          +\frac{-6 {\bf a}_0 + 6 {\bf a}_p }{\Delta t^2} \label{a2Int} \\
{\bf a}^{(3)}_0 &=  \frac{6 {\bf\dot a}_0 + 6{\bf\dot a}_p }{\Delta t^2}
                          +\frac{12 {\bf a}_0 - 12 {\bf a}_p }{\Delta t^3} \label{a3Int}
\end{align}
 Using the additional derivatives one can improve the prediction:
\begin{align}
{\bf x}_c &= {\bf x}_p + \frac{1}{24}{\bf a}^{(2)}_0(\Delta t)^4
                    + \frac{1}{120} {\bf a}^{(3)}_0(\Delta t)^5+ \mathcal{O}(\Delta t^6) \nonumber \\
          &= {\bf x}_p + \left(-\frac{3}{20}{\bf a}_0 +\frac{3}{20}{\bf a}_p \right) (\Delta t)^2 -
                \left( \frac{7}{60}{\bf\dot a}_0 +\frac{1}{30}{\bf\dot a}_p \right) \nonumber \\
          & (\Delta t)^3 + \mathcal{O}(\Delta t^6)  \\
{\bf v}_c &= {\bf v}_p + \frac{1}{6}{\bf a}^{(2)}_0(\Delta t)^3 +
                  \frac{1}{24} {\bf a}^{(3)}_0(\Delta t)^4 + \mathcal{O}(\Delta t^5)  \nonumber \\
          &= {\bf v}_p + \frac{1}{2}(-{\bf a_0}+{\bf a_p} )\Delta t
              + \left(-\frac{5}{12}{\bf\dot a}_0-\frac{1}{12}{\bf\dot a}_p\right)(\Delta t)^2 + \nonumber \\ 
          & \mathcal{O}(\Delta t^5)
\end{align}
The corrected positions are fourth order accurate. While the Hermite scheme is robust and stable,
even in combination with the neighbour scheme, it is not accurate enough to integrate planetesimal
orbits efficiently.

\subsection{Hermite Iteration}

The Hermite scheme bears the potential for a powerful extension, since it is a predictor--corrector
scheme. An essential part of this scheme is the calculation of the new forces with the predicted
positions, but it should improve the accuracy if they are recalculated using the corrected positions.
The new forces are readily used to improve the corrected values, which closes the scheme to an iteration loop --
the {Hermite iteration} \citep[][]{Koku1998}.

There are only few modifications necessary to obtain the iterated scheme.
The particle prediction remains second-order accurate:
\begin{align}
{\bf x}_p &=  {\bf x}_0 + {\bf v}_0\Delta t + \frac{1}{2}{\bf a}_0(\Delta t)^2
                  +\frac{1}{6}{\bf\dot a}_0(\Delta t)^3 \nonumber \\
{\bf v}_p &=  {\bf v}_0 + {\bf a}_0\Delta t + \frac{1}{2}{\bf\dot a}_0(\Delta t)^2
\end{align}
The force and its first time derivative are calculated to derive higher derivatives according to Eq.~\ref{a2Int} and \ref{a3Int}:
\begin{align}
 {\bf a}_p &=  {\bf f}({\bf x}_p,{\bf v}_p) \label{a_p}  \\
{\bf\dot a}_p &=  {\bf f}({\bf x}_p,{\bf v}_p)  \label{a_pdot}
\end{align}
The new corrector omits the highest order term in the position, making the velocity and the position
to the same order accurate:
\begin{align}
{\bf x}_c &=  {\bf x}_p + \frac{1}{24} {\bf a}^{(2)}_0(\Delta t)^4 +  \mathcal{O}(\Delta t^5)  \nonumber \\
      &=  {\bf x}_p + \frac{1}{4}( {\bf a}_p - {\bf a}_0 ) (\Delta t)^2 +
                \left( -\frac{1}{6}{\bf\dot a}_0 -\frac{1}{12}{\bf\dot a}_p \right)\nonumber \\ 
      & (\Delta t)^3 + \mathcal{O}(\Delta t^5) \\
{\bf v}_c &=  {\bf v}_p + \frac{1}{2}(-{\bf a}_0+{\bf a}_p)\Delta t
              + \left(-\frac{5}{12}{\bf\dot a}_0-\frac{1}{12}{\bf\dot a}_p\right)\nonumber \\
          & (\Delta t)^2 + \mathcal{O}(\Delta t^5)
\end{align}
It seems unreasonable to drop one order in the position, but a reformulation of the predictor--corrector
step reveals that this slight change yields a time symmetric scheme:
\begin{align}
{\bf v}_c &=  {\bf v}_0 +\frac{1}{2}({\bf a}_p+{\bf a}_0)\Delta t
                   -\frac{1}{12}({\bf\dot a}_p-{\bf\dot a}_0 )(\Delta t)^2 \nonumber \\
{\bf x}_c &=  {\bf x}_0 +\frac{1}{2}({\bf v}_c+{\bf v}_0)\Delta t -\frac{1}{12}({\bf a}_p-{\bf a}_0)(\Delta t)^2
\end{align}
The iteration is achieved by returning to Eq.~\ref{a_p}--\ref{a_pdot} with the predicted positions replaced
by the more accurate values ${\bf x}_c,{\bf v}_c$. Although the integration does not need the second and third time
derivative of the forces explicitly, it is useful to provide them at the end of the iteration for the
calculation of the new time step:
\begin{align}
{\bf a}^{(2)} (t+\Delta t ) &=  \frac{2 {\bf\dot a}_0 + 4{\bf\dot a}_p }{\Delta t}
                           +\frac{6 {\bf a}_0 - 6 {\bf a}_p }{\Delta t^2} \label{F2} \\
{\bf a}^{(3)} (t+\Delta t ) &=  \frac{6 {\bf\dot a}_0 + 6{\bf\dot a}_p }{\Delta t^2}
                           +\frac{12 {\bf a}_0 - 12 {\bf a}_p }{\Delta t^3} \label{F3} 
\end{align}
Numerical tests show that convergence is reached after one or two iterations, making this
approach very efficient. It needs less force evaluations than the Hermite scheme for
the same accuracy. Moreover, its time symmetry suppresses a secular drift of the energy
error.

\subsection{Extended Hermite Scheme}

The Hermite scheme is an integral part of {\sc Nbody6++} and proved its value
in many applications.  It would have been natural to improve the performance
with an additional iteration, but our first tentative implementations showed
rather negative results: The iterated scheme was more unstable, slower and even
less accurate than the plain Hermite scheme. An inspection of the code
structure revealed that the  Ahmad--Cohen neighbour scheme is the cause of this.

Each particle integration is composed of two parts -- frequent neighbour force
evaluations and less frequent total force evaluations including derivative corrections.
Every regular correction leads to an additional change in the position of a
particle, which introduces a spurious discontinuity in the neighbour force and
its derivatives.  The Hermite iteration reacts to this artificial jump in two
ways: It increases the regular correction, and -- what is more important -- it
amplifies any spurious error during the iteration which leads to an extreme
unstable behaviour.

Since the Hermite iteration is a key element in the efficient integration of
planetesimal orbits, we sought for a modification of the Hermite scheme that
circumvents the depicted instability.  The problem gives already an indication
of a possible solution. A scheme with much smaller corrections would not suffer
from the feedback of spurious errors.

{\sc Nbody6++} stores already the second and third time derivative of the forces for the time step
calculation. A manifest application of these derivatives at hand is the improvement
of the predictions to fourth order:

\begin{align}
{\bf x}_p &= {\bf x}_0 + {\bf v}_0\Delta t + \frac{1}{2}{\bf a}_0(\Delta t)^2
                +\frac{1}{6}{\bf\dot a}_0(\Delta t)^3 \nonumber \\
      &  + \frac{1}{24} {\bf a}_0^{(2)}(\Delta t)^4 + \frac{1}{120} {\bf a}_0^{(3)}(\Delta t)^5 \\
{\bf v}_p &= {\bf v}_0 + {\bf a}_0\Delta t + \frac{1}{2}{\bf\dot a}_0(\Delta t)^2  \nonumber  \\
     & +   \frac{1}{6} {\bf a}_0^{(2)}(\Delta t)^3 + \frac{1}{24} {\bf a}_0^{(3)}(\Delta t)^4
\end{align}

\noindent
The prediction to fourth order was used in the iterative schemes of
\cite{Koku1998,Mikko98}.  Again, the new forces ${\bf a}_p$ and ${\bf\dot a}_p$ 
are calculated to improve ${\bf x}_p$ and ${\bf v}_p$ -- but with a modified
corrector:

\begin{align}
 {\bf a}_p          & = {\bf f}({\bf x}_p,{\bf v}_p) \\
{\bf\dot a}_p       & =  {\bf f}({\bf x}_p,{\bf v}_p) \\
{\bf a}^{(2)}_n (t) & =  \frac{-4 {\bf\dot a}_0 -2{\bf\dot a}_p }{\Delta t}
                         +\frac{-6 {\bf a}_0 + 6 {\bf a}_p }{\Delta t^2} \\
{\bf a}^{(3)}_n (t) & = \frac{6 {\bf\dot a}_0 + 6{\bf\dot a}_p }{\Delta t^2}
                         +\frac{12 {\bf a}_0 - 12 {\bf a}_p }{\Delta t^3} \\
{\bf x}_c           & = {\bf x}_p + \frac{1}{24}({\bf a}^{(2)}_n
                         -{ {\bf a}_0}^{(2)})(\Delta t)^4 + \nonumber \\
                    &   \frac{1}{120}({\bf a}^{(3)}_n
                         -{{\bf a}_0}^{(3)})(\Delta t)^5 \\
{\bf v}_c           & = {\bf v}_p + \frac{1}{6}({\bf a}^{(2)}_n
                         -{\bf a_0}^{(2)})(\Delta t)^3 + \nonumber \\
                    &   \frac{1}{24}({\bf a}^{(3)}_n
                         -{\bf a_0}^{(3)})(\Delta t)^4
\end{align}

\noindent
Finally, the derivatives are updated:

\begin{align}
 {\bf a}_0(t+\Delta t)       &= {\bf a}_p \\
 {\bf\dot a}_0(t+\Delta t)   &= {\bf\dot a}_p \\
 {\bf a_0}^{(2)}(t+\Delta t) &= {\bf a}^{(2)}_n +\Delta t \, {\bf a}^{(3)}_n \\
 {\bf a_0}^{(3)}(t+\Delta t) &= {\bf a}^{(3)}_n
\end{align}
The new scheme has an appealing property, which is related to the usage of the higher
force derivatives. As the predictor is fourth-order accurate, it is equivalent to
one full Hermite step. Since the corrector uses new forces to improve the two highest
orders, it is equivalent to a first iteration step. {Thus we obtained a one-fold iterated
Hermite scheme at no extra cost. This {extended Hermite scheme} reduces the energy
error by three orders of magnitude, compared to the plain {\sc Nbody6++} with the same number
of force evaluations}.

\subsection{KS--Regularisation}

Two bodies undergoing a close encounter are integrated in a special set of
regular coordinates that separates the relative motion from the motion of the
centre-of-mass.  A close encounter poses a numerical problem due to the
singularity of the gravitational forces at zero separation. While the growing
force amplifies roundoff errors as the two bodies approach each other closely,
the collision is only an apparent singularity since the analytic solution stays
well-defined. This opens the possibility of a proper coordinate transformation
which removes the singularity from the equations of motion.  The
Kustaanheimo--Stiefel regularisation takes advantage of a four-dimensional set
of variables to transform the Kepler problem into a harmonic oscillator
\citep{KS65}.  Perturbations are readily included in the new set of
equations of motion.

The centre-of-mass is added as a pseudo-particle, the {CM-body}, which is
integrated as a normal particle plus a perturbation force due to the deviation
from a point mass. See \cite{Mikko1997} or \cite{Mikko98} for more details.

\section{Additional forces for planetesimal disc dynamics}

{\sc Nbody6++} only includes the gravitational interaction of all particles,
therefore additional forces have to be added ``per hand''. A planetesimal disc
requires two new forces: The presence of a central star introduces an
additional central potential, while the gaseous component of the protoplanetary
disc is the source of a friction force. It is important that the new forces are
properly included in the neighbour scheme to assure that regular steps remain
larger than irregular steps. Since a dissipative force breaks the energy
conservation, one has to integrate the energy loss as well to maintain a valid
energy error control. In the next subsections we describe how we have done this.

\subsection{Central Potential}

A star is much heavier than a planetesimal. Thus, the central star is introduced as
a spatially fixed Keplerian potential:
\begin{equation}
  \Phi_{c} = -\frac{GM_{c}}{r} \qquad  {\bf F} = -\frac{GmM_{c}\,{\bf x}}{|{\bf x}|^3}
\end{equation}
Since the orbital motion of the planetesimals sets the dominant (and largest) dynamical timescale in
the system, we included the central force as a component of the regular force. Moreover,
the central potential also introduces a strong synchronisation, since
planetesimals in a narrow ring share virtually the same regular block time step.

\subsection{Drag Force}

As the whole planetesimal system is embedded in a dilute gaseous disc, each planetesimal is subjected to a small,
but noticeable drag force. The drag regime\footnote{The main drag regimes are Stokes (laminar flow),
Epstein (mean free molecular path larger than object size) and Newton's drag law (turbulent flow).  \cite{Weidenschilling77}
provides a nice review on the different drag regimes.} depends on the gas density and the size of the
planetesimals. Kilometer-sized planetesimals are subjected to the deceleration
\begin{align}
  \frac{d{\bf v}}{dt} &=  - \frac{\pi C_D}{2m} \rho_{\mathrm{gas}} R^2
     |{\bf v} -{\bf v}_g|({\bf v} -{\bf v}_g) \label{DragEq} \\
    C_D &=  0.5
\end{align}
which is inversely proportional to the radius $R(m)$ of the planetesimal in this drag regime.
$v_g$ is the rotational velocity of the gaseous disc, which rotates slower than the planetesimal
system as it is partially pressure supported. The drag force leads to an orbital decay $\dot a$ of the
semimajor axis of a planetesimals:
\begin{equation}
\dot a  = -\frac{3}{4}C_D\frac{\rho_{\mathrm{gas}}}{\rho_{\mathrm{Body}}}\frac{\langle (\Delta v)^2 \rangle}{R(m)\Omega}  \label{MigSpeed}
\end{equation}
Thus
smaller particles migrate faster, with a maximum at $R \approx 1$ m. Even smaller bodies
couple to the gas, which reduces the effective drag force.
The dissipation rate and its time derivative are:
\begin{align}
 W_{\mathrm{drag}} &=  {\bf F}_{\mathrm{drag}} \cdot {\bf v}     \nonumber \\
\dot {W}_{\mathrm{drag}}  &=  {\bf\dot F}_{\mathrm{drag}} \cdot {\bf v}
       + {\bf F}_{\mathrm{drag}} \cdot {\bf F}_{\mathrm{tot}}
\end{align}
We integrate the dissipation rate $W_{\mathrm{drag}}$ to maintain a valid energy error:

\begin{align}
  \Delta E &= \int_{t_1}^{t_2} W_{\mathrm{drag}} \, dt \\
  \Delta t &= t_2-t_1 \\
  \Delta E &= \frac{1}{2}\left(W_{\mathrm{drag,1}}+ W_{\mathrm{drag,2}} \right)\Delta t \nonumber \\
           & +\frac{1}{12}\left(\dot{W}_{\mathrm{drag,1}}-\dot{W}_{\mathrm{drag,2}}\right) {\Delta t}^2 + \mathcal{O}({\Delta t}^5)
\end{align}

\noindent
The expression is fourth order accurate in accordance with the order of the extended Hermite scheme.

\subsection{Accurate integration of close encounters: Tidal perturbations of
KS--Pairs and impact of the gaseous disc}

Both new forces also demand a modification of the regularisation treatment.
They perturb the relative motion of a KS--pair and modify the orbit of the
centre-of-mass. While the modification of the equations of motion is rather
clear, the neighbour scheme requires some additional work.

\noindent
Let ${\bf r}_1, {\bf r}_2$ be the positions of the two regularised particles. The
equations of motion read ($G=1$)
\begin{align}
 {\bf r} &=  {\bf r}_2 - {\bf r}_1 \nonumber \\
 {\bf\ddot r}_1 &=  -M_{c}\frac{{\bf r}_1}{r_1^3}+m_2\frac{{\bf r}}{r^3} + {\bf F}_{\mathrm{drag},1} \nonumber \\
 {\bf\ddot r}_2 &=  -M_{c}\frac{{\bf r}_2}{r_2^3}-m_1\frac{{\bf r}}{r^3} + {\bf F}_{\mathrm{drag},2}
\end{align}
where the perturbations by other particles have been omitted for clarity.
Centre-of-mass motion and the orbital motion are separated:

\begin{align}
 {\bf\ddot r} &= -M\frac{{\bf r} }{r^3}-M_{c}\frac{{\bf r}_2}{r_2^3}+
          M_{c}\frac{{\bf r}_1}{r_1^3} +  {\bf F}_{\mathrm{drag},2} - {\bf F}_{\mathrm{drag},1} \label{Eq_Rel} \\
 {\bf\ddot R} &= -M_{c}\frac{m_1}{M}\frac{{\bf r}_1}{r_1^3}-M_{c}\frac{m_2}{M}\frac{{\bf r}_2}{r_2^3}
  +\frac{m_1}{M} {\bf F}_{\mathrm{drag},1}+\nonumber \\
              & \frac{m_2}{M} {\bf F}_{\mathrm{drag},2} \label{Eq_CM} \\
 {\bf r} &= {\bf r}_2 - {\bf r}_1 \qquad   M = m_1+m_2   \\
 {\bf R} &= \frac{1}{M}\left( m_1{\bf r}_1+ m_2{\bf r}_2 \right)
\end{align}

Two new contributions show up due to the external forces: The KS--pair is
tidally perturbed by the central star and influenced by the gaseous disc. While
the aerodynamic properties of a single particle are well understood, two bodies
revolving about each other may induce complex gas flows in their vicinity,
which could invalidate the linear combination of the drag forces on each
component. Therefore we drop the drag force term to avoid spurious dissipation.
Since the dynamic environment allows virtually no stable
binaries\footnote{Tidal capturing of moons starts in the late stages of planet
formation, but is limited to the planets or their precursors.  However, the
quiescent conditions in an early Kuiper belt lead to a more prominent role of
binaries.  See the summary of  \cite{Astakhov2005}.} in a planetesimal disc,
the influence of the drag force on the encounter dynamics is negligible.

We further decompose the additional acceleration of the centre-of-mass motion,
since the neighbour scheme benefits from a clear separation of the timescales.
Therefore, the tidal perturbation is split into a smooth mean force and a
perturbation force:

\begin{align}
 {\bf\ddot R}        &= {\bf F}_{\mathrm{mean}}   + {\bf F}_{\mathrm{pert}} \nonumber \\
 {\bf F}_{\mathrm{mean}} &=    -M_{c}\frac{{\bf R}}{R^3} \nonumber \\
 {\bf F}_{\mathrm{pert}} &=  M_{c}\frac{{\bf R}}{R^3}-M_{c}\frac{m_1}{M}\frac{{\bf r}_1}{r_1^3}-
                   M_{c}\frac{m_2}{M}\frac{{\bf r}_2}{r_2^3} \label{FPert} \\
               &=  0 + \mathcal{O}(r^2)
\end{align}
The mean forces varies on the orbital timescale and is hence included as a regular force
component, while the perturbation is treated as an irregular force as it changes with
the internal orbital period of the pair.

\section{Further tuning the dial: Optimising $N-$body to the disc}
\label{ChapOpt}

An astrophysical simulation is a tool to analyse problems and predict
dynamical systems which are not accessible to experiments. The design
of a new simulation tool does not only require the careful implementation
of the invoked physics, but also an analysis of the code performance to
make best use of the available hardware.

We want to apply {\sc Nbody6++} for the first time to planet formation, a
subject that is quite different to stellar clusters. The central star forces
the planetesimals on regular orbits which need higher accuracy than the motion
of stars in a cluster.  In addition, the orbital motion also introduces a
strong synchronisation among the planetesimals, thus allowing a more efficient
integration.

We examine the differences due to the integration of a disc system in the following sections.
In particular, we will address in detail the role of
the geometry of the problem and the neighbour scheme, the prediction of the
number of neighbouring particles, the communication, the block size
distribution and the optimal neighbour particle number for the direct-summation
of the massive particles in the protoplanetary system

\subsection{Disc Geometry and Neighbour Scheme\label{DiscNS}}

\begin{figure}
\resizebox{\hsize}{!}{\includegraphics[scale=1,clip]
{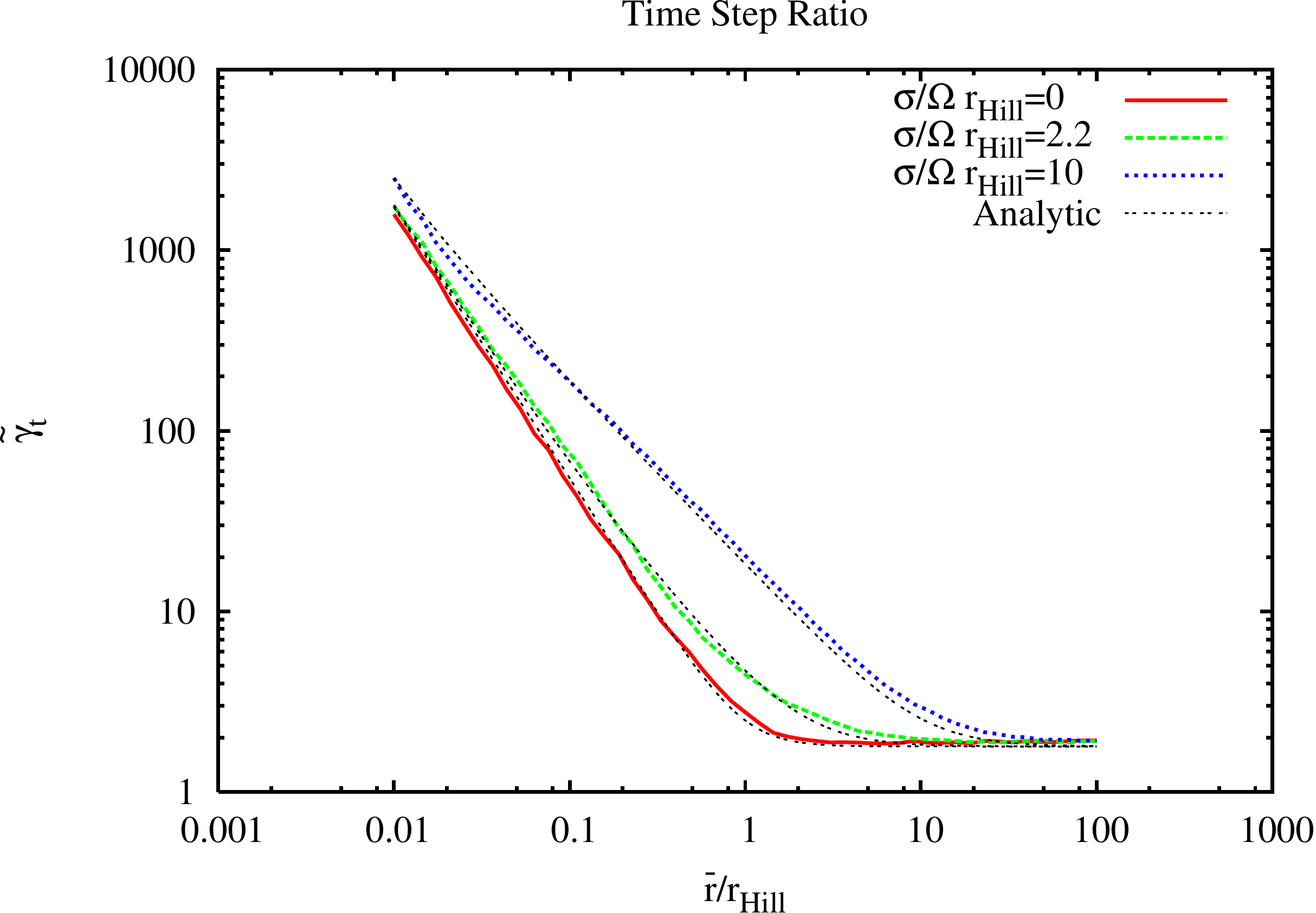}}
\caption{ Time step ratio for $ N_{\mathrm{nb}}=100 $. Curves are plotted for
different values of $\sigma_v/(r_{\mathrm{Hill}}\Omega)$. The dotted line is approximation Eq.~\ref{Apr_R}. \label{plotdt} 
\label{plotdt}
}
\end{figure}

The introduction of the neighbour scheme by \cite{AhmadCohen73} has provided us
with a technique to save a considerable amount of computational time in star
cluster simulations. Since the average ratio of the regular to the irregular
time step $\gamma_t$ is of the order of 10, the integration is speeded up by
the same factor. One may expect a similar speedup for planetesimal systems, but
in this case the time step ratio is roughly three. The time step is calculated
with the standard Aarseth time step criterion (it should be mentioned, however, that
the relation of regular and irregular costs is more complicated with GPU technology)

\begin{equation}
 \Delta t  =  \sqrt{\eta \frac{FF^{(2)}+(F^{(1)})^2 }{F^{(1)}F^{(3)}+ (F^{(2)})^2}  }  \label{tstep}
\end{equation}
where $F^{(i)}$ are the force and its time derivatives. It is applied to the calculation of
the regular step using the regular force and accordingly to the irregular step based on the irregular force.
The regular time step of a particle orbiting the central mass $M_c$ at a distance $r_0$ is:
\begin{equation}
 \Delta t_r  =  \sqrt{\eta_r}\frac{1}{\Omega} \qquad \Omega = \sqrt{ \frac{GM_c}{r_0^3} }
\end{equation}
For simplicity, we introduce the scaled timestep ratio
$ \tilde \gamma_t = \gamma_t\sqrt{\eta_{\mathrm{irr}}/\eta_{\mathrm{reg}}}$.
The free parameters of the problem are the mean particle
distance $\bar r$, the velocity dispersion $\sigma_v$ (additional to the Keplerian shear),
the particle mass $m_i$ and the neighbour number $N_{\mathrm{nb}}$. We employ Hill's approximation
for the central potential and obtain:

\begin{align}
 \Delta t_i        & \approx \frac{\sqrt{\eta_i}}{\Omega}
                     f(\Omega,r_{\mathrm{Hill}},\bar r,\sigma_v,N_{\mathrm{nb}} ) \\
 r_{\mathrm{Hill}} & = r_0 \sqrt[3]{ \frac{2 m_i}{3 M_c} } \label{D1}
\end{align}

\noindent
$f$ is a  yet unknown function. Dimensional analysis leads to
\begin{align}
 \Delta t_i & \approx   \frac{\sqrt{\eta_i}}{\Omega}f\left(\frac{\sigma_v}{r_{\mathrm{Hill}} \Omega},\frac{\bar r}{r_{\mathrm{Hill}}},
                N_{\mathrm{nb}}\right) \label{D2} \\
 \tilde \gamma_t &\approx   f\left(\frac{\sigma_v}{r_{\mathrm{Hill}} \Omega},\frac{\bar r}{r_{\mathrm{Hill}}},
               N_{\mathrm{nb}}\right)
\end{align}
which shows that the time step ratio is essentially controlled by the interparticle distance and the
velocity dispersion. We generated different random realisations of planetesimals discs with different
densities and velocity dispersions to cover the range of possible values. The neighbour number is
fixed to $N_{\mathrm{nb}} = 100 $ to reduce the noise due to small number statistics, but $\gamma_t$ converges
 to a value independently of the neighbour number already for $N_{\mathrm{nb}} > 10 $.
Fig.~\ref{plotdt} shows the numerical calculation of the time step ratio for various values of $\bar r$ and $\sigma_v$.
A good approximation to the calculated values of  $\tilde \gamma_t$ is:
\begin{equation}
 \tilde \gamma_t  \approx   1.79 \times \sqrt{1+1.03\frac{ \sigma_v^2}{\bar r^2\Omega^2}+0.94\frac{r_{\mathrm{Hill}}^3}{\bar r^3}  } \label{Apr_R}
\end{equation}
Planetesimal discs have usually a small velocity dispersion (compared to the
orbital velocity) and a low density in terms of the Hill radius, which leaves a
major influence to the Keplerian shear. Since the shear motion is directly
linked to the local Keplerian frequency, this synchronisation reduces
$\gamma_t$ to values smaller than ten. The numerical calculations show larger
time step ratios with increasing velocity dispersion and for high
densities\footnote{$\bar r/r_{\mathrm{Hill}}<1$ corresponds to an unstable
self-gravitating disc.}, but planetesimal discs are far from these extreme
parameter values.

\subsection{Optimal Neighbour Criterion}

\begin{figure}
\resizebox{\hsize}{!}{\includegraphics[scale=1,clip]
{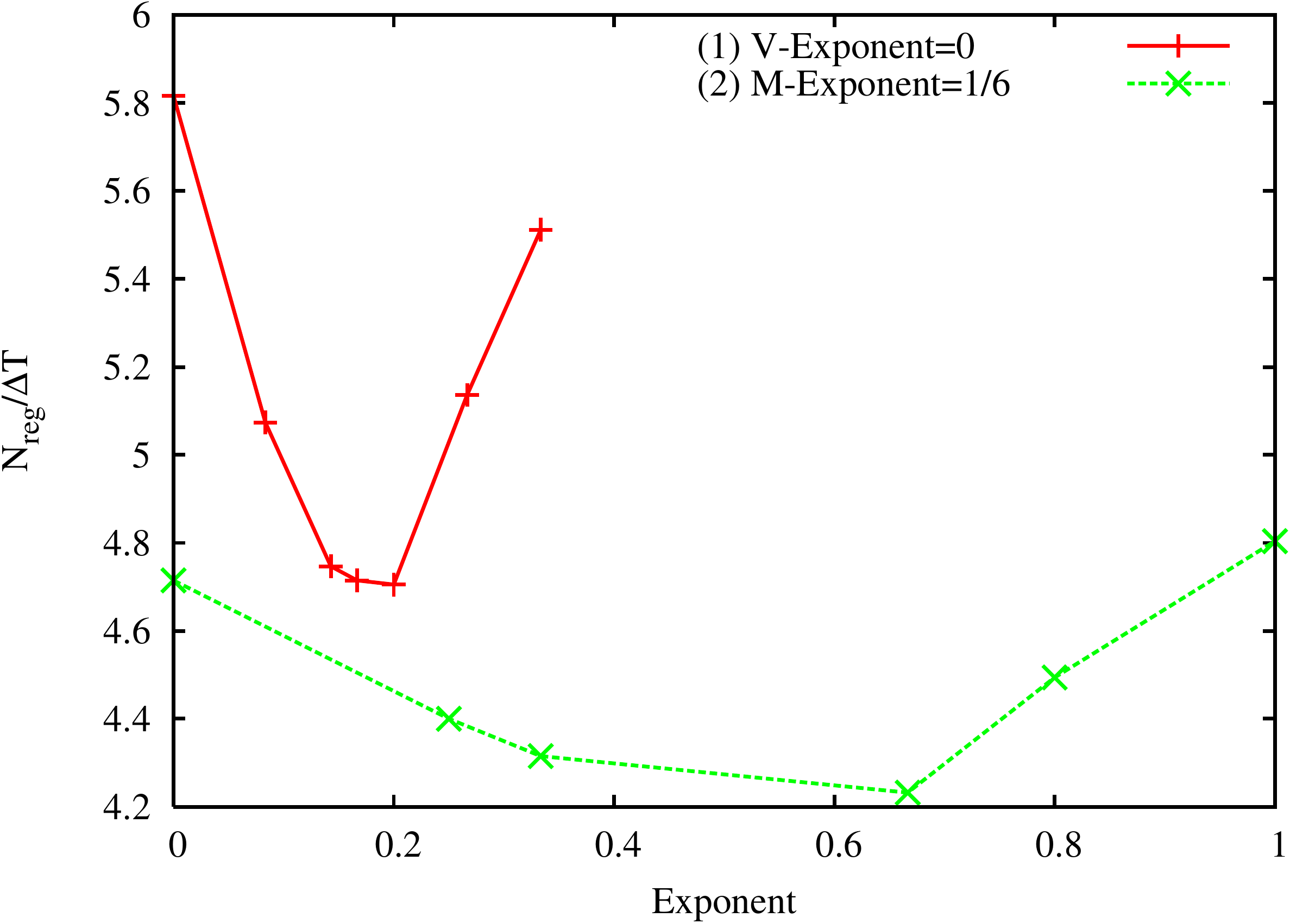}}
\caption{ Regular steps per particle and per 1 $N$--body time in the inner core ($R_i<0.5$) of
a 5000 particle plummer model. Plotted are (1) different mass exponents with velocity exponent $0$
and (2) different velocity exponents with mass exponent $1/6$. 
\label{NBex}
}
\end{figure}

The standard neighbour criterion uses the geometrical distance: Particles are
neighbours if their distance to the reference particle is smaller than a limit
$R_s$. This criterion is simple and probably the best choice for an equal mass
system. However, a multi-mass system may require a different criterion, since a
massive particle outside the neighbour sphere could have a stronger influence
than  lighter particles inside the neighbour sphere. Also, relative effects are
smaller at large distances. A more appropriate selection should rely on some
``perturbation strength'' of a particle.

It turned out that a better criterion is the magnitude of the fourth time derivative of the pairwise force $F^{(4)}_{ij}$,
 i.\,e. those particles are selected as neighbours which produce the largest integration error
in accordance with the Hermite scheme.
$F^{(4)}_{ij}$ is a complicated expression (compare Appendix~\ref{Force23Ana}), but the leading term can be estimated via
dimensional analysis:
\begin{equation}
 F^{(4)}_{ij}  \propto  \frac{m_j v_{ij}^4}{r_{ij}^6}
\end{equation}
We use this expression to define a new {apparent distance} between the integrated particle $i$ and a neighbour $j$:
\begin{equation}
  r_{\mathrm{app}}  =  r_{ij} \left( \frac{m_i}{m_j} \right)^{1/6} \left( \frac{v_s}{v_{ij}} \right)^{2/3} \label{NewRn}
\end{equation}
$v_s$ is an arbitrary scaling velocity to obtain a distance with dimension length.
This new distance definition moves massive or fast neighbours to an apparently smaller distance, thus
enforcing that these particles are preferentially included in the neighbour list. In addition,
the modified distance is readily included in the conventional neighbour scheme.
We tested
different mass and velocity exponents to verify that Eq.~\ref{NewRn} is the optimal choice.
Figure~\ref{NBex} shows that these exponents are indeed the optimal choice for a Plummer
model \citep{Plummer1911} with mass spectrum.
The new scheme saves $25\%$ of the force evaluations in the core, but
the impact on a planetesimals system is smaller, as it is the case for the neighbour scheme.
While a velocity dependent distance reduces the number of necessary full force
evaluations, it introduces a distance changing with time which destabilises the
integration. The result is a much larger energy error compared to the achieved
speedup.  Therefore we only recommend the mass modification of the apparent
distance.

\subsection{Neighbour Changes}

The rate at which the neighbours of a given particle change has a noticeable influence on the
accuracy of the code. During the course of an integration the second and third time
derivative of the regular and irregular force are calculated from an interpolation
formula (see Eq.~\ref{a2Int} and Eq.~\ref{a3Int}). Whenever a particle leaves (or enters) the neighbour sphere, these
derivatives are corrected by analytic expressions\footnote{Appendix~\ref{Force23Ana} gives a
complete set of the force derivatives up to third order.}. Hence many neighbour
changes lead to a pronounced spurious difference.

We estimated the rate at which particles cross the neighbour sphere boundary to
quantify this effect. Neighbour changes are due to the Keplerian shear and the superimposed random
velocities of the particles. The two effects lead to:
\begin{align}
N^{+/-} &=   \Delta t_r \frac{N_{\mathrm{nb}}}{T_{\mathrm{orb}}}\qquad\qquad\ \ \mbox{Shear} \nonumber \\
N^{+/-} &=  \frac{3}{2}\Delta t_r N_{\mathrm{nb}}\frac{\sigma_v}{R_{\mathrm{nb}}} \nonumber             \nonumber \\
 & = \Delta t_r \frac{N_{\mathrm{nb}}}{T_{\mathrm{orb}}} \frac{3\pi \sigma_v}{R_{\mathrm{nb}}\Omega}\qquad \mbox{Dispersion}
\end{align}
$R_{\mathrm{nb}}$ and  $N_{\mathrm{nb}}$ are the neighbour sphere radius and the neighbour number, respectively.
In practice, the neighbour changes due to the shear account for up to 80\% of
the total neighbour changes. The standard regular time step $\Delta t_r =2^{-5}$ and 50 neighbours
yield a change of one particle per regular step, which is fairly safe.

\subsection{Neighbour Prediction}

Each integration step is preceded by the prediction of all neighbours of the
particles that are due.  A regular step requires the full prediction of all
particles, so there is no possibility to save computing time. In contrast, an
irregular step calculates only neighbour forces, which requires the prediction
of less particles. Thus the prediction of all particles to prepare an irregular
step is a simple, but, depending on the block size, computational costly
solution.  It seems to be more efficient to predict only the required
particles, but random access to the particle data and the complete check of all
neighbour list entries introduces an additional overhead.  Therefore large
block sizes should favour the first approach, whereas the second approach is
more suitable for small block sizes.

Both regimes are separated by a critical block size $N_{\mathrm{irr}}^*$.  If
$N_{\mathrm{irr}}$ particles with $\langle N_{\mathrm{nb}}\rangle$ neighbours
are due, then only $N_{\mathrm{merge}}$ particles need to be predicted:

\begin{align}
 N_{\mathrm{merge}} & \approx N_{\mathrm{tot}} \left(1-\exp\left(-\frac{ N_{\mathrm{irr}}\langle N_{\mathrm{nb}}\rangle }{N_{\mathrm{tot}}}\right)\right) \le  \nonumber \\
                    & N_{\mathrm{irr}}\langle N_{\mathrm{nb}}\rangle \label{EqNMerge}
\end{align}

The size  $N_{\mathrm{merge}}$ of the merged neighbour lists is smaller than the total number of
neighbour list entries, since some particles are by chance members of more than one neighbour list.
Performance measurements show that the prediction of the merged neighbour lists is 10 \% more
costly (per particle) than the full prediction, mainly due to additional sorting and a random
memory access. Thus $N_{\mathrm{irr}}^*$ satisfies:
\begin{equation}
  N_{\mathrm{tot}}   =   1.1 \times N_{\mathrm{merge}}
\end{equation}
Inserting Eq.~\ref{EqNMerge} yields the critical block size:
\begin{equation}
  N_{\mathrm{irr}}^*   \approx   2.4 \frac{N_{\mathrm{tot}}}{\langle N_{\mathrm{nb}} \rangle} 
\end{equation}
The prediction mode is chosen according to the actual block size.

\subsection{Communication Scheme}

\begin{table}
\begin{center}
\begin{tabular}{|l|c|c|c|c|c|c|c|c|}  \hline
Process & 0 & 1 & 2 & 3 & 4 & 5 & 6 & 7 \\ \hline
Send to & 1 & 2 & 3 & 4 & 5 & 6 & 7 & 0 \\  \hline
Receive from & 7 & 0 & 1 & 2 & 3 & 4 & 5 & 6 \\ \hline
\end{tabular}
\caption[Ring communication]
{\label{RingC} Ring Communication. Communication partners are fixed, while the
 exchanged data varies. $n_p-1$ cycles are needed.}
\end{center}
\end{table}

\begin{table}
\begin{center}
\begin{tabular}{|l|l|c|c|c|c|c|c|c|c|}  \hline
Cycle & Process       & 0 & 1 & 2 & 3 & 4 & 5 & 6 & 7 \\ \hline
1     & Exchange with & 1 & 0 & 3 & 2 & 5 & 4 & 7 & 6 \\  \hline
2     & Exchange with & 2 & 3 & 0 & 1 & 6 & 7 & 4 & 5 \\  \hline
3     & Exchange with & 4 & 5 & 6 & 7 & 0 & 1 & 2 & 3 \\  \hline
\end{tabular}
\caption[Hierarchical communication]{\label{HierC} Hierarchical Communication.
Communication partners change after every cycle.
The exchanged data amount doubles with every new cycle, hence only $\ln_2(n_p)$ cycles are needed.}
\end{center}
\end{table}

{\sc Nbody6++} is parallelised using a {copy algorithm}. A complete copy of
the particle data is located on each node, so the integration step of one
particle does not need any communication. Therefore a block of
$N_{\mathrm{bl}}$ particles is divided in $n_p$ parts ($n_p$ is the processor
number), which are integrated by different processors in parallel. The
integration step is completed by an all--to--all communication of the different
subblocks to synchronise the particle data on all nodes.  Hence the amount of
communicated data is proportional to $N_{\mathrm{bl}}\times n_p$.  A
communication in a ring-like fashion (see table~\ref{RingC}) needs $n_p-1$
communication cycles, but a hierarchical scheme (see table~\ref{HierC}) sends
the same amount of data with only $\ln_2 (n_p)$ communication cycles. The
difference between the two approaches remains small, as long as the
communication is bandwidth limited, i.\,e. the blocks are large. Small block
sizes shift the bottleneck to the latency, which is significantly reduced by
the second scheme -- especially if the code runs on many processors.

\begin{table}
\begin{center}
\begin{tabular}{|l|c|c|r|r|c|c|c|c|}  \hline
 Block    & $n_p$ & $\alpha   [\mu\mathrm{s}] $ & $ \tau_l [\mu\mathrm{s}]$  & A & B \\ \hline
Irregular & 10    & 0.35 &  51 & 145 & 4.5  \\  \hline
Regular   & 10    & 0.22 & 113 & 512 & 40  \\  \hline
Irregular & 20    & 0.35 & 308 & 877 & 8.8  \\  \hline
Regular   & 20    & 0.22 & 368 & 1668 & 75  \\  \hline
\end{tabular}
\caption[Timings on Hydra]{\label{Tim1}Timings on a Beowulf cluster (Hydra, see table \ref{CompRes}). See text for an explanation of the variables. Timings
are obtained for a maximal neighbour number LMAX=64. In practice, $B$ is twice as large due to
storage rearrangements in {\sc Nbody6++}. See Appendix~\ref{CompRes} for details on the computers. }
\end{center}
\end{table}

\begin{table*}
\begin{center}
\begin{tabular}{|l|c|c|c|} \hline
  Name      & Beowulf (Hydra) & Titan & JUMP    \\ \hline
 Institute    &   ARI/ZAH & ARI/ZAH &  Forschungszentrum J\"ulich   \\ \hline
 Location   & Heidelberg & Heidelberg & J\"ulich  \\ \hline
Processors  &  20 & 64 & 1248 \\ \hline
Speed       &  2.2GHz & 3.2GHz   & 1.7GHz  \\ \hline
Processors/Node &  2 & 2 & 32 \\ \hline
Network     &  Myrinet & Infiniband & Gigabit--Ethernet \\ \hline
Bandwidth   & 2Gbit/sec & 20 Gbit/sec & 10 Gbit/sec  \\ \hline
\end{tabular}
\caption{Specs of the different supercomputers used in running the algorithm
\label{CompRes}
}
\end{center}
\end{table*}

\begin{table}
\begin{center}
\begin{tabular}{|l|c|c|r|r|c|c|c|c|}  \hline
 Block    & $n_p$ & $\alpha   [\mu\mathrm{s}] $ & $ \tau_l [\mu\mathrm{s}]$  & A & B \\ \hline
Irregular & 8    & 0.29 & 255 & 837  & 1.7  \\  \hline
Regular   & 8     & 0.60 & 981 & 1763 & 4.1  \\  \hline
Irregular & 16    & 0.28 & 188 & 700  & 1.7  \\  \hline
Regular   & 16    & 0.60 & 306 & 561  & 6.7  \\  \hline \hline
Irregular & 64    & 0.27 & 241 & 887  & 7.7  \\  \hline
Regular   & 64    & 0.46 & 401 & 871  & 21.7  \\  \hline
\end{tabular}
\caption[Timings on JUMP]{\label{Tim2}Timings on the IBM. More than 32 processors require more
than one node.}
\end{center}
\end{table}

A hierarchical scheme reduces the latency, but nevertheless it is possible that
the parallel integration is actually slower than a single CPU integration. We estimated
both the runtime on one CPU and on a parallel machine to explore the transition between
these two regimes. The latency time $\tau_l$ per communication is included in the wallclock time expressions
for one regular/irregular step:
\begin{align}
 \tau_l &=   \alpha A   \nonumber \\
 t_{\mathrm{single}}&=  \alpha N_{\mathrm{bl}} N_{\mathrm{nb}}  \label{TimeSing} \\
 t_{\mathrm{par}}   &=  \alpha \Bigl( \underbrace{\frac{N_{\mathrm{bl}} N_{\mathrm{nb}}}{n_p}}_{\mathrm{Arithmetic}}+
                 \underbrace{ A \ln_2(n_p)}_{\mathrm{Latency}}+
                 \underbrace{ B N_{\mathrm{bl}}}_{\mathrm{Communication}} \Bigr) \label{TimePar}
\end{align}
If $t_{\mathrm{single}}$ (runtime on a single CPU) is equal to $t_{\mathrm{par}}$ (parallel computation),
one can deduce the critical block size $N_{\min}$ which gives the minimal block size for efficient
parallelisation:
\begin{align}
 t_{\mathrm{single}} &=   t_{\mathrm{par}} \nonumber \\
 N_{\min} &=  \frac{A\ln_2(n_p)n_p}{N_{\mathrm{nb}}(n_p-1)-Bn_p}
\end{align}
The hierarchical communication gives a minimal block size that increases logarithmically with the processor number.
Eq.~\ref{TimePar} gives immediately the speedup $S$ and the optimal processor number for a certain block size
$N_{\mathrm{bl}}$:
\begin{align}
 S &=  \frac{n_p}{1 +A n_p\frac{\ln_2(n_p)}{N_{\mathrm{bl}}N_{\mathrm{nb}}}+B\frac{n_p}{N_{\mathrm{nb}} }} \nonumber \\
 n_{p,\mathrm{opt}}(N_{\mathrm{bl}}) &=  \frac{\ln(2)N_{\mathrm{bl}}\times N_{\mathrm{nb}}}{A} ~ - {\rm Hierarchical} \label{EqOptH}
\end{align}
A comparison to the optimal processor number for a ring communication
\begin{align}
n_{p,\mathrm{opt}}(N_{\mathrm{bl}}) &=  \sqrt{\frac{N_{\mathrm{bl}} \times N_{\mathrm{nb}}}{A}} ~ - {\rm Ring}  \nonumber \\
t_{\mathrm{par}}   &=  \alpha \Bigl( \frac{N_{\mathrm{bl}} N_{\mathrm{nb}}}{n_p}+A n_p+ B N_{\mathrm{bl}} \Bigr) \label{TimeParOld}
\end{align}
stresses the efficiency of the hierarchical communication, since it allows a much larger
processor number for a given problem size. Equation~\ref{TimeSing} and \ref{TimePar} are also useful
to derive the total wallclock time, since the total runtime scales with the number of regular and irregular
blocks:
\begin{align}
  N_{\mathrm{reg}} & \approx   T \frac{N^{1/3}}{N_{\mathrm{nb}}^{1/3}\sqrt{\eta_{\mathrm{reg}}}} \label{EqB_R}  \\
  N_{\mathrm{irr}} &\approx  T \frac{N^{1/3}}{\sqrt{\eta_{\mathrm{irr}}}} \label{EqB_I}
\end{align}
These equations are only approximate expressions, but they give the right order of magnitude
without detailed calculations that need a precise knowledge of the $N$--body model.
Table~\ref{Tim1} and table~\ref{Tim2} summarise the timing parameters drawn from our experience
with the Hydra and JUMP parallel supercomputers.

\subsection{Block Size Distribution}

\begin{figure}
\resizebox{\hsize}{!}{\includegraphics[scale=1,clip]
{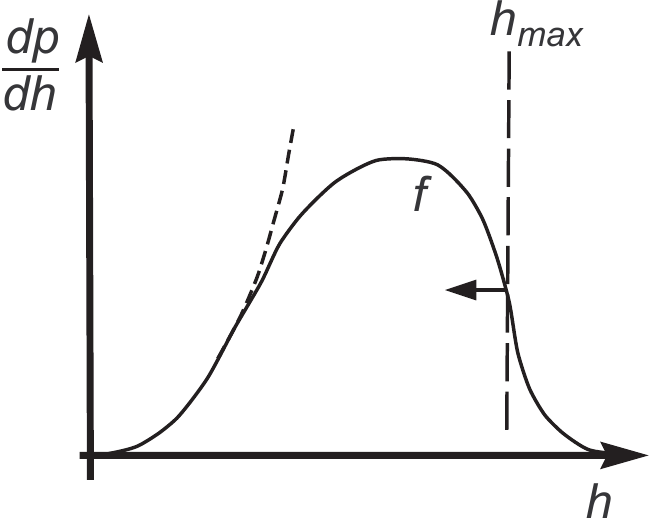}}
\caption{Timestep distribution $f=dp/dh$. The short-dashed line on the left
indicates approximation Eq.~\ref{Apprdp_dh}, whereas the dashed line on the right
defines a reasonable upper limit $h_{\max }$. 
\label{SketchBlock}
}
\end{figure}

The preceding section showed that the block size is closely related to the efficiency of
the parallelisation. Small blocks are dominated by the latency and the parallelisation
could be even slower than a single CPU calculation. Therefore we derive the block size distribution
for the block time step scheme to asses its influence on the efficiency.

\begin{figure}
\resizebox{\hsize}{!}{\includegraphics[scale=1,clip]
{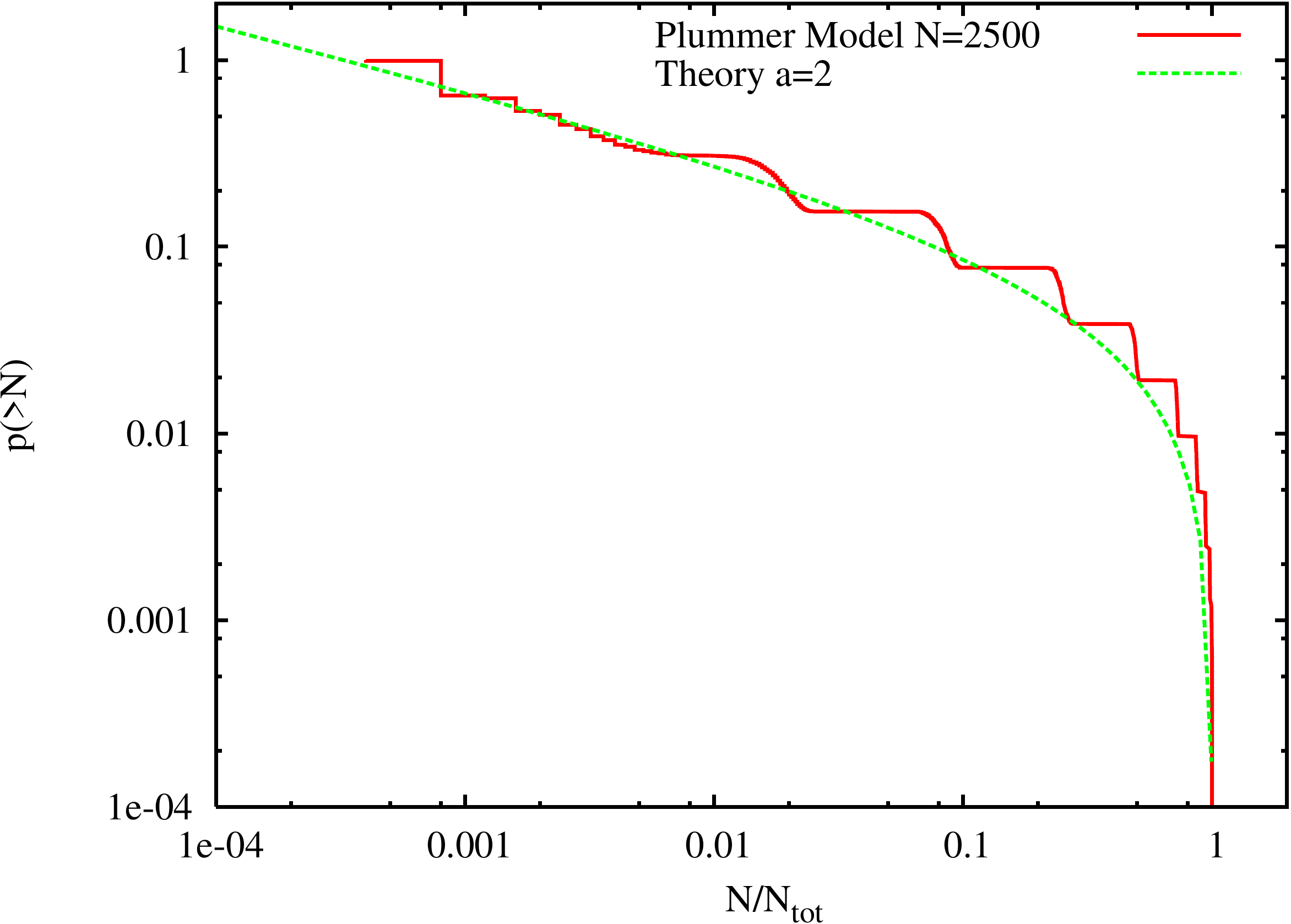}}
\caption{Cumulative irregular block size distribution for a $N=2500$ particle Plummer model.  
\label{CumBlockDis}
}
\end{figure}

\noindent
Suppose that the time steps\footnote{We use $h$ instead of $\Delta t$ in this section to avoid unclear notation.} $h$ of all $N$ particles in the model are distributed according to some known function $f$:
\begin{equation}
  dp  =  f(N,h)dh
\end{equation}
$f$ is in most cases a complicated function. It involves spatial averaging and integration over the velocity
distribution, which could be quite complicated even for simple time step formul{\ae}. Nevertheless there is a
constraint on the time step distribution, simply because every particle has a neighbour within a finite
distance: There is some upper limit $h_{\max}$ that restricts the major fraction of the time steps to a finite interval.
Thus it is possible to capture the main features of the time step distribution with an expansion
around $h=0$ (Fig.~\ref{SketchBlock} sketches this approximation):
\begin{equation}
  f  \approx   C(N)h^{a}  \qquad h \leq h_{\max} \label{Apprdp_dh}
\end{equation}
$a$ is the lowest non-vanishing order of the expansion. Now we consider a block level with the largest possible time step  $h_k$.
The number of particles $N_{\mathrm{bl}}$ in this block is:
\begin{equation}
   N_{\mathrm{bl}} = N \int_0^{h_k} f dh \approx  \frac{C(N)}{a+1}(h_k)^{a+1}
\end{equation}
According to the block time step scheme the number of blocks per time with the largest possible time step $h_k$ is
proportional to $(h_k)^{-1}$.
Therefore the probability that a block size is in the range $[N_{\mathrm{bl}},N_{\mathrm{bl}}+dN_{\mathrm{bl}}]$ is
\begin{equation}
  dp   \propto   \sum_k \delta\left(N_{\mathrm{bl}}-\frac{C(N)}{a+1}h_k^{a+1} \right) \frac{1}{h_k} dN_{\mathrm{bl}}
\end{equation}
where $\delta$ is Dirac's delta function.
The sum over the logarithmically equidistant time steps $h_k$ is approximated by an integral:
\begin{align}
  dp & \propto   \int_0^{\infty} \delta\left(N_{\mathrm{bl}}-\frac{C(N)}{a+1}h^{a+1} \right)\frac{d \ln(h)}{h} dN_{\mathrm{bl}}\nonumber \\
     & \approx   \frac{1}{a+1} N_{\mathrm{bl}}^{-(a+2)/(a+1)} dN_{\mathrm{bl}}  \label{DistNbl}
\end{align}
Thus the average block size and the median of the block size distribution are:
\begin{align}
 \langle N_{\mathrm{bl}} \rangle &\approx   \frac{1}{a} N^{a/(a+1)} \nonumber \\
 \mbox{median}(N_{\mathrm{bl}}) & \approx   2^{a+1}
\end{align}
Special expressions for the average block size were already derived by \cite{Makino1988}, but
the general relation of the time step distribution to the block size distribution is a new result.
The median is surprisingly independent of the particle number, i.\,e. 50 \% of all blocks are always smaller than
a fixed value. It seems that this is a threat
to the efficiency of the method, but the median of the wallclock time
\begin{equation}
 \mbox{median}(N_{\mathrm{bl}}^2)   \approx   \frac{N}{2^{(a+1)/a}}
\end{equation}
demonstrates that these small blocks account only for a small fraction of the total CPU time.
We confirmed the derived block size distribution  (Eq.~\ref{DistNbl}) by numerical calculations (see Fig.~\ref{CumBlockDis}).
The order parameter  $a$ is roughly two in (at least locally)  homogenous systems, while an additional
Keplerian potential reduces the order to $a=1$. A planetesimal disc -- or more precisely, a narrow ring
of planetesimals -- has a very narrow distribution of time steps since all particles share nearly the
same orbital period. Thus the regular block size is always equal to the total particle number making the
parallelisation very efficient.

\subsection{Optimal Neighbour Number}

We treated the mean neighbour number $N_{\mathrm{nb}}$ so far as some fixed
value. But it is also a mean to optimise the speed of the integration. Large
neighbour spheres reduce fluctuations in the regular forces allowing larger
regular steps, which reduces the total number of force evaluations.  But larger
neighbour lists also imply a larger communication overhead, as all the
neighbour lists have to be broadcast to synchronise the different nodes. The
best choice balances these two extremes, thus maximising the speed.

Before we derive the optimal neighbour number on a parallel machine, we briefly summarise the
known solution for a single CPU run \citep[][]{Makino1988} for an extensive derivation. The computational
effort of the irregular steps is proportional to the neighbour number, while the number of force evaluations for
the regular steps is proportional to the total number of particles, reduced by the time step
ratio $\gamma_t$:
\begin{align}
\gamma_t &:=  \frac{\Delta t_{\mathrm{reg}}}{\Delta t_{\mathrm{irr}}} \nonumber \\
 T_{\mathrm{CPU}} & =   f(N)\left( N_{\mathrm{nb}}+\frac{N}{\gamma_t(N_{\mathrm{nb}})}\right)\label{Eq_1} \\
  \gamma_t(N_{\mathrm{nb}}) &\approx   N_{\mathrm{nb}}^{1/3} \label{gam}
\end{align}
$f(N)$ collects all factors depending only on the total number of particles. Optimisation with respect
to the neighbour number $ N_{\mathrm{nb}}$ yields the well known result:
\begin{align}
  0 &=  \frac{d}{d N_{\mathrm{nb}}} T_{\mathrm{CPU}} \nonumber \\
  N_{\mathrm{nb},\mathrm{opt}} & \propto   N^{3/4} \label{NbOptM}
\end{align}
The calculation of the elapsed time for {\sc Nbody6++} on a PC cluster includes more terms.
For clearness, we restrict ourselves to a rather simple model that involves only the dominant
terms to show how parallelisation influences the optimal neighbour number.
We make the following approximations:
\begin{enumerate}
\item We only take the force calculation and communication  into account.
\item We use the same time constants for regular and irregular expressions.
\item We neglect all numerical factors that are comparable to unity.
\end{enumerate}

\begin{figure}
\resizebox{\hsize}{!}{\includegraphics[scale=1,clip]
{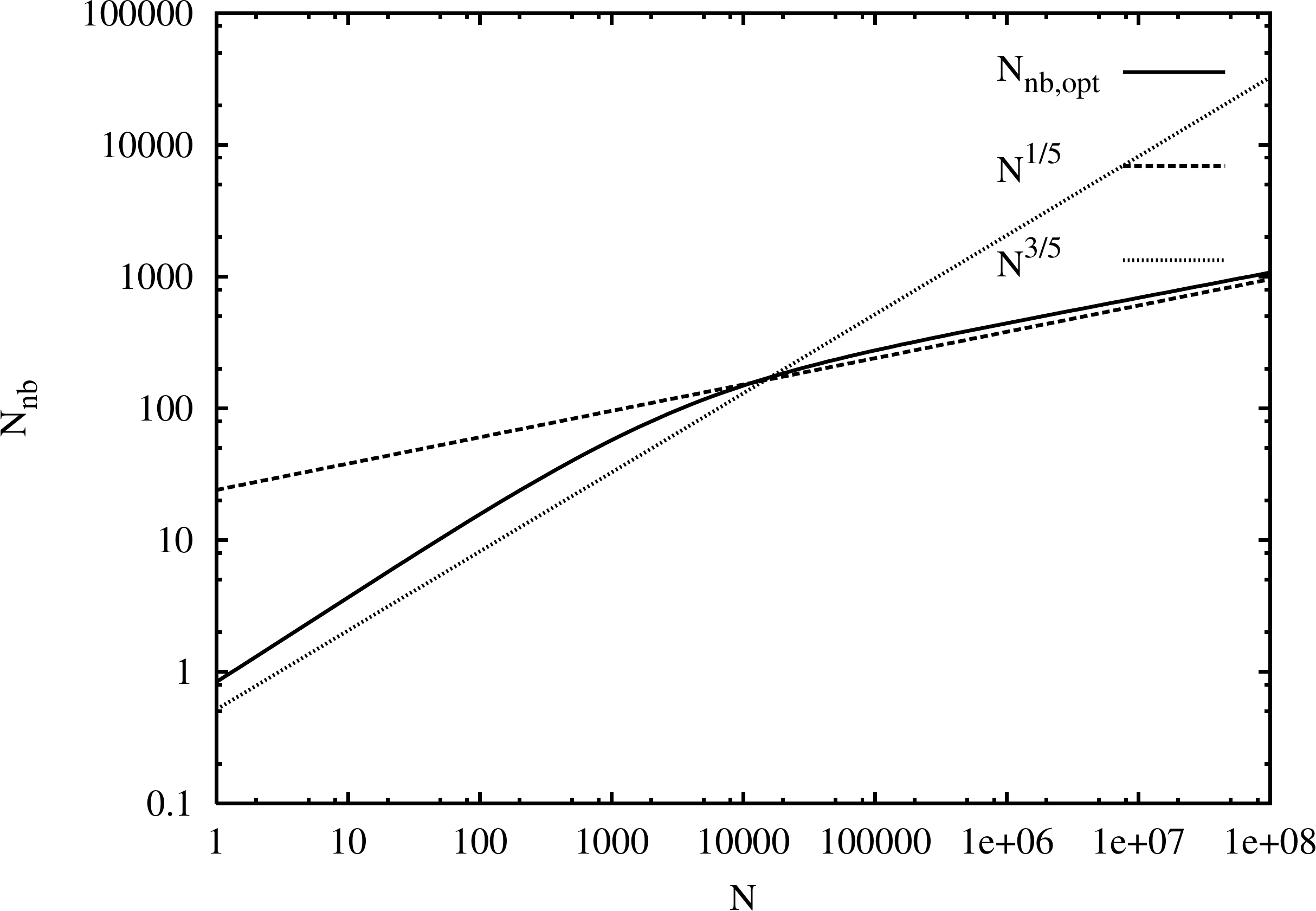}}
\caption{ Optimal neighbour number as function of particle number $N$.
The plot includes the numerical solution of Eq.~\ref{nbopt_1} and the two
asymptotic solutions. Timing constants are taken from a Beowulf cluster. 
\label{NBopt}
}
\end{figure}

The total CPU time is an extension of Eq.~\ref{TimeParOld}, which is applied to the regular and the irregular
step. A new constant $B_n$ includes the neighbour list communication separately, while all factors depending on $N$ are
represented by $f(N)$:
\begin{align}
    N_{\mathrm{bl}} &\approx   N^{2/3} \nonumber \\
    \gamma_t &\approx   N_{\mathrm{nb}}^{1/3} \nonumber \\
 T_{\mathrm{irr}} &=  f(N)\left(  \frac{N_{\mathrm{bl}}N_{\mathrm{nb}}}{p} +Ap +B N_{\mathrm{bl}} \right)   \nonumber \\
 T_{\mathrm{reg}} &=  \frac{1}{\gamma_t}f(N)\left(  \frac{N_{\mathrm{bl}}N}{p} +Ap +(B+ B_n N_{\mathrm{nb}}) N_{\mathrm{bl}} \right) \nonumber \\
 T_{\mathrm{tot}} &=  T_{\mathrm{irr}}+T_{\mathrm{reg}}
\end{align}
Optimisation with respect to the processor number $p$ leads to:
\begin{align}
  0 &=  \frac{\partial}{\partial p}T_{\mathrm{tot}}  \nonumber \\
  p_{\mathrm{opt}} &= \sqrt{ \frac{N^{2/3}N_{\mathrm{nb}}^{4/3}+N^{5/3} }{A(N_{\mathrm{nb}}^{1/3}+1)}} \nonumber \\
        &\approx   \frac{N^{5/6}}{\sqrt{A} N_{\mathrm{nb}}^{1/6}} \label{Popt_1}
\end{align}
Further optimisation with respect to the neighbour number gives the expression:
\begin{align}
 0 &= \frac{\partial}{\partial N_{\mathrm{nb}} }T_{\mathrm{tot}}  \\
 0 &= N_{\mathrm{nb}}^{4/3} - \frac{1}{3} N - \frac{1}{3} A N^{-2/3} p^2 - \frac{1}{3}Bp \,+ \nonumber \\
   & \frac{2}{3} N_{\mathrm{nb}} B_n p \label{Nbopt_1}
\end{align}
For a fixed $p$ or $B_n=0$ (very fast neighbour list communication), we recover
for large $N$:
\begin{equation}
  N_{\mathrm{nb},\mathrm{opt}}  \propto  N^{3/4}
\end{equation}
In general, one can not neglect the neighbour list communication. Therefore we seek for the optimal choice
of $p$ and $N_{\mathrm{nb}}$, thus combining Eq.~\ref{Popt_1} and Eq.~\ref{Nbopt_1}:

\begin{align}
& A N_{\mathrm{nb}}^{5/3} +\left( \frac{2}{3} N_{\mathrm{nb}} -\frac{B}{3B_n}\right)B_n\sqrt{A}N_{\mathrm{nb}}^{1/6}N^{5/6} = \nonumber \\
& \frac{1}{3}(N_{\mathrm{nb}}^{1/3}+1)N A \label{nbopt_1}
\end{align}

Since this equation has no closed solution, we identify the dominant terms in Eq.~\ref{nbopt_1}
to calculate the asymptotic solution for large $N$:
\begin{equation}
 N_{\mathrm{nb},\mathrm{opt}}  \approx  \left(\frac{A}{4B_n^2} \right)^{3/5} N^{1/5} \qquad N \gg \left(\frac{3A}{4B_n^2}\right)^{3/2}
\end{equation}
For small $N$  we get the approximated solution:
\begin{equation}
 N_{\mathrm{nb},\mathrm{opt}}  \approx  \left( \frac{N}{3} \right)^{3/5} \qquad N<\left(\frac{3A}{4B_n^2}\right)^{3/2}
\end{equation}
Fig.~\ref{NBopt} compares the approximate expressions with the numerical solution of equation~\ref{nbopt_1}.
In spite of the complicated structure of Eq.~\ref{nbopt_1}, both approximate expressions are reliable solutions.
The example uses timing constants
derived form our local Beowulf cluster:
\begin{equation}
 A \approx 200 \quad B \approx 5 \quad B_n \approx 0.5
\end{equation}
If we compare the new optimal neighbour number to the single CPU expression (Eq.~\ref{NbOptM}), we find that the influence
of the neighbour list communication favours much smaller neighbour numbers.  $N_{\mathrm{nb}}$ increases so
slowly with the particle number that a neighbour number around 100 is a safe choice.

\section{Collisional and fragmentation Model}
\label{CollModel}

The growth of planetesimals proceeds through collisions among planetesimals
which form (at least in a sufficient fraction of incidents) larger bodies with
a net gain of accreted matter. But some collisions are mere destructive events
that shatter and disperse the colliding planetesimals. Small bodies are more
susceptible to destruction, but they are also driven to high relative
velocities due to the global energy equipartition making them even more
vulnerable. A model that attempts to cover the full size range from objects
ranging between a kilometer and the size of Mars needs a realistic collision
algorithm that covers both {fragmentation} and accretion. Some examples in
the literature are \cite{CazenaveEtAl82,BeaugeAarseth90}.
In our algorithm we use the approach of \cite{Glaschke2003}, which was applied
to asteroid families.  The following section describes the approach to
fragmentation that is implemented in the code.

\subsection{Handy quantities for quantifying the models}

Two colliding bodies are equal in the sense that their intrinsic properties are
not different.  Only the comparison of two bodies defines the larger body --
usually denoted as {target} -- and the smaller one denoted as {\em
projectile}. The two terms stem from laboratory experiments where they indicate
much more than different sizes. A small projectile is shot on a target at rest
to study the various parameters related to fragmentation.  In the following,
projectile and target only indicate the relative size of the two bodies.

The collision of two bodies initiates a sequence of complex phenomena. Shock
waves run through the material, flaws start to grow rapidly breaking the bodies
apart in many pieces. Some kinetic energy is transferred to the fragments,
which leads to the ejection of  fragments at different velocities in various
directions. If the fragment cloud is massive enough, some of the larger
fragments may capture debris. This post-collisional accretion is denoted as
{reaccumulation}.

Although the depicted scenario is quite complex, there are a few measures
that describe the most important aspects:
\begin{enumerate}
\item Mass of the largest fragment $M_L$, or dimensionless $f_l=M_L/M$ where $M$
  is the combined mass of the two colliding bodies.
\item $f_l<\frac{1}{2}$  refers to {fragmentation}, whereas $f_l > \frac{1}{2}$ is denoted
    as {cratering}.
\item
 Energy per volume $S$ that yields $f_l=\frac{1}{2}$ is denoted as {impact strength}.
\item
 $f_{\mathrm{KE}}:=2E_{\mathrm{kin}}^{\mathrm{frag}}/E_{\mathrm{kin}}$:
 Fraction of the impact energy that is converted into kinetic energy of the fragments.
\end{enumerate}
Different fragment sizes and velocities are summarised  by appropriate distribution functions.
$m_i$, $D_i$ and $v_i$ are  mass, diameter and modulus of the velocity of a given
fragment, respectively.
\begin{enumerate}
\item Fragment size distribution:
\begin{enumerate}
\item $N_{m}(m)$ : Number of all fragments  with a mass $m_{i}\geq m$,
\item $M(m)$ \  : Mass of all fragments with a mass $m_{i}\geq m$,
\item $N_{D}(D)$ : Number of all fragments with a diameter $D_{i}\geq D$.
\end{enumerate}
The distribution functions are related to each other:
\begin{eqnarray*}
 N_{m}(m) & = & N_{D}(D(m)) \\
 M(m)     & = & \int_{m}^{\infty} x\left|\frac{\textrm{d}N_{m}(x)}{\textrm{d}x}\right| dx \\
 N_{m}(m) & = & \int_{m}^{\infty} \frac{1}{x}\left|\frac{\textrm{d}M(x)}{\textrm{d}x}\right| dx \\
\end{eqnarray*}
$D(m)$ is the size--mass relation.
\item Velocity distribution:
\begin{enumerate} \item
$ \bar v(m)$: mean velocity as a function of mass.
\end{enumerate}
\end{enumerate}

\begin{figure}
\resizebox{\hsize}{!}{\includegraphics[scale=1,clip]{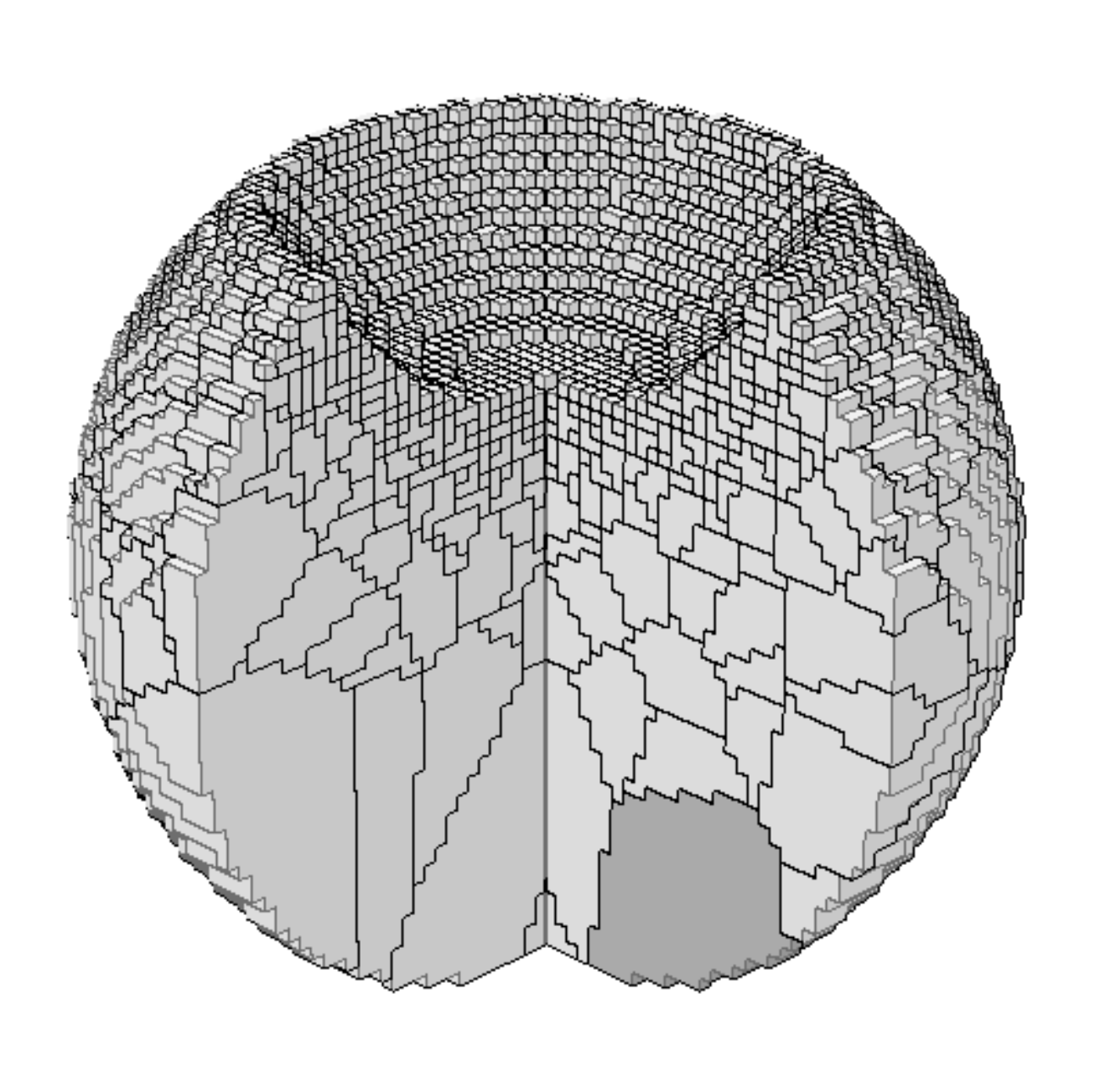}}
\caption{Section for $f_l=0.04$ and $n=3$. The largest fragment is coloured in
dark-grey. In this calculation $60 \times 60 \times 60$ grid cells are used. Note the decomposition
in grid cells and the Voronoy polyhedra which form the fragments.  
\label{FragB}
}
\end{figure}

\subsection{Prediction of collisional outcomes: Derivation from a Voronoy tessellation}

Any theoretical or empirical prescription of a collision has to relate the
afore mentioned parameters, namely the impact energy, to the sizes and
velocities of the produced fragments.  The central quantity is the impact
strength, which is a measure for the overall stability of a body. Objects
smaller than 1 m are accessible to laboratory experiments, while collisions
of larger bodies up to asteroid size have to be analysed by complex computer
simulations. Asteroid families, which are remnants of giant collisions in the
asteroid belt, provide independent insight, although the data is difficult to
interpret.

We selected two different impact strength models as reference for our work. The
first was obtained by \cite{Housen1990} through the
combination of asteroid family data and laboratory experiments via scaling
laws:

\begin{align}
S               &= S_0\left(\frac{R}{1\,\mathrm{m}} \right)^{-0.24}\left[1+1.6612 \times10^{-7}\left(\frac{R}{1\,\mathrm{m}} \right)^{1.89}\right] \nonumber \\ 
f_{\mathrm{KE}} &=0.1  \label{HousenS} \\
S_0 &= 1.726 \times 10^6 \textrm{ J}\textrm{m}^{-3}= 1.726 \times 10^7 \textrm{erg}\textrm{ cm}^{-3}
\end{align}

\begin{figure}
\resizebox{\hsize}{!}{\includegraphics[scale=1,clip]{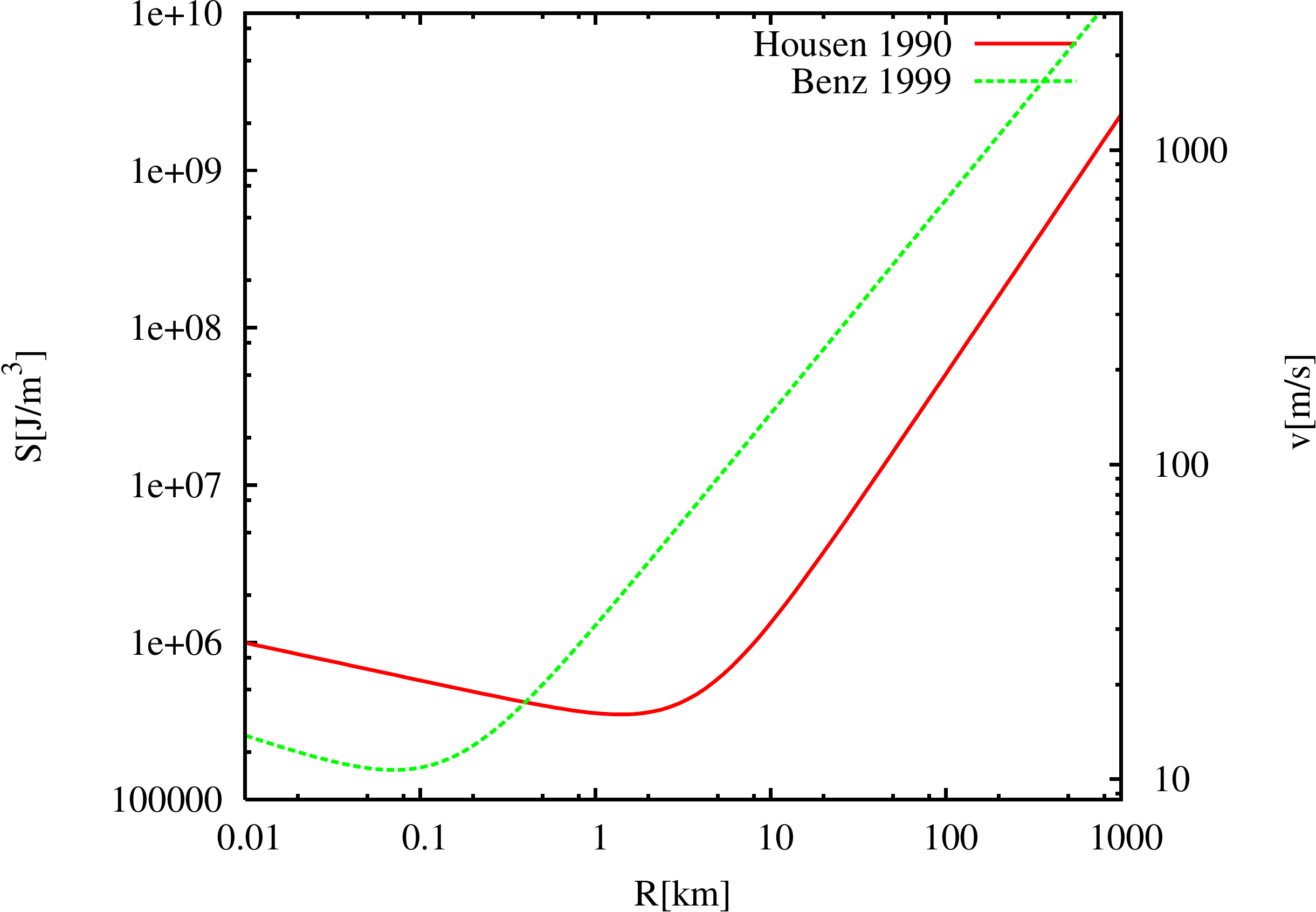}}
\caption{Impact strength according to Eq.~\ref{HousenS} and Eq.~\ref{BenzS}. The right
axis gives the corresponding impact velocity according to $S=1/2\rho v^2$ with $\rho = 2.7$ g$/$cm$^3$.  
\label{StrengthP}
}
\end{figure}

Later, \cite{Benz1999} obtained another result through SPH simulations (for basalt, $v=$3 km$/$s):

\begin{align}
S   &= S_0\left(\frac{R}{1\,\mathrm{m}} \right)^{-0.38}\left[1+6.989 \times10^{-5}\left(\frac{R}{1\,\mathrm{m}} \right)^{1.74}\right]\nonumber \\
f_{\mathrm{KE}}& \approx 0.01 \label{BenzS} \\
S_0 &= 6.082 \times 10^5 \textrm{ J}\textrm{m}^{-3} \\
\rho &= 2.7 \frac{\mathrm{g}}{\mathrm{cm}^3}
\end{align}

\noindent
$f_{\mathrm{KE}}$ is a measure of the kinetic energy that is transferred to the fragments:
\begin{equation}
 E_{\mathrm{kin}}^{\mathrm{frag}}   =   \frac{f_{\mathrm{KE}}}{2} E_{\mathrm{kin}}
\end{equation}
We introduce a dimensionless measure $\gamma$ of the relative importance of gravity
for the result of a collision. It is defined as the ratio of the energy per volume $S_G$
that is necessary to disperse the fragments to the impact strength $S_0$:
\begin{align}
S_G &=  2\pi \frac{4-2\sqrt[3]{2}}{5 f_{\mathrm{KE}} } G R^2 \rho^2 \nonumber \\
\gamma &:=  S_G/S
\end{align}
The first step towards the prediction of a collisional outcome is to relate the impact energy
and the impact strength to ascertain the size of the largest fragment $f_l$.
Laboratory experiments and simulations indicate the functional form
\begin{align}
\renewcommand{\arraystretch}{1.5}
 \epsilon(f_l) &=  \left\{ \begin{array}{l} 2(1-f_l) \quad \textrm{for}\quad f_l>\frac{1}{2} \nonumber \\
                           (2 f_l)^{-\frac{1}{K} } \quad \textrm{otherwise.}
                       \end{array} \right.
\renewcommand{\arraystretch}{1.0}  \label{calcfl} \\
\epsilon   &=    \frac{E_{\mathrm{kin}} \rho}{2 S M  }
\end{align}
which is both valid in the fragmentation regime and the cratering limit. The size
of the largest fragment is used to derive the full size distribution.
To accomplish the decomposition ``seed fragments'' are distributed inside the target according
to the largest desired fragment. The full set of fragment is derived from
a {Voronoy tessellation}\,\footnote{The Voronoy tessellation assigns every volume element
to the closest seed point. First applications date back to the 17th century, but the Russian mathematician of Ukrainian origin Georgy Feodosevich Voronoy
put it on a general base in 1908.} using these seed points.
Fig.~\ref{FragB} depicts the result of such a decomposition. The fragment velocities
are calculated from the total kinetic energy
after the collision to initiate a post-collisional $N$--body calculation to treat
reaccumulation.

We conducted a large set of such calculations to cover a sufficient range in $f_l^i$
(i.\,e. impact energy) and $\gamma$ (i.\,e. body size). Table~\ref{DataCom} summarises
the derived values of the largest and second largest fragment including reaccumulation.

\begin{table*}
\begin{center}
Largest Fragment
\vspace*{5mm}

\begin{tabular}{|l|l|l|l|l|l|l|l|l|l|}\hline
$\gamma$ \ $f_l^i$ & 0.1 & 0.2 & 0.3 & 0.4 & 0.5 & 0.6 & 0.7 & 0.8 \\ \hline
0.02552 & 0.10000 & 0.20000 & 0.30000 & 0.40000 & 0.50000 & 0.60000 & 0.70000 & 0.80000\\ \hline
0.19897 & 0.10000 & 0.20000 & 0.30000 & 0.40000 & 0.50000 & 0.60000 & 0.70000 & 0.80000\\ \hline
0.67985 & 0.10000 & 0.20000 & 0.30000 & 0.40000 & 0.57612 & 0.67315 & 0.82006 & 0.96073\\ \hline
1.14050 & 0.10000 & 0.31526 & 0.35708 & 0.61362 & 0.83511 & 0.92832 & 0.94380 & 0.97572\\ \hline
1.77057 & 0.10884 & 0.58883 & 0.75974 & 0.87922 & 0.92755 & 0.93662 & 0.97107 & 0.97924\\ \hline
2.26021 & 0.15895 & 0.68891 & 0.87217 & 0.89592 & 0.92965 & 0.96089 & 0.96701 & 0.98727\\ \hline
3.11626 & 0.30954 & 0.83774 & 0.90272 & 0.92682 & 0.95943 & 0.95565 & 0.97677 & 0.98791\\ \hline
\end{tabular}
\end{center}
\begin{center}
Second largest Fragment
\vspace*{5mm}

\begin{tabular}{|l|l|l|l|l|l|l|l|l|l|}  \hline
$\gamma$ \ $f_l^i$ & 0.1 & 0.2 & 0.3 & 0.4 & 0.5 & 0.6 & 0.7 & 0.8 \\ \hline
0.02552 & 0.08171 & 0.10770 & 0.06471 & 0.08107 & 0.04976 & 0.03982 & 0.03171 & 0.02155 \\  \hline
0.19897 & 0.09174 & 0.09533 & 0.08410 & 0.06967 & 0.04930 & 0.04825 & 0.03510 & 0.02077 \\ \hline
0.67985 & 0.07713 & 0.08365 & 0.07791 & 0.08387 & 0.07026 & 0.06147 & 0.06082 & 0.00847 \\ \hline
1.14050 & 0.08621 & 0.07256 & 0.10331 & 0.09640 & 0.02467 & 0.00675 & 0.00783 & 0.00265 \\ \hline
1.77057 & 0.07909 & 0.05549 & 0.06961 & 0.02035 & 0.00329 & 0.00719 & 0.00273 & 0.00161 \\ \hline
2.26021 & 0.06693 & 0.02288 & 0.00528 & 0.00882 & 0.00584 & 0.00268 & 0.00664 & 0.00126 \\ \hline
3.11626 & 0.06940 & 0.00884 & 0.00384 & 0.00488 & 0.00064 & 0.01007 & 0.00225 & 0.00162 \\ \hline
\end{tabular}
\end{center}
\caption[Fragmentation look-up tables]{Data compilation of the fragmentation calculations. \label{DataCom} }
\end{table*}

\section{Collisional Cascades}
\label{CollCasc}

\newcommand{\difft}{ \frac{\partial}{\partial t} }
\newcommand{\diffm}{ \frac{\partial}{\partial m} }
\newcommand{\diffu}{ \frac{\partial}{\partial u} }
\newcommand{\dint }{ \int \!\! \int }
\newcommand{\vrel }{ v_{\mathrm{rel}} }

A first well-defined application of the fragmentation model is a {collisional cascade}.
The term {\em cascade} denotes that fragments of one collision in a many-body system may
hit other bodies, whose fragments further shatter even more bodies. Thus the particle number
increases exponentially with every subsequent collision.

Although the formation of planets requires a net growth due to collisions, this
destructive process plays a role in the formation of larger bodies as the
overall size distribution controls the accretion rate of the protoplanets.
Therefore it is worth to have a closer look into this mechanism.

\subsection{Self--similar collisions}

A system of colliding bodies is usually embedded in a broader context, like
stars moving in a galaxy or asteroids orbiting in our own solar system. First,
we simplify this dynamical background as well as some aspects of the collisions
to make the problem tractable.

The first step is to decompose an inhomogeneous system into smaller subvolumes
which are locally homogenous. Furthermore, it is assumed that these subvolumes
hardly interact with each other. Hence it is possible to apply the {\em
particle--in--a--box--method} \citep{Safronov1969} to analyse collisions within
the small subvolumes:

\begin{enumerate}
 \item All particles are contained in a constant volume.
 \item The particle sizes are described by a distribution function $n(m)$, i.\,e.
   the particle number per volume and mass interval.
 \item For convenience, we assume a constant (or typical) relative velocity for a given pair
  of colliding bodies.
\end{enumerate}

\noindent
The distribution function is evolved by the {coagulation equation}.
We modified the equation given by \cite{Tana1996b} by introducing a new function $M_{\mathrm{red}}$ to arrive
at a more concise expression:
\begin{equation}
0  =  \difft mn(t,m)+  \diffm F_m(t,m) \label{ME1}
\end{equation}
The mass flux $F_m$ is given by:

\begin{align}
F_m &=  - \dint
n(t,m_1)n(t,m_2) \xi 
dm_1 dm_2 \nonumber \\
\xi & \equiv \sigma(m_1,m_2)v_{\mathrm{rel}}M_{\mathrm{red}}(m,m_1,m_2)\label{Flux1}\\
M_{\mathrm{tot}} &=  \int n(t,m)m dm \\
\difft M_{\mathrm{tot}} &=  F_m(m_{\mathrm{min}}),
\end{align}

\noindent
where $\xi$ is the coagulation kernel, $n$ is the already introduced size
distribution,$v_{\mathrm{rel}}$ is the mean relative velocity, $\sigma$ is the
cross section for colliding bodies ($m_1$, $m_2$) and $M_{\mathrm{red}}$ is the
newly introduced  {fragment redistribution function}.  $M_{\mathrm{red}}$
contains all information on the fragments arising from the breakup of body
$m_1$ due to the impact of body $m_2$. Its definition avoids double counting of
collisions in the above integral. The redistribution function is related to the
differential number distribution function $n_{\mathrm{coll}}(m_1,m_2,m)$,
i.\,e. the number of fragments produced by a collision per mass interval. Since
the target $m_1$ formally disappears, it is included as a negative
contribution:

\begin{align}
 & M_{\mathrm{red}}(m,m_1,m_2) := \nonumber \\
 & \int_0^m \big(
  n_{\mathrm{coll}}(m_1,m_2,\tilde m)-\delta(\tilde m-m_1) \big) \tilde m d \tilde m \label{Mredncoll}
\end{align}

Mass conservation in each collision is reflected by $  M_{\mathrm{red}}(0,m_1,m_2)= M_{\mathrm{red}}(\infty,m_1,m_2)=0$.
The cross section $\sigma$ depends on the velocities and radii $R_i$ of the particles. A simple
approach is the {geometric cross section}:
\begin{equation}
 \sigma(m_1,m_2)  =  \pi(R_1+R_2)^2
\end{equation}
If gravity plays an important role during encounters, two colliding bodies move on hyperbolic orbits
with a pericentre distance that is smaller than the impact parameter. This leads to an additional
enlargement of the cross section, denoted as {gravitational focusing}:
\begin{equation}
 \sigma(m_1,m_2)  =  \pi(R_1+R_2)^2 \left(1+\frac{2G(m_1+m_2)}{v_{\mathrm{rel}}^2(R_1+R_2)}\right)
\end{equation}
A special class of collisional models are self-similar collisions. Self-similarity implies
an invariance of the collisional outcome with respect to the scale of the colliding bodies.
If the target mass as well as the projectile mass are enlarged by a factor of two, then only
the masses of all fragments doubles without further changes in the collisional outcome.
They allow us to introduce the convenient dimensionless fragment redistribution function $f_m$:
\begin{align}
 M_{\mathrm{red}}(m,m_1,m_2) &=   m f_m(m_1/m,m_2/m)
\end{align}
We follow \cite{Tana1996b} and employ the
substitution\footnote{A similar approach to the solution of the coagulation equation
is the Zakharov transformation, see \cite{Connaughton2004}.}
$m_1=mx_1$, $m_2=mx_2$ to simplify Eq.~\ref{Flux1}:

\begin{align}
F_m & = - \dint
n(t,mx_1)n(t,mx_2)m^{11/3} \nonumber \\
& \sigma(x_1,x_2)v_{\mathrm{rel}}f_m(x_1,x_2) dx_1 dx_2 \label{EqFluxS}
\end{align}

\noindent
A simple solution is a steady-state cascade with $F_m=\mbox{const}$. The loss of bodies of a given
size is balanced by the fragment supply from larger bodies, hence the system maintains a steady-state $\difft
n(t,m)=0 $. Eq.~\ref{EqFluxS} inspires the ansatz $n(m) \propto m^{-k}$, which yields $k=11/6$. This is the
well known equilibrium slope in self-similar collisional cascades, which was already found by \cite{Dohnanyi1969}.
Strong gravitational focusing changes the
exponent\footnote{\cite{Tana1996b}
state that $k<2$ is a necessary condition for a finite mass flux. However, their analysis is not valid
for all possible collisional models.} to $k=13/6$. Both steady-state
solutions seem to be rather artificial, as they contain an infinite amount of mass and require a
steady mass influx from infinity. However, they provide an appropriate description for the relaxed
fragment tail of a size distribution, as long as the largest bodies provide a sufficient flux of new fragments.
Once the largest bodies start to decay, the finite amount of mass in the system leads to an overall
decay of the collisional cascade.
Thus we seek for a more general solution to Eq.~\ref{ME1} using the ansatz $n(t,m)=a(t)n_0(m)$:
\begin{align}
 \difft a(t) &= - C a(t)^2 \\
  m n_0(m) &= \frac{1}{C} \diffm F_m \label{steS}
\end{align}
$C$ is determined by fixing $n_0$ at an arbitrary value $m^{*}$.
$a(t)$ is independent of the collision model:
\begin{align}
a(t) & = \frac{1}{1 +C t } \\
 C   & \propto n(m^*)  \label{Cn}
\end{align}
A power law solution is $n_0(m)\propto -C m^{-k+1}$ which is only valid for $C<0$ (agglomeration dominates).
To examine $C>0$, we perturb the already
known equilibrium solution:
\begin{align}
  n_0(m) &= N_0 m^{-k} - C N_1 m^{-2k+2}+\mathcal{O}(C^2) \label{FinitMR}\\
 \frac{1}{N_1} & = (2-k) \dint
x_1^{-k}x_2^{-2k+2}\sigma(x_1,x_2) \nonumber \\
& v_{\mathrm{rel}}\left(f_m(x_1,x_2)+f_m(x_2,x_1)\right) dx_1
dx_2
\end{align}
$N_1$ is small if the integral on the right hand side is
large. This is the case for a sufficiently large impact strength. Eq.~\ref{FinitMR} has the
interesting property that $n(m')=0$ for some mass $m'$, given that $k<2$. This mass $m'$ represents the
largest body in the system, e.\,g. the largest asteroid in a fictitious  asteroid belt.

\section{Size--dependent Strength}

Self-similarity is an enormous help in analysing the coagulation equation. It
releases completely the need to know any specific details of the collisional
process and provides valuable insight at the same time. But self-similarity is
also a strong limitation on the underlying collisional physics.

A major component of a fragmentation model is the knowledge of the impact
strength as a function of size. Simulations as well as asteroid families
establish that it is not some fixed value, but changes with size which
immediately breaks the self-similarity.  Larger bodies are weaker due to an
increasing number of flaws (there are no big monocrystals), but then gravity
leads to a turnover and increases the strength.

We model the size dependent strength $S$ with a power law to examine the
influence on the equilibrium solution. The velocity dispersion $v$ and the
collisional cross section $\sigma$ are also modelled with power laws to account
for relaxation processes:

\begin{align}
 v & =  v_0 \left( \frac{m}{m_0} \right)^w  \\
 \sigma & =  \sigma_0 \left( \frac{m}{m_0}\right)^s  \\
 S &= S_0 \left(\frac{m}{m_0}\right)^{\alpha}
\end{align}

\noindent
The subscript ``0'' denotes values for an arbitrarily chosen scaling mass.
Since smaller bodies are more abundant than larger ones, we safely assume
that most collisions involve a large mass ratio. In addition, we assume $w<0$, since
we expect energy equipartition to some degree in most cases. These restrictions
lead to the following simplifications $(m_1>m_2)$:
\begin{align}
    \sigma(m_1,m_2) &\approx  \sigma(m_1) \label{AprxScale1} \\
        v_{\mathrm{rel}} &\approx  v(m_2) \\
 \epsilon &\approx  \frac{1}{2} \frac{m_2 \rho v_{\mathrm{rel}}^2}{m_1S_1} \label{AprxScale3}
\end{align}
Therefore the smaller body $m_2$ enters only through the specific energy $\epsilon$:
\begin{align}
F_m & \approx  - \dint
n(t,m_1)n(t,m_2)\sigma(m_1)v_{\mathrm{rel}}(m_2)m_1 \nonumber \\
& f_m(m_1/m,\epsilon)
dm_1 dm_2
\end{align}
We introduce  new dimensionless quantities with the help of Eq.~\ref{AprxScale1}--\ref{AprxScale3}
 to simplify the integral:
\begin{align}
  m_1 &=  m x_1 \nonumber \\
  m_2 &=  m_0 \left( \frac{m_1}{m_0}\right)^{\frac{1+\alpha}{1+2w}}\left(\frac{2S_0}{\rho v_0^2}\right)^{\frac{1}{1+2w}}
  \epsilon^{\frac{1}{1+2w}}
\end{align}
Again we assume a power law for the density $n\propto m^{-k}$ and change the integration parameters
to $(x_1,\epsilon)$. Applying the constant--flux condition yields the equilibrium exponent
\begin{equation}
 k   \approx   \frac{s+3+\alpha+w(2s+\alpha+5) }{2+\alpha+2w } \label{kgen}
\end{equation}
 and the scaling exponent $k'$ of the total mass loss:
\begin{align}
  k' & \approx    \frac{s-w+1}{2+\alpha+2w}  \nonumber \\
\difft M_{\mathrm{tot}} &\propto  - {\tilde S}^{-k'} \qquad \tilde S = \frac{2S_0}{\rho v_0^2}  \label{MSdot}
\end{align}
The exponent $k'$ in Eq.~\ref{MSdot} is close to unity for realistic values of the free parameters. Thus
the mass loss is roughly inversely proportional to the strength of the bodies. The general
formula Eq.~\ref{kgen} contains the special solution of  \cite{Brien2003},
who concentrated on the parameters \mbox{$s=2/3$}, \mbox{$w= 0$} and a
special collisional model. In fact, the derivation applies to a much wider class of collisional
models that we denote as {scalable collisional models}. Scalable indicates that the model is self-similar
except a scaling of the impactor mass.

\section{Perturbation of Equilibrium\label{PertEquilib}}

The derived scaling relations provide insight into the overall
properties of a collisional cascade, which is in (or close to)
equilibrium. However, they do not provide information on how the
equilibrium is attained or how the system responds to various
external perturbations. A rigorous approach would be the approximate
solution of the coagulation equation\footnote{Appendix~\ref{AprCoagEq} highlights a possible approach.},
which is by no means simple since it requires a careful analysis
of the collision model.

Hence we turn to perturbations of the equilibrium size distribution, as it is
easier to asses the quality of the derived expression for a variety of collision
models. In addition, all equations are linear in the perturbation, allowing the
detailed analysis of the solution.

\begin{figure}
\resizebox{\hsize}{!}{\includegraphics[scale=1,clip]{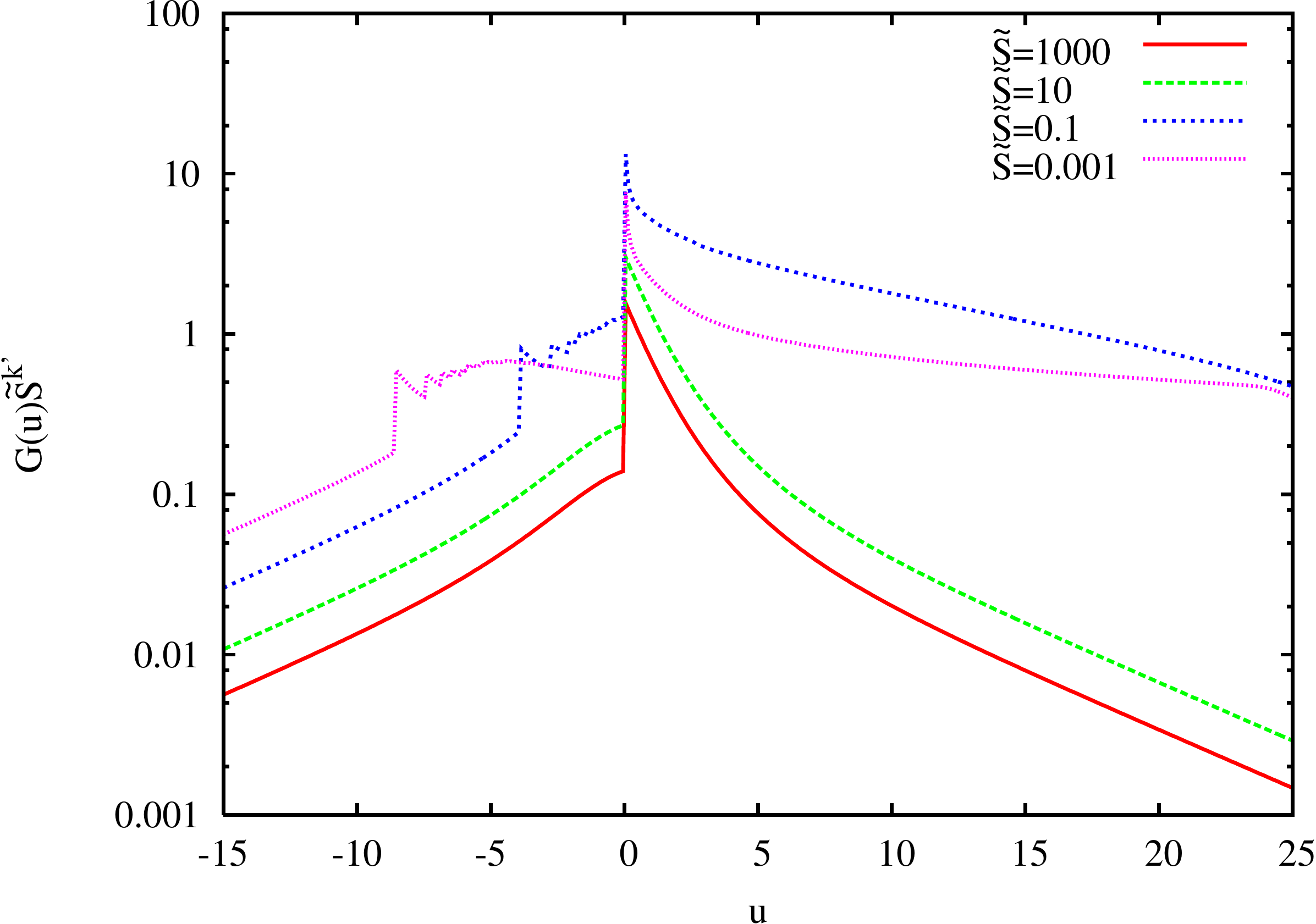}}
\caption{Scaled fragmentation kernel $G(u)$ for a simple fragmentation model (see Eq.~\ref{TestFrag}) and different scaled impact
strength $\tilde S$.  
\label{FigGKernel}
}
\end{figure}

If the equilibrium solution $ n(m)=n_0(m/m_0)^{-k} $ is perturbed with a small
deviation $ \Delta n(m) $, we get to first order:
\begin{align}
 0 &=  \difft m \Delta n(m) + \diffm F_p(t,m)  \label{EqDeltan} \\
 F_p &=  -\dint \Delta n(m_1)n(t,m_2)\sigma(m_1,m_2)v_{\mathrm{rel}} \nonumber \\
     &  \times \left(
         M_{\mathrm{red}}(m,m_1,m_2)+M_{\mathrm{red}}(m,m_2,m_1) \right) dm_1dm_2 \nonumber
\end{align}
Despite of the expansion in $\Delta n$, Eq.~\ref{EqDeltan} is still a complicated
{integro-differential equation}. Thus it is not possible to obtain a solution
without further information about the problem. While there is no general solution,
we restrict our attention to self-similar collisional processes.
In virtue of this assumption it is possible to simplify Eq.~\ref{EqDeltan}, as we can
see in Eq.~\ref{PEq} and \ref{PEq}. In those expressions 
\begin{figure*}
\begin{align}
0 &= \difft m \Delta n(m)-n_0 m_0^3\sigma_0 v_0 \diffm \int \Delta n(t,mx_1) F(x_1) (mx_1/m_0)^{k} dx_1 \label{PEq} \\
  F(x_1) &= \int m_0^{2k-3}x_1^{-k}x_2^{-k} \frac{\sigma(x_1,x_2)}{\sigma_0}
   \frac{v_{\mathrm{rel}}}{v_0} ( f_m(x_1,x_2)+f_m(x_2,x_1) ) dx_2 \label{PEq2}
\end{align}
\end{figure*}
$\sigma_0$ and $v_0$ are velocity and cross section of an arbitrarily chosen scaling mass $m_0$.
$F(x_1)$ contains all information about the collisional process. If collisions do not result
in extreme outcomes, like cratering or a complete destruction of the target, most of
the fragment mass is contained in bodies with similar size as the parent body. Hence
we expect that $F(x_1)$ peaks around $x_1 \approx 1$ and drops to zero as $x_1$ gets
larger (or smaller).
We introduce the dimensionless relative perturbation $g(m)$:
\begin{equation}
    g(m) = \frac{\Delta n(m)}{n(m)} =\frac{\Delta n(m) m^k}{n_0 m_0^k}
\end{equation}
Thus the new differential equation reads:
\begin{align}
 0 &= \difft (m/m_0)^{1-k} \Delta g(m)- \nonumber \\
 & n_0 m_0^2\sigma_0 v_0 \diffm \int \Delta g(t,mx_1) F(x_1) dx_1
\end{align}
We change to logarithmic coordinates to arrive at a convolution integral:
\begin{equation}
    u   =  \ln(m/m_0)\quad\qquad \quad \quad u_1 = \ln(x_1)
\end{equation}
Furthermore we define a collisional timescale $\tau_0$
\begin{equation}
    \tau_0  =  (n_0 m_0\sigma_0 v_0)^{-1} \label{tau-coll}
\end{equation}
to obtain a more concise expression. The transformed equation is:
\begin{align}
 0   =& \difft g(t,u)e^{u(2-k)} -  \frac{1}{\tau_0} \diffu \int g(t,u+u_1)G(u_1) d u_1  \label{gEq} \\
G(u) =& F(e^u)e^{u}
\end{align}
If $g(u)$ is varying on a scale larger than the width of the kernel $G(u)$ (compare Fig.~\ref{FigGKernel}),
it is justified to expand $g(u)$ under the integral. We retain the first two moments of $G(u)$:
\begin{align}
 0 &= \difft g(t,u)e^{u(2-k)} - \frac{G_0}{\tau_0}  \diffu g(t,u) - \frac{G_1}{\tau_0}  \frac{\partial^2}{\partial u^2} g(t,u) \label{AprgEq} \\
G_k & = \int u^k G(u) d u
\end{align}

The first order moment $G_1$, which introduces a diffusive term, is omitted in the following for
clarity\footnote{The study of wave-like structures in the size distribution \citep[see e.g.]{Bagatin1994} requires
even the second order moment~$G_2$.}.
We introduce a fragmentation time $ \tau_\mathrm{frag}(u) $ and transform Eq.~\ref{AprgEq} back to $m$:
\begin{align}
 0 &=  \difft g(t,m) - \frac{m}{ \tau_{\mathrm{frag}}(m)}  \diffm g(t,m) \label{EqgmFin} \\
 \tau_\mathrm{frag} &=  \frac{\tau_0}{G_0} e^{u(2-k)} \nonumber \\
                          &=  \frac{\tau_0}{G_0} (m/m_0)^{2-k} \label{Eq_Tau}
\end{align}
Eq.~\ref{EqgmFin} is a modified {advection equation}, which conserves the total mass. It is possible
to derive equations similar to Eq.~\ref{EqgmFin} for any collisional model. However, the general approach
is less fruitful, as it lacks a robust frame of a known equilibrium solution and reliable scaling
relations. Therefore we provide only the extension to
scalable collisional models in Appendix~\ref{ScalFlux}.
We readily obtain the general solution:
\begin{align}
 g(t,m) &=  f\left(t+  \tau_0 \frac{(m/m_0)^{(2-k)}}{G_0(2-k)} \right) \nonumber \\
\Delta n(t,m) &=  n(m) f\left(t+  \tau_0 \frac{(m/m_0)^{(2-k)}}{G_0(2-k)} \right)
\end{align}
The function $f$ is determined by the initial value $g(0,m)$ of the perturbation. As the
collisional cascade evolves, the initial perturbation function is shifted as a whole to smaller
masses. This evolution becomes clearer if we attach labels $M(0)$ to the initial
perturbation function and follow the time evolution of these tags.
The functions $M(t)$ are the {characteristics}\,\footnote{In general, characteristics of a partial differential
equation are paths along which the solution is constant.}
of the differential equation~\ref{EqgmFin}:
\begin{equation}
  M(t) = m_0\left[ (M(0)/m_0)^{(2-k)} -t/\tau_0 G_0(2-k) \right]^{1/(2-k)}
\end{equation}
The meaning of the fragmentation time $ \tau_{\mathrm{frag}}$ becomes clear
by the relation
\begin{equation}
   \frac{M}{\dot M} = -\tau_{\mathrm{frag}}
\end{equation}
which is the time until a body has lost a significant fraction of its mass due to destructive collisions.
A comparison of the perturbation equation~\ref{EqgmFin} with the scaling relations from
the previous section gives the scaling of the zeroth order moment $G_0$ with respect to the impact strength:
\begin{equation}
  G_0 = G_0' {\tilde S}^{-k'}
\end{equation}
$G_0'$ should only depend on the fragmentation model (i.\,e. fragment size distribution as a function
of the largest fragment $f_l$) within the limits of this approximation. Fig.~\ref{FigGKernel} shows
that the scaling with the impact strength works quite well, except slight variations which are small
compared to the covered range of impact strengths.
Likewise, it is possible to restate the total equilibrium flux $F_{\mathrm{eq}}$ in terms of $G_0'$:
\begin{equation}
   F_{\mathrm{eq}}(m)  \approx   -\frac{G_0'}{2} n(m)^2\sigma(m) m^3 v_{\mathrm{rel}}  {\tilde S}^{-k'}
\end{equation}
The fragmentation timescale $\tau_{\mathrm{frag}}(m)$ allows a more intuitive expression:
\begin{equation}
 F_{\mathrm{eq}}(m)  \approx   -\frac{1}{2}\frac{n(m)m^2}{\tau_{\mathrm{frag}}(m)}
\end{equation}
Our simple collisional model (see Fig.~\ref{FigGKernel} and Eq.~\ref{TestFrag}) refers to:
\begin{equation}
   F_{\mathrm{eq}}(m)   =   - (1 \dots 30)\times n(m)^2\sigma(m)m^3v_{\mathrm{rel}} {\tilde S}^{-k'}
\end{equation}

\section{Migration and Collisions}

The local perturbation analysis is only applicable to a planetesimal disc, if the migration velocity of the
planetesimals is negligible small. This assures that collisional cascades at different
radial distances do not couple to each other, so that the whole disc is composed of many
local cascades. While this assumption is justified for larger bodies, migration is strongly
influencing bodies below 1 km in size. Hence we extend our analysis to examine the influence of
migration on the (no longer) local collisional processes.

We assume that the collisional evolution of the system leads to an equilibrium planetesimal
distribution everywhere in the disc:
\begin{equation}
 \Sigma_0(r,m)  =  \Sigma_{r,0}(r)C_0(m) \label{NoMig}
\end{equation}
$\Sigma_r(r)$ is the total surface density at a given distance $r$, while $C_0(m)$ is the
universal equilibrium distribution. Though the planetesimal distribution at larger
sizes is likely different at different locations in the disc, we only demand a
universal function at smaller sizes, where migration is important. The power law
exponent $k$ depends on the details of the invoked physics, but numerical simulations
show that $k\approx 2$ is a fiducial value. Eq.~\ref{NoMig}
does not yet include migration effects. If we include migration, the surface
density is modified to
\begin{equation}
 \Sigma(r,m)  =   g(r,m)\Sigma_0(r,m)
\end{equation}
where the dimensionless function $g$ contains the changes due to migration.
The collisional evolution is governed by the continuity equation with an additional collisional
term
\begin{equation}
\frac{ \partial \Sigma(r,m)}{\partial t} - \frac{1}{r}\frac{\partial}{\partial r}\left(v(r,m)r\Sigma(r,m)\right)  =   \dot \Sigma_{\mathrm{coll}}
\end{equation}
where $v(r,m)$ is the migration velocity (see Eq.~\ref{MigSpeed}), defined such that positive $v$
imply an inward migration. We express the collisional term with the help of Eq.~\ref{EqgmFin}
and seek for a steady-state solution $\dot \Sigma=0$:
\begin{equation}
\frac{1}{\tau_{\mathrm{frag}}(m,r)}\frac{\partial g}{\partial m}m\Sigma_{r,0}(r) +\frac{1}{r}\frac{\partial}{\partial r}\left(gvr\Sigma_{r,0}(r)\right)  =  0
\label{Eq_Sig1}
\end{equation}
$\tau_{\mathrm{frag}}(m,r)$ is the fragmentation timescale of a mass $m$ at a distance $r$. Since the surface density $\Sigma$ and the various
contributions to the drag force are well described by a power law (with respect to radius),
Eq.~\ref{Eq_Sig1} further simplifies to:
\begin{equation}
\frac{1}{\tau_{\mathrm{frag}}(m,r)}\frac{\partial g}{\partial m}m +\frac{\partial g}{\partial r} v-  \frac{b}{r} gv =  0
\end{equation}

\begin{figure}
\resizebox{\hsize}{!}
          {\includegraphics[scale=1,clip]{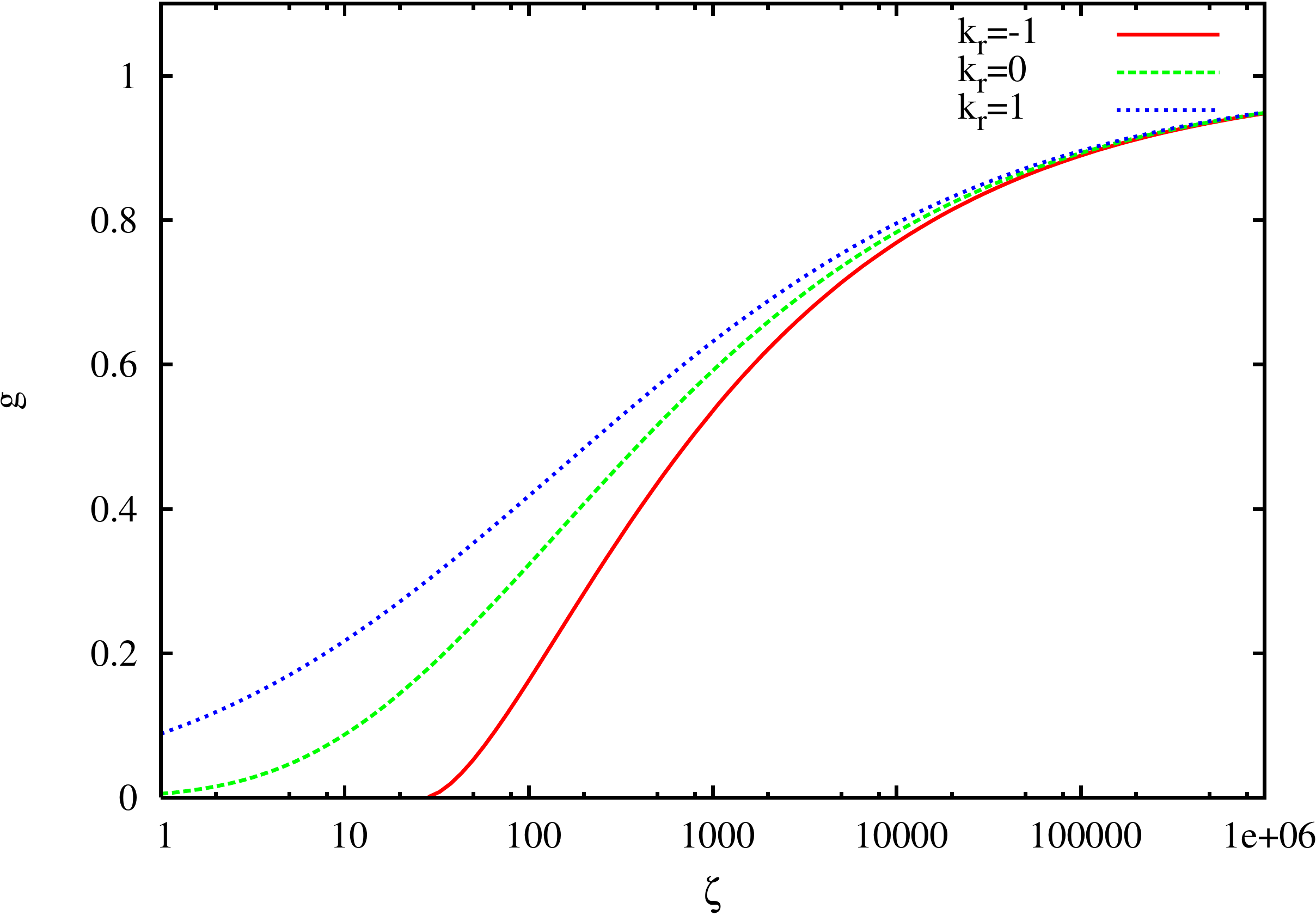}}
\caption
   {  
Cut-off function $g$ according to Eq.~\ref{SolvefZeta}. The mass exponent is $k_m=1/3$,
while the mass influx exponent is $b=1.75$ according to the minimum solar nebula.  
   }
\label{CutOff}
\end{figure}

$b$ is a combination of the various invoked power law exponents.  As the surface density $\Sigma$ and the
gas density drop with increasing radius in any realistic disc model, it is safe to assume  $b>0$.
We choose a self-similar ansatz for $g$:
\begin{equation}
 g(r,m) = g(\zeta) \quad, \quad  \zeta =m g_m(r)
\end{equation}
The new differential equation is
\begin{equation}
\frac{1}{\tau_{\mathrm{frag}}(m,r)}\frac{d g}{d \zeta}mg_m(r) +m \frac{d g}{d \zeta}\frac{d g_m}{d r} v-  \frac{b}{r} gv =  0
\end{equation}
which is equivalent to the more concise expression:
\begin{equation}
 \frac{d \ln(g) }{d \ln(\zeta)}\left(\frac{r}{v\tau_{\mathrm{frag}}}+\frac{d \ln(g_m)}{d\ln(r)}    \right)  =  b
\end{equation}
We assume a power-law dependence for the timescale ratio  $\tau_{\mathrm{mig}}/\tau_{\mathrm{frag}}$:
\begin{equation}
 \frac{r}{v\tau_{\mathrm{frag}}} = \frac{\tau_{\mathrm{mig}}}{\tau_{\mathrm{frag}}} = (m/m_0)^{k_m} (r/r_0)^{k_r}
\end{equation}
The cut-off mass $m_0$ at a distance $r_0$ has a timescale ratio $\tau_{\mathrm{mig}}/\tau_{\mathrm{frag}}=1$,
which defines a proper lower cut-off within this context. Hence the solution is:
\begin{align}
 b  =& \frac{d \ln(g) }{d \ln(\zeta)}\left(\zeta^{k_m}+\frac{k_r}{k_m}    \right) \label{EqFinZeta} \\
  g_m(r) =& \frac{(r/r_0)^{k_r/k_m}}{m_0} \\
 g(\zeta) =&  \left(1+\frac{k_r}{k_m \zeta^{k_m} }\right)^{-b/k_r} \label{SolvefZeta}
\end{align}
Though the analytical solution Eq.~\ref{SolvefZeta} provides a complete description of the
lower cut-off of the size distribution, it is more appropriate
within the frame of this discussion to translate the equilibrium solution to
an equilibrium mass loss due to migration:
\begin{align}
 \dot \Sigma_{\mathrm{mig}}(r,m) &= -\frac{bv}{r}\Sigma+\frac{bv}{r}\Sigma\frac{k_r/k_m}{\zeta^{k_m}+k_r/k_m} \nonumber \\ 
 &= - \frac{b\Sigma}{\tau_{\mathrm{mig}}+k_r/k_m\tau_{\mathrm{frag}}} \label{EqMassLoss}
\end{align}
An inspection of the timescale ratio shows that the mass exponent $k_m$ should be positive,
whereas simple estimations of $k_r$ on the basis of the minimum mass solar nebula are somewhat
inconclusive. The value of $k_r$ is so close to zero that any change in the assumed
equilibrium slope or the impact strength scaling gives easily both positive and negative
values. Moreover, Eq.~\ref{EqMassLoss} requires a globally relaxed planetesimal disc,
but the huge spread in the various involved timescales at different radii inhibits
any significant relaxation in the early stages.

However, it is possible to gain valuable information from the two limiting
cases $k_r>0$ and $k_r<0$.  Both values of $k_r$ give the proper limit
$g\rightarrow 1$ at large masses, where the migration timescale is much larger
than the fragmentation timescale and we recover the steady-state collisional
cascade.

A positive exponent $k_r$ reduces the effective mass loss due to migration, as
fragments from the outer part of the disc replenish the local mass loss. Hence
the fragmentation timescale controls the net loss of smaller planetesimals. In
contrast, a negative exponent $k_r$ leads to a pronounced cut-off in the size
distribution, since only larger planetesimals are replenished through inward
migration.  Though the mass loss rate is singular at some mass $m'$, this sharp
cut-off is an artifact due to the perturbation approximation.

Our analysis is subjected to several restrictions. We applied the perturbation
equation to values of $g$ that exceed the limit for a safe application (i.\,e.
$g \not\approx 1$) of the perturbation expansion.  Furthermore, the
steady-state solution requires a global relaxation of the collisional
processes, which is practically never obtained during the disc evolution.
Despite of these restrictions, we gained insight on a more qualitative level.
Numerical calculations indicate that the perturbation approximation is
inappropriate close to the lower cut-off of the size distribution. However, a
comparison of different exponents $k_r$ (see Fig.~\ref{CutOff}) attributes only
a minor role to the replenishment of fragments due to inward migration. Only
unrealistic small slopes $b$ of the migrational mass influx would strengthen
the importance of this process. Though temporally non-equilibrium phenomena are
not ruled out by the previous derivation, their study would require the global
simulation of the system.

\section{Coagulation}

While most coagulation kernels are only restricted to a limited analytical
analysis (e.\,g. scaling relations), there exist some special kernels that
allow the closed solution of the coagulation equation. All rely on the assumption
of perfect mergers, which allows the reformulation of the general equation~\ref{ME1}
to
\begin{align}
 \frac{\partial n(m,t)}{\partial t }  &=  \frac{1}{2}\int_0^m \mathcal{K}(m-m',m')n(m-m',t)n(m',t) dm'\nonumber \\
 &  -n(m,t)\int_0^{\infty}\mathcal{K}(m,m')n(m',t)dm' \nonumber
\end{align}
where $\mathcal{K}$ is the coagulation kernel. One of these particular kernels was introduced by \cite{Safronov1969}:
\begin{align}
 \mathcal{K}(m_1,m_2) = A_1(m_1+m_2)
\end{align}
This coagulation kernel implies perfect mergers, where the coalescence rate of two particles $m_1$ and $m_2$ is assumed to be proportional
to the sum of their masses. It seems that this is an artificial choice, devised to allow an analytic
solution. However, the Safronov cross section provides an intermediate case between a geometric
cross section ($\sigma \propto m^{2/3}$) and strong gravitational focusing ($\sigma \propto m^{4/3}$).
A special solution to the initial condition
\begin{equation}
 n(m,0)  =  \frac{n_0}{m_0}\exp\left(-m/m_0 \right)
\end{equation}
is the function \citep[see e.g.]{Ohtsuki1990}
\begin{align}
 n(m,\tau) &=  \frac{n_0 \tilde g}{m\sqrt{1-\tilde  g}}\exp(-(2-\tilde  g)m/m_0)I_1(2m/m_0\sqrt{1-\tilde  g})\label{SavSolut} \\
 \tilde  g &= \exp(-\tau) \qquad \tau = A_1\rho t \nonumber \\
 \rho &=  \int_0^{\infty} mn(m)dm = n_0 m_0
\end{align}
where $\tau$ is the dimensionless time and $I_1$ is a modified Bessel function of the first kind.

\begin{table}
\begin{center}
\begin{tabular}{|c|c|} \hline
$\rho$  & 2,700 kg/m$^3$ \\ \hline
$k$     & $1/6 $ \\ \hline
Model  & Gaussian Scatter \\ \hline
$f_{KE}$  &  0.1 \\ \hline
$K$       & 1.24 \\ \hline
\end{tabular}
\end{center}
\caption[Parameters of the fragmentation model]{\label{CollPar} Main parameters of the collisional model.}
\end{table}

\section{Models for $M_{\mathrm{red}}$ }

Though we already obtained insight into the nature of collisional cascades without a detailed
specification of the coagulation kernel, any detailed study of a collisional system requires
the \mbox{specification} of a realistic collisional model.

First, we restate the well-known perfect accretion model. While it is a gross oversimplification for
collisions among kilometre--sized planetesimals, its simplicity allows a reliable code testing and eases
the comparison with other works:
\begin{align}
 M_{\mathrm{red}}(m,m_1,m_2)&= -m_1\Theta(m-m_1)-m_2\Theta(m-m_2)+\nonumber \\
 &(m_1+m_2)\Theta(m-m_1-m_2)
\end{align}

Although our fragmentation model (see section~\ref{CollModel}) provides a very detailed
description of the outcome of a collision, we abandon most of the details for the
following reasons. The computational effort of the numerical solution of the coagulation
equation scales with the third power of the number of mass bins. Hence we chose a
mass grid whose resolution is by far smaller than the information provided by
the detailed collisional model. As a mismatch of the mass resolution could
produce undesired artifacts, a lower resolution of the collisional model
is needed for consistence. Thus only the largest fragment $f_l(f_l^i,\gamma)$ and the
second fragment $f_l^{(2)}(f_l^i,\gamma)$(which contains information on reaccumulation) enter
the fragment size distribution:
\begin{align}
\frac{M_{\mathrm{red}}(xM)}{M} &= \left\{\begin{array}{ll}
              1  & \mbox{if~} x \ge f_l \\
              1-f_l &\mbox{if~}f_l > x\ge f_l^{(2)}  \\
              (1-f_l-f_l^{(2)})(x/f_l^{(2)})^{f_l} & \mbox{otherwise} \end{array} \right.
\end{align}
Both values $f_l$ and $f_l^{(2)}$ are interpolated from table~\ref{DataCom}, where
the initial fragment size $f_l^i$ is calculated from the dimensionless impact energy $\epsilon$.
We used a reduced fragmentation model for test purposes:
\begin{equation}
 M_{\mathrm{red}}(xM)/M  =  \left\{\begin{array}{ll}
              1    \mbox{if } x \ge f_l \\
              (1-f_l)(x/f_l)^{f_l}   \mbox{otherwise} \end{array} \right.  \label{TestFrag}
\end{equation}
Table~\ref{CollPar} summarises the most important model parameters.

\section{Statistical Model}
\label{StatModel}

The direct approach to the integration of an $N$--body system is, in principle,
possible for any particle number. While this procedure becomes computationally too expensive
for very large particle numbers, a by far more efficient approach is applicable in this regime.
Instead of tracking all particle
orbits, a {distribution function} $f$ (also {phase-space density }),
which gives the probability to find a particle at a position ${\bf x}$ with a velocity ${\bf v}$, contains
the state of the system:
\begin{equation}
    dp = f({\bf x},{\bf v}) d^3x d^3v
\end{equation}
As long as only dynamical interactions are taken into account, the number of all particles (e.\,g. stars, planetesimals) is conserved.
The continuity equation reads:
\begin{equation}
 0  =  \frac{\partial f}{\partial t}+ {\bf v}\cdot \nabla f - \nabla \Phi\cdot\frac{\partial f}{\partial {\bf v}} \label{Eq_Boltz}
\end{equation}
This is the {collisionless Boltzmann equation}. Collisions lead to an additional term
\begin{equation}
\left(\frac{\partial f}{\partial t}\right)_{\mathrm{coll}}  =  \frac{\partial f}{\partial t}+ {\bf v}\cdot \nabla f
- \nabla \Phi\cdot\frac{\partial f}{\partial {\bf v} } \label{Eq_BColl}
\end{equation}
which will be discussed later. $f$ is a function of six variables, so an exact solution
is usually very complicated or even impossible. However, it is possible to gain valuable
insight into the problem by taking the {moments} of the distribution function:
\begin{equation}
   \langle  x_i^n v_j^m \rangle  =   \int f({\bf x},{\bf v}) x_i^n v_j^m d^3x d^3v \qquad n,m > 0
\end{equation}
The spatial density (particles per volume) is related to distribution function:
\begin{equation}
\nu({\bf x})  =  \int f({\bf x},{\bf v}) d^3v
\end{equation}
Integration of Eq.~\ref{Eq_BColl} over all velocities yields the corresponding
continuity equation:
\begin{equation}
 \frac{\partial \nu }{\partial t} + \frac{\partial \nu \bar v_i }{\partial x_i}    =   \left(\frac{\partial \nu}{\partial t}\right)_{\mathrm{coll}}
  \label{ContEq}
\end{equation}
The first order moment with respect to velocity gives the time evolution of the mean velocity $\bar v$
\begin{align}
 \nu \frac{\partial \bar v_j }{\partial t} +\nu  \bar v_i \frac{\partial  \bar v_j }{\partial x_i}
  &=  -\nu \frac{\partial \Phi}{\partial x_j} - \frac{\partial(\nu \sigma_{ij}^2)}{\partial x_i}
  + \nu \left(\frac{\partial \bar v_j}{\partial t}\right)_{\mathrm{coll}} \label{MomEq} \\
\bar v_i &=  \frac{1}{\nu} \int f({\bf x},{\bf v}) v_i d^3v \nonumber \\
\sigma_{ij}^2 &=  \overline{v_i v_j} -\bar v_i \bar v_j
\end{align}
where $\sigma_{ij}$ is the anisotropic velocity dispersion and the continuity equation was used to arrive
at a more concise formulation. Equations~\ref{ContEq} and \ref{MomEq} are the {Jeans equations}. While
the structure of the moment equations is already familiar from hydrodynamics, they do not provide a closed set of
differential equations, since each differential equation of a given moment is related to (yet unknown) higher order
moments. Hence any finite set of momenta needs a {closure relation} -- additional constraints
that relate the highest order moments to known quantities. The choice of this relation is a key element
in the validity of the equations, but it is not unique and depends well on the problem at hand
\citep[compare e.g.]{Larson1970}.

Owing to the geometry of a planetesimal disc, it is useful to express the Boltzmann equation
in cylindrical coordinates
\begin{align}
&\frac{\partial f}{\partial t}+
v_r \frac{\partial f}{\partial r} +v_z \frac{\partial f}{\partial z}
 +\left(\frac{v_{\phi}^2}{r}-\frac{\partial \Phi}{\partial r} \right)  \frac{\partial f}{\partial v_r}
 -\nonumber \\
 &\frac{v_rv_{\phi}}{r}\frac{\partial f}{\partial v_{\phi} } -\frac{\partial \Phi}{\partial z} \frac{\partial f}{\partial v_z}  = 0
 \label{Eq_BoltzCyl}
\end{align}
where all derivatives with respect to $\phi$ have been dropped due to the assumed axisymmetry
of the disc.

\subsection{Distribution Function}

Any statistical description of a planetesimal disc requires the knowledge of the distribution
function. Since the full problem including collisions, encounters and gas drag has no analytic
solution, a collisionless planetesimal disc (i.\,e. no perturbations) is a natural basis
for further investigations. The distribution function that describes such a simplified
system is a solution of the Boltzmann equation. A special solution to Eq.~\ref{Eq_BoltzCyl}
is a thin homogenous planetesimal disc
\begin{equation}
 f(z,v)  =  \frac{\Omega\Sigma}{2\pi^2T_r T_z m}
  \exp\left( -\frac{v_r^2+4v_{\phi}^2}{2T_r} -\frac{v_z^2+\Omega^2z^2}{2T_z}  \right) \label{Fdis}
\end{equation}
provided that the radial velocity dispersion $T_r$ and the vertical dispersion $T_z$ are small
compared the mean orbital velocity $v_K$.
The azimuthal velocity dispersion $T_{\phi}$ is locked to $T_r$ by the local epicyclic frequency $\kappa$
in a central potential, where the ratio $1:4$ is a special solution of \citep[see e.g]{Binney1994}
\begin{equation}
  \kappa^2 T_r = 4\Omega^2 T_{\phi}
\end{equation}
All velocities $v_r$, $v_{\phi}$ and $v_z$ refer to the local Keplerian velocity.
The normalisation is the same as in \cite{Stew2000}:
\begin{equation}
 \int d^3v dz f(z,v)  =  \frac{\Sigma}{m}
\end{equation}
A planetesimal disc is a slowly evolving  system compared to the orbital time, hence it is reasonable
to use Eq.~\ref{Fdis} as a general solution of the perturbed problem. $\Sigma$, $T_z$ and
$T_r$ are now functions of time and of the radial distance to the star. All information on
the system is contained in these three momenta of the distribution function, where higher
order moments can be deduced from Eq.~\ref{Fdis}. Thus the functional form of the distribution
function represents an implicit closure relation.

The validity of this approximation can be further assessed by a closer examination
of the Boltzmann equation. We summarise all perturbations in an evolution timescale
$T_{\mathrm{evol}} $ and reduce the radial structure to some
typical length scale $\Delta r$ to estimate the deviation from the functional
form Eq.~\ref{Fdis}. A comparison with Eq.~\ref{Eq_BoltzCyl} shows that the difference is small if the migration timescale and the
evolution timescale are large compared to the orbital time $T_0$:
\begin{align}
     T_0  & \ll   \Delta r / \langle v_r \rangle \label{CondV1}  \\
     T_0  & \ll   T_{\mathrm{evol}}   \label{CondV2}
\end{align}
An order--of--magnitude estimate of the evolution time supports  condition
\ref{CondV1} and \ref{CondV2}. Furthermore, numerical calculations confirm that the
velocity distribution stays triaxial Gaussian \citep[see][]{Ida1992}.

The distribution function is equivalent to an isothermal vertical density structure
with scale height $h$:
\begin{align}
   h &=   \sqrt{\frac{T_z}{\Omega^2}} \label{ScaleHeight} \\
  \rho(z) &=   \rho_0 \exp\left( -\frac{z^2}{2h^2} \right) \label{DensPDisc}
\end{align}
Thus the central density $\rho_0$ and the mean density $ \langle \rho \rangle$ are related to the
surface density in a simple way:
\begin{align}
    \rho_0 &=  \frac{\Sigma}{\sqrt{2\pi}h} \nonumber \\
  \langle \rho \rangle  &=  \frac{\rho_0}{\sqrt{2}}
\end{align}
The triaxial Gaussian velocity distribution is equivalent to a {Rayleigh distribution} of the
orbital elements $e$ and $i$\footnote{Eq.~\ref{Approx_a}--\ref{Approx_i} provide the coordinate transformation.}:
\begin{align}
 dn(e^2,i^2) &=  \frac{1}{\langle e^2\rangle\langle i^2\rangle}
   \exp\left(-\frac{e^2}{\langle e^2\rangle }-\frac{i^2}{\langle i^2 \rangle}\right) de^2 di^2 \nonumber \\
  \langle e^2\rangle &=  \frac{2T_r}{(\Omega r_0)^2} \qquad   \langle i^2\rangle = \frac{2T_z}{(\Omega r_0)^2}
\end{align}
Planetesimal encounters couple the time evolution of eccentricity and inclination, so that
the ratio $i^2/e^2$ tends to an equilibrium value after a few relaxation times.
It is close to $1/4$ in a Keplerian potential, but the precise value also
depends on the potential itself \citep{Ida1993}.

\subsection{Dynamical Friction}

Planetesimal--planetesimal scatterings change the velocity distribution through
two different processes. Firstly, it is unlikely that two planetesimals scatter
each other on circular orbits. Thus we expect a steady increase of the velocity
dispersion due to this {viscous stirring}. Secondly, encounters between
unequal masses lead successively to energy equipartition, slowing down the
larger bodies through {dynamical friction}.
The later mechanism is not related to the disc geometry at all, but operates
in any multi-mass system. A special case is the systematic deceleration of a massive
body $M$ in a homogeneous sea of lighter particles $m$ with density $n_0$, which is given by
the {Chandrasekhar dynamical friction formula} \citep{Chandra1942}
\begin{align}
\frac{d{\bf v_M}}{dt}&=  -{\bf v_M}\frac{4\pi\ln\Lambda G^2(M+m)n_0m}{v_M^3}\left(\mbox{erf}(X)-\frac{2X}{\sqrt{\pi}}e^{-X^2}\right) \nonumber \\
  X&=  \frac{v_M}{\sqrt{2}\sigma_v}
\end{align}
where $\sigma_v$ is the velocity dispersion of the lighter particles.
The {Coulomb logarithm} $\Lambda$ arises from an integration over all impact parameters smaller than an upper limit
 $l_{\max}$ and is given by
\begin{equation}
\Lambda   \approx   \frac{\sigma_v^2 l_{\max}}{G(m+M)}
\end{equation}
Although encounters in the gravitational field of the sun deviate from pure two-body scatterings,
it is safe to neglect the presence of the sun if the encounter velocity is large compared
to the {Hill velocity}\,\footnote{Whenever relative velocities are classified as ``high'' or ``low'' in the following
sections, a comparison with the Hill velocity is implied.} $\Omega R_{\mathrm{Hill}}$.
Thus the classical dynamical friction formula is also applicable to planetesimal encounters
in the high velocity regime, though a generalisation to triaxial velocity distributions $\sigma_i$
is necessary \citep[see e.g.]{Bin1977}:
\begin{align}
 \frac{dv_{M,i}}{dt} &=  - v_{M,i}  \sqrt{2\pi} G^2 \ln(\Lambda)(M+m)n_0 m B_i \label{TriDynF} \\
  B_i &=  \int_0^{\infty} \exp\left(-\frac{1}{2}\sum\frac{v_j^2}{\sigma_j^2+u}\right) \nonumber \\
    &\frac{du}{\sqrt{(\sigma_1^2+u)(\sigma_2^2+u)(\sigma_3^2+u)}(\sigma_i^2+u)}
\end{align}
An additional complication is the choice of $l_{\max}$ (i.\,e. the choice of the Coulomb logarithm).
There are several scale lengths, which could determine the largest impact parameter $l_{\max}$:
The scale height of the planetesimal disc, the radial excursion due to the excentric motion of the
planetesimals and the Hill radius of the planetesimals. As it is not possible to derive a unique
expression for $l_{\max}$ from first principles, a proper formula is often fitted to $N$--body
calculations (compare Eq.~\ref{EqLambda}).
The velocity dispersion of a planetesimal disc is triaxial with $T_{\phi}/T_r=1/4$ and $T_z/T_r\approx 1/4$.
We take these values and expand Eq.~\ref{TriDynF} for small velocities $v_M$:
\begin{align}
 \frac{dv_{M,r}}{dt} & \approx   -1.389 \ v_{M,r} \frac{ \sqrt{2\pi} G^2 \ln(\Lambda) (M+m)n_0 m }{T_r^{3/2}} \nonumber \\
 \frac{dv_{M,\phi}}{dt}&  \approx   -3.306 \ v_{M,\phi} \frac{ \sqrt{2\pi} G^2 \ln(\Lambda) (M+m)n_0 m }{T_r^{3/2}} \nonumber \\
 \frac{dv_{M,z}}{dt} &  \approx   -3.306 \ v_{M,z} \frac{ \sqrt{2\pi} G^2 \ln(\Lambda) (M+m)n_0 m }{T_r^{3/2}}
\end{align}
The derived expressions provide a compact tool to analyse dynamical friction
in disc systems. However, the involved approximations are too severe compared
to the needs of an accurate description. While these concise expressions are valuable for basic
estimations, the following sections derive viscous stirring and dynamical
friction formul{\ae} for a planetesimal system in a rigorous way.

\subsection{High Speed Encounters}

We return to the Boltzmann equation as a starting point for the derivation of
the scattering coefficients:
\begin{equation}
\left(\frac{\partial f}{\partial t}\right)_{\mathrm{coll}}  =  \frac{\partial f}{\partial t}+ {\bf v}\cdot \nabla f
 - \nabla \Phi\cdot\frac{\partial f}{\partial {\bf v}}
\end{equation}
In virtue of the ansatz for the distribution function (see Eq.~\ref{Fdis}), it is sufficient
to derive the time derivative of the second order velocity moments $T_r$ and $T_z$.
Since the distribution function is time independent in the absence of encounters,
only the collisional term contributes to the time derivative of the velocity
dispersions $T_k$ ($k \in (r,z,\phi)$ in the following):
\begin{equation}
 \frac{d\rho T_k}{dt}  =  \int d^3vmv_k^2\left(\frac{\partial f}{\partial t}\right)_{\mathrm{coll}}
\end{equation}
The collisional term invokes the averaging over many different scattering trajectories
and is, given that the underlying encounter model is analytically solvable, still too complex
to derive an exact expression. If most of the encounters are weak -- a realistic assumption
in a planetesimal disc -- it is possible to expand the collisional contribution in terms of the velocity change $\Delta v_i$.
This is the {Fokker-Planck approximation} \citep[see e.g.][]{Binney1994})
\begin{align}
\left(\frac{\partial f}{\partial t}\right)_{\mathrm{coll}}  & = 
 -\sum_i \frac{\partial}{\partial v_i} [fD(\Delta v_i)]+ \nonumber \\
& \frac{1}{2} \sum_{i,j} \frac{\partial^2}{\partial v_i\partial v_j}[D(\Delta v_i,\Delta v_j)] \label{FPTerm}
\end{align}
where the {diffusion coefficients} $D$ contain all information on the underlying
scattering process. Next we consider two interacting planetesimal populations $m, m^*$ with distribution functions
\begin{align}
 f &=  \frac{\Omega\Sigma}{2\pi^2T_r T_z m}
  \exp\left( -\frac{v_r^2+4v_{\phi}^2}{2T_r} -\frac{v_z^2+\Omega^2z^2}{2T_z}  \right) \nonumber \\
 f^* &=  \frac{\Omega\Sigma^*}{2\pi^2T_r^* T_z^* m^*}
  \exp\left( -\frac{v_r^2+4v_{\phi}^2}{2T_r^*} -\frac{v_z^2+\Omega^2z^2}{2T_z^*}  \right)
\end{align}
to evaluate the terms in equation~\ref{FPTerm}. We follow \cite{Stew2000} except some
minor changes in the notation. The collisional term requires an averaging over the velocities
of the two interacting planetesimals $m$ and $m^*$:
\begin{align}
\frac{d \langle \rho v_k^2 \rangle }{dt} &=  2\pi G^2m{m^*}^2  \int d^3v \int d^3v^* ff^*\times   \nonumber \\
    &  \left[ -\frac{2A(m+m^*)u_kv_k}{m^*u^3}+\frac{Bu^2+(2C-B)3u_k^2)}{u^3}
   \right]     \nonumber   \nonumber \\
 u_k &=  v_k-v_k^*   \nonumber \\
   A &=  \ln(\Lambda^2+1) \quad C=\frac{\Lambda^2}{\Lambda^2+1} \quad B=A-C
\end{align}
A coordinate transformation to the relative velocity $u$ and the modified centre-of-mass velocity $w$
\begin{align}
 w_k &=  \left\{\begin{array}{l} \displaystyle
        V_k +\frac{(m^* T_r^* -m T_r)u_k}{(m+m^*)(T_r+T_r^*)} \qquad \mbox{ for } k \in \{r,\phi\} \nonumber \\
\displaystyle V_k +\frac{(m^* T_z^* -m T_z)u_k}{(m+m^*)(T_z+T_z^*)} \qquad \mbox{ for } k =z \nonumber \\
        \end{array} \right. \nonumber \\
   {\bf V}&=  \frac{m{\bf v} +m^* {\bf v^*}}{m+m^*}
\end{align}
further simplifies the double integral. Thus the integration separates in a simple integral over $w$ and
a more demanding $u$--integration:
\begin{align}
 \frac{d \langle \rho v_r^2 \rangle }{dt} &=  2\pi G^2m{m^*}^2  \int d^3w \int d^3u ff^*\times   \nonumber \\
    &  \left[ \frac{2A(m^*T^*_r-mT_r)u_r^2}{m^*(T_r+T_r^*)u^3}+\frac{B(u^2-3u_r^2)}{u^3}
   \right]     \nonumber   \nonumber \\
 \frac{d \langle \rho v_{\phi}^2 \rangle }{dt} &=  2\pi G^2m{m^*}^2  \int d^3w \int d^3u ff^*\times   \nonumber \\
    &  \left[ \frac{2A(m^*T^*_r-mT_r)u_{\phi}^2}{m^*(T_r+T_r^*)u^3}+\frac{B(u^2-3u_{\phi}^2)}{u^3}
   \right]     \nonumber   \nonumber \\
 \frac{d \langle \rho v_z^2 \rangle }{dt} &=  2\pi G^2m{m^*}^2  \int d^3w \int d^3u ff^*\times   \nonumber \\
    &  \left[ \frac{2A(m^*T^*_z-mT_z)u_z^2}{m^*(T_z+T_z^*)u^3}+\frac{B(u^2-3u_z^2)}{u^3}
   \right]     \nonumber
\end{align}
All integrals are solvable and give the result
\begin{align}
\frac{d \langle v_{r,\phi}^2 \rangle }{dt} &=  \frac{G^2\rho^*}{2\sqrt{2}(T_r+T_r^*)^{3/2}} \nonumber \\
    &  \left[  B(T_r^*+T_r)m^*J_{r,\phi}(\beta)+2A(T_r^*m^*-mT_r)H_{r,\phi}(\beta) \right] \nonumber \nonumber \\
\frac{d \langle v_{z}^2 \rangle }{dt} &=  \frac{G^2\rho^*}{2\sqrt{2}(T_r+T_r^*)^{1/2}(T_z+T_z^*)} \nonumber \\
    &  \left[  B(T_z^*+T_z)m^*J_{z}(\beta)+2A(T_z^*m^*-mT_z)H_{z}(\beta) \right] \nonumber \nonumber \\
 \beta^2 &:=  \frac{T_z+T_z^*}{T_r+T_r^*}
\end{align}
where six auxiliary functions are introduced to arrive at a more compact notation:
\begin{align}
  a &=  \sqrt{4-3x^2} \quad b=\sqrt{1-(1-\beta^2)x^2} \nonumber \\
 H_r &:=  8\sqrt{\pi} \int_0^1 \frac{x^2}{ab} dx \nonumber \\
 H_{\phi} &:= 8\sqrt{\pi} \int_0^1 \frac{1-x^2}{a(\beta a+b)} dx \nonumber \\
 H_{z} &:= 8\sqrt{\pi} \int_0^1 \frac{\beta(1-x^2)}{b(\beta a+b)} dx \nonumber \\
 J_r &:=  -2H_r+\,H_{\phi} +H_z\nonumber \\
 J_{\phi} &:=  \quad H_r-2H_{\phi} +H_z\nonumber \\
 J_{z} &:=  \quad H_r\,+H_{\phi} -2H_z
\end{align}
Since these are non-trivial functions, we apply a standard Chebyshev approximation
for $\beta \in [0,1]$:
\begin{equation}
 f(x)  \approx   \sum_{k=0}^{5} c_kT_k(x)-\frac{1}{2}c_0
\end{equation}
Table~\ref{ChebTab} summarises the Chebyshev coefficients.
\begin{table*}
\begin{center}
\begin{tabular}{|r||r|r|r||r|r|r|} \hline
 $f$ & $ J_r$ & $J_{\phi}$ & $J_z$ & $ H_r$ & $H_{\phi}$ & $H_z$ \\ \hline
 $c_0$ &-10.34660733 & 1.81674741 & 8.52985992 & 11.00434580 &  6.94989422 &  4.71219005 \\ \hline
 $c_1$ &  4.69990443 & 2.95397208 &-7.65387651 & -2.64707927 & -2.06510182 &  1.47084771 \\ \hline
 $c_2$ & -1.25533220 &-1.18724874 & 2.44258094 &  0.60969641 &  0.58700192 & -0.62294130 \\ \hline
 $c_3$ &  0.30288875 & 0.37775788 &-0.68064662 & -0.13815856 & -0.16311494 &  0.18968657 \\ \hline
 $c_4$ & -0.07040537 &-0.11070339 & 0.18110876 &  0.03112047 &  0.04455314 & -0.05271757 \\ \hline
 $c_5$ &  0.01540098 & 0.02922947 &-0.04463045 & -0.00669979 & -0.01130929 &  0.01331068 \\ \hline
 $\Delta$ & 0.006  & 0.015 & 0.022  & 0.0025 & 0.0058 & 0.0066\\ \hline
\end{tabular}
\end{center}
\caption[Chebyshev coefficients]
{Chebyshev coefficients of the auxiliary functions $J_k$ and $H_k$.\label{ChebTab}}
\end{table*}
A final $z$--averaging yields the expressions:
\begin{align}
\frac{d \langle v_{r,\phi}^2 \rangle }{dt} &=  \frac{G^2\Omega\Sigma^*}{4\sqrt{\pi}(T_r+T_r^*)^{3/2}(T_z+T_z^*)^{1/2}} \times \label{StirrHigh1} \\
    &  \left[  B(T_r^*+T_r)m^*J_{r,\phi}(\beta)+2A(T_r^*m^*-mT_r)H_{r,\phi}(\beta) \right] \nonumber \nonumber \\
\frac{d \langle v_{z}^2 \rangle }{dt} &=  \frac{G^2\Omega\Sigma^*}{4\sqrt{\pi}(T_r+T_r^*)^{1/2}(T_z+T_z^*)^{3/2}} \times \label{StirrHigh2} \\
    &  \left[  B(T_z^*+T_z)m^*J_{z}(\beta)+2A(T_z^*m^*-mT_z)H_{z}(\beta) \right] \nonumber
\end{align}
The determination of a proper Coulomb logarithm $\Lambda$ leaves room for further optimisation.
 \noindent
A careful comparison with $N$--body models gives rise to the empirical choice \citep{Ohtsuki2002}:
\begin{align}
  \Lambda &=  \frac{1}{12}(\langle \tilde e^2 \rangle +\langle \tilde i^2 \rangle ){\langle\tilde i^2\rangle}^{1/2} \label{EqLambda}\\
  \tilde e &=  \frac{\sqrt{2T_r}}{\Omega R_{\mathrm{Hill}}} \qquad   \tilde i = \frac{\sqrt{2T_z}}{\Omega R_{\mathrm{Hill}}}
\end{align}
Ohtsuki et al. 2002 also report a further improvement by setting $B\equiv A$.

\subsection{Low Speed Encounters}

Encounters in the low velocity regime exhibit a wealth of different orbits, as
the solar gravity field perturbs the two-body scattering. Only a small subset
of the trajectories represents simple, regular orbits like Tadpole or Horseshoe
orbits\footnote{The most famous example of such a regular orbit are
the two saturnian moons Janus and Epimetheus which share nearly the same orbit.}. Hence an examination
of this velocity regime is done best with a numerical study of the parameter space
by integrating the equations of motions numerically (see Eq.~\ref{EqX1}).

\cite{Ohtsuki2002} integrated a large set of planetesimal encounters
and extracted fitting formul{\ae} that cover the low velocity regime.
Their expressions for viscous stirring are:
\begin{align}
\frac{d T_r}{dt} &=  \frac{G r_0 \Omega h \Sigma^*}{6(m+m^*)}73C_1 m^* \nonumber \\
\frac{d T_z}{dt} &=  \frac{G r_0 \Omega  h \Sigma^*}{6(m+m^*)}
     C_2 m^*\left(4\langle\tilde i^2\rangle+0.2(\langle \tilde e^2\rangle )^{3/2} \sqrt{\langle \tilde i^2\rangle} \right) \nonumber \\
  \tilde e &=  e/h \qquad \tilde i = i/h \nonumber \\
  C_1 &:=  \frac{\ln(10\Lambda^2/\tilde e^2+1)}{10\Lambda^2/\tilde e^2} \nonumber \\
  C_2 &:=  \frac{\ln(10\Lambda^2\sqrt{\tilde e^2}+1)}{10\Lambda^2\sqrt{\tilde e^2}}
\end{align}
The stirring rate of the radial velocity dispersion approaches a finite value for
very low velocity dispersions, while the stirring rate for the vertical velocity dispersion
drops to zero as the velocity dispersion decreases. This different behaviour of the two
limits is due to the encounter geometry: If two planetesimals
have zero inclination, they may still excite higher eccentricities during an encounter,
but they remain confined to the initial orbital plane preventing any excitation of
inclinations.

The respective expressions for the dynamical friction rates are:
\begin{align}
\frac{d T_r}{dt} &=  \frac{G r_0\Omega  h \Sigma^*}{6(m+m^*)(T_r+T_r^*)}10C_3\langle
     \tilde e^2 \rangle (T_r^*m^*-T_r m)  \nonumber \\
\frac{d T_z}{dt} &=  \frac{G r_0\Omega h \Sigma^*}{6(m+m^*)(T_z+T_z^*)}10C_3\langle
     \tilde i^2\rangle (T_z^*m^*-T_z m) \nonumber \\
 C_3 &:=  \frac{\ln(10\Lambda^2+1)}{10\Lambda^2}
\end{align}
As the stirring rates are only valid in the low velocity regime,
\cite{Ohtsuki2002} introduced special interpolation coefficients $C_i$.
These coefficients tend to unity for very small velocity dispersions, and drop to
zero in the high velocity regime. Thus the interpolation formul{\ae} are properly
``switched off'' in the high velocity regime, so they do not interfere
with the known high velocity stirring rates.

\subsection{Distant Encounters}

All formul{\ae} include only the stirring rates due to close encounters, but non-crossing
orbits also contribute to the overall change of the velocity distribution. As these distant
encounters lead to small changes of the orbital elements, the problem is accessible to
perturbation theory; see \cite{Hase1990} for a detailed treatment. \cite{Stew2000} 
integrated the perturbation solution over all impact parameters to 
derive the collective effect of all distant encounters:
\begin{align}
 \frac{d\langle e^2 \rangle}{dt} &=  \frac{\Omega m^* \Sigma^* r_0^2}{(m+m^*)^2}\langle P_{\mathrm{VS,dist}} \rangle \nonumber \\
 \langle P_{\mathrm{VS,dist}} \rangle &=  7.6\frac{\alpha(m+m^*)^2}{M_{c}^2} \nonumber \\
   & \frac{\mbox{EXINT}\left(\alpha\frac{h^2}{(\langle e^2\rangle +\langle{e^*}^2\rangle)}\right)-
       \mbox{EXINT}\left(\alpha\frac{h^2}{(\langle i^2\rangle+\langle{i^*}^2\rangle)}\right)   }
        {\langle e^2\rangle +\langle{e^*}^2\rangle-\langle i^2\rangle-\langle{i^*}^2\rangle} \nonumber \nonumber \\
 \mbox{EXINT}(x) &:=  \exp(x)\Gamma(0,x) \qquad h=\sqrt[3]{\frac{m+m^*}{3M_c}}\nonumber \\
 \alpha &\approx   1
\end{align}
$\alpha $ accounts for the uncertainty in the smallest impact parameter that is regarded as a distant
encounter. While distant encounters are already included in the interpolation formula
of the low--velocity regime, we use the modified expression:
\begin{align}
  \left( \frac{dT_r}{dt} \right)_{\mathrm{dist}} &=  \frac{1}{2}(\Omega r_0)^2
  \left(\frac{d\langle e^2\rangle}{dt}\right)_{\mathrm{dist}}(1-C_1) \nonumber \\
&=  \frac{GM_{c}r_0\Omega m^* \Sigma^* }{2(m+m^*)^2}\langle P_{\mathrm{VS,dist}} \rangle(1-C_1)
\end{align}
\cite{Stew2000} omitted the change in the inclination, as it is small due to the encounter
geometry. Nevertheless we derived the integrated stirring rate for completeness, which we give in
\ref{StirringRate}
\begin{figure*}
\begin{align}
 \frac{d\langle i^2 \rangle}{dt} =  \frac{\Omega m^* \Sigma^* r_0^2}{(m+m^*)^2}\langle Q_{\mathrm{VS,dist}} \rangle 
 \langle Q_{\mathrm{VS,dist}} \rangle & =  0.4\frac{\alpha^2(m+m^*)^2}{M_{c}^2} \times
    \frac{1}{\langle e^2\rangle +\langle{e^*}^2\rangle-\langle i^2\rangle-\langle{i^*}^2\rangle } \times  
    \Bigg[ 1   -\frac{\alpha h^2}{\langle i^2\rangle+\langle{i^*}^2\rangle}\mbox{EXINT}\left(\alpha\frac{h^2}{(\langle i^2\rangle+\langle{i^*}^2\rangle)}\right) \nonumber \\
&    -(\langle i^2\rangle+\langle{i^*}^2\rangle)\frac{\mbox{EXINT}\left(\alpha\frac{h^2}{(\langle e^2\rangle +\langle{e^*}^2\rangle)}\right)-
       \mbox{EXINT}\left(\alpha\frac{h^2}{(\langle i^2\rangle+\langle{i^*}^2\rangle)}\right)   }
        {\langle e^2\rangle +\langle{e^*}^2\rangle-\langle i^2\rangle-\langle{i^*}^2\rangle }  \Bigg] \label{StirringRate}
\end{align}
\end{figure*}
A close inspection of the integrated perturbation shows that the above formula
is roughly a factor $\langle i^2\rangle+\langle{i^*}^2\rangle$ smaller than the corresponding
changes in the eccentricity.

\subsection{Gas Damping}

The presence of a gaseous disc damps the velocity dispersion of the planetesimals and introduces
a slow inward migration. \cite{Adachi1976} used the drag law Eq.~\ref{DragEq} to
approximate\footnote{A formal expansion at $e=0$, $i=0$, $\eta_g=0$ is not possible, since the drag law involves the modulus
of the relative velocity. \cite{Kary1993} corrected a missing factor $3/2$ in Eq.~\ref{edotm}.}
the average change of the orbital elements:
\begin{align}
\tau_0 &=  \frac{2m}{\pi C_D \rho_g R^2 v_K} \qquad \eta_g = \frac{|v_K-v_g|}{v_K} \nonumber \\
\frac{d}{dt} e^2 &\approx   -\frac{2e^2}{\tau_0}\left(0.77\, e+0.64\,i+\frac{3}{2}\eta_g  \right) \label{edotm} \\
\frac{d}{dt} i^2 &\approx   -\frac{2i^2}{\tau_0}\left(0.39\, e + 0.43\, i +\frac{1}{2} \eta_g  \right) \nonumber \\
\frac{d}{dt} a   &\approx   -\frac{2a}{\tau_0}\eta_g \left(0.97\,e+0.64\, i + \eta_g   \right)
\end{align}
$\eta_g$ is the dimensionless velocity lag of the sub--Keplerian rotating gaseous disc.

\subsection{Unified Expressions}

All expressions for the different velocity regimes are constructed such that
a smooth transition between the different regimes is assured.
Thus, a simple addition of all contributions yields already
the unified expressions
\begin{align}
 \frac{dT_r}{dt} &=   \left( \frac{dT_r}{dt} \right)_{\mathrm{high}}+
                    \left( \frac{dT_r}{dt} \right)_{\mathrm{low}}+
          \left( \frac{dT_r}{dt} \right)_{\mathrm{gas}}+
          \left( \frac{dT_r}{dt} \right)_{\mathrm{dist}}  \label{Trdot_F} \\
 \frac{dT_z}{dt} &=   \left( \frac{dT_z}{dt} \right)_{\mathrm{high}}+
                    \left( \frac{dT_z}{dt} \right)_{\mathrm{low}}+
                    \left( \frac{dT_z}{dt} \right)_{\mathrm{gas}}
             +\left( \frac{dT_z}{dt} \right)_{\mathrm{dist}}\label{Tzdot_F}
\end{align}
which cover the full range of relative velocities. Although only two
populations $m$ and $m^*$ were assumed, Eq.~\ref{Trdot_F} and Eq.~\ref{Tzdot_F}
are readily generalised to a multi-mass system by adding a summation over
all masses.

\subsection{Inhomogeneous Disc}

The preceding derivations assumed a homogeneous disc, which simplified the calculation,
since the integration over all impact parameters needed no special precaution. A more
sophisticated consequence is that the spatial density and the density in semimajor
axis space are equal:
\begin{equation}
  \Sigma(r) = \Sigma(a) =\Sigma_0
\end{equation}
Density inhomogeneities break this simple relation, as particles at the same
radial distance could have different semimajor axes, and particles with the same
semimajor axis are located at different positions at a given time. While both representations
are equivalent (i.\,e. describe the same system in different ways), we chose
the density in semimajor axis space as the primary density\footnote{We denote $\Sigma(a)$ also as ``surface density''
and refer to $a$ as a radial coordinate. However, all formul{\ae} are precise in discriminating both representations in $r$ and $a$.}.
The spatial density is derived as:
\begin{equation}
  \Sigma(r)  =   \int \frac{1}{\sqrt{2\pi a^2 \langle  e^2(a)\rangle} }
      \exp\left( -\frac{(a-r)^2}{2 a^2\langle  e^2(a)\rangle} \right) \Sigma(a) da
\end{equation}
Likewise, $T_r$ and $T_z$ are also functions of the semimajor axis.

Furthermore, an inhomogeneous surface density invalidates the averaging
over all impact parameters. Planetesimal encounter are most efficient for
impact parameters smaller than a few Hill radii, so the derivation is still valid if
the surface density is roughly constant on that length scale. However, a planetesimal
that is large enough will ``feel'' the spatial inhomogeneities or even generates density
fluctuations. Hence it is essential to extend the validity of the averaged expressions
to inhomogeneous systems.
We use the averaged expressions
\begin{equation}
 \left< \frac{dT_{r,z}}{dt} \right>  =  \Sigma(a) \int_{-\infty}^{\infty} \frac{d\tilde T_{r,z}(b)}{dt} db
\end{equation}
as a starting point ($d\tilde T_{r,z}/dt$ excludes the surface density, as opposed to the averaged
expressions). The (yet unknown) scattering contribution $d\tilde T_{r,z}/dt$ as a function of the impact parameter $b$ is our starting point
for a general expression for a varying surface density:
\begin{equation}
  \frac{dT(a_0)_{r,z}}{dt}   =  \int_{-\infty}^{\infty} \Sigma(a_0+b) \frac{d\tilde T_{r,z}(b)}{dt} db \label{IntSigT}
\end{equation}
We restate Eq.~\ref{IntSigT} in terms of a weight function $w(b)$:
\begin{align}
\frac{dT(a_0)_{r,z}}{dt}  &=    \left< \frac{dT_{r,z}}{dt} \right> \frac{1}{\Sigma(a_0)} \int_{-\infty}^{\infty} \Sigma(a_0+b) w(b) db \label{IntScat} \\
 w(b) &=  \Sigma(a_0)\frac{d\tilde T_{r,z}(b)}{dt}  \left< \frac{dT_{r,z}}{dt} \right>^{-1}
\end{align}

\begin{figure}
\resizebox{\hsize}{!}
          {\includegraphics[scale=1,clip]{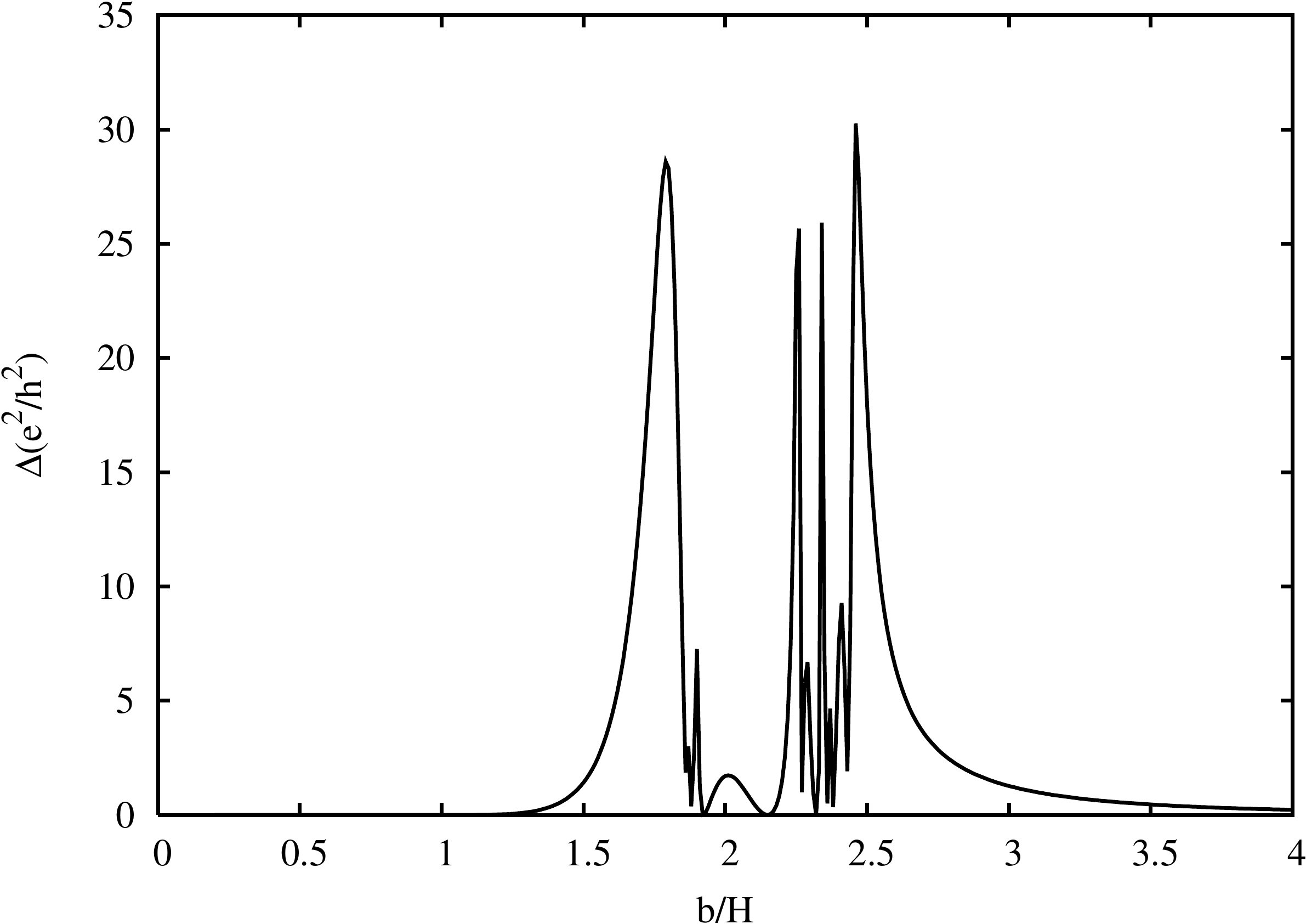}}
\caption
   {  
 Change of the relative eccentricity $e^2$ due to an encounter of two bodies initially on  circular orbits.
$b/H$ is the impact parameter in units of the Hill radius. The plot was obtained by integrating
Eq.~\ref{EqX1}.  
   }
\label{plotde2}
\end{figure}

The numerical solution of the Hill problem (see Eq.~\ref{EqX1}) gives some insight into how the weight function $w(b)$ changes
with the impact parameter. Fig.~\ref{plotde2} illustrates the change in $e^2$ of the relative
motion during an encounter of two planetesimals that were initially on circular orbits.
While the details depend on the initial inclination and eccentricity as well as on the
selected orbital element, all result share some basic features. Small (compared to the Hill radius) impact
parameters allow for a horseshoe orbit and the change in the orbital elements is small except a change
in the semimajor axis. Intermediate impact parameters which lead to close encounters provide
the strongest perturbation, but they are also more susceptible to complicated dynamics
(compare the resonant structures in Fig.~\ref{plotde2}).
As the gravitational attraction drops with increasing distance, non-crossing orbits
yield ever smaller perturbations with increasing impact parameter.
Aside from this qualitative behaviour, it is very difficult to derive precise expressions.
While the limit of high velocities is accessible through the two-body approximation,
any general formula involves some empiric interpolation to cover the full parameter
space \citep[see the approximations of]{Rafikov2003, Rafikov2003b}.
Therefore we decided to approximate the weight function such that
the main features of the true weight function $w(b)$ are reproduced. While this approach
is less accurate, it provides better insight into the involved approximations.
We expand the surface density under the integral in Eq.~\ref{IntScat}  and compare the expansion coefficients
for $w(b)$ and the approximation $\tilde w(b)$ to derive constraints on the choice of  $\tilde w(b)$. The lowest non-vanishing order is:
\begin{equation}
  l^2  =  \int_{-\infty}^{\infty} b^2 w(b) db = \int_{-\infty}^{\infty} b^2 \tilde w(b) db  \label{Cond_l2}
\end{equation}
$l$ can be interpreted as the width of the heating zone. Condition~\ref{Cond_l2}
inspires our choice of the weight function $\tilde w(b)$
\begin{align}
  \tilde w(b) &=  \frac{1}{\sqrt{2\pi}l}\exp\left(-\frac{b^2}{2l^2}\right)  \nonumber \\
   l^2 &=  \frac{1}{\Omega^2}(T_r^{(i)}+T_r^{(j)})+R_{\mathrm{Hill}}^2
\end{align}
where $T_r^{(i)}$ and $T_r^{(j)}$ are the radial velocity dispersions of the interacting
radial bins and $l$ is adjusted to the findings of \cite{Ida1993b}.
The advantage of the bell curve is that it has a discrete counterpart
\begin{equation}
\tilde w(b)    \approx   \frac{1}{\Delta a 2^N} {N \choose b/\Delta a+N/2 } \qquad N = 4 l^2/(\Delta a)^2
\end{equation}
which makes the weight function readily applicable to the summation on an equidistant
radial grid with spacing $\Delta a $.

\subsection{Diffusion Coefficient }

We concentrated on the evolution of the velocity dispersion so far, but
scatterings among planetesimals also change the semimajor axis of the
disc particles, inducing a diffusive evolution of the surface density:
\begin{equation}
 \frac{\partial \Sigma}{\partial t}  =  \Delta_a (D \Sigma)
\end{equation}
The {diffusion coefficient} $D$ is related to the typical change in semimajor
axis $\Delta a$ and the timescale $T_{\mathrm{2Body}}$ on which planetesimal
encounters operate:
\begin{equation}
 D   \approx   \frac{(\Delta a)^2}{T_{\mathrm{2Body}}}
\end{equation}
If we neglect the radial displacement during an encounter, the change in semimajor axis
is solely due to the change of the velocity:
\begin{align}
 -\frac{GM}{2a} &=  -\frac{GM}{r}+\frac{1}{2} v^2 \nonumber \\
 \Delta a & \approx   \frac{2a^2}{GM} {\bf v}\cdot \Delta {\bf v}
\end{align}
An average over all orientations of the velocity ${\bf v}$ and the velocity change $\Delta {\bf v}$
yields the mean square change in semimajor axis:
\begin{align}
  \langle (\Delta a)^2\rangle & \approx    \frac{4a^3}{3GM} \langle (\Delta v)^2 \rangle \nonumber \\
                               &= \frac{4}{3\Omega^2} ( \Delta T_r +\Delta T_{\phi}+\Delta T_z )
\end{align}
This yields the mean diffusion coefficient
\begin{equation}
 D   \approx   \frac{4}{3\Omega^2}\left(\frac{5}{4}\frac{d}{dt} T_r +\frac{d}{dt} T_z \right) \label{Apr_D}
\end{equation}
where the time derivatives of the velocity dispersions $T_r$ and $T_z$ are taken with respect to encounters.

\subsection{Coagulation Equation}

We already stated the coagulation equation for a multi-mass system:
\begin{align}
0 &=  \difft mn(t,m)+  \diffm F_m(t,m) \label{Eqn1} \\
F_m &=  - \dint
n_1(t,m_1)n_2(t,m_2)\sigma(m_1,m_2)v_{\mathrm{rel}} \nonumber \\
& M_{\mathrm{red}}(m,m_1,m_2) dm_1 dm_2
\end{align}
Since the vertical density profile of a planetesimal disc is
specified by the known distribution function,
we insert the isothermal density profile (see Eq.~\ref{DensPDisc})
in the coagulation equation~\ref{Eqn1} and integrate over $z$:
\begin{align}
0 &=  \difft \Sigma(t,m)+  \diffm F_m(t,m)  \label{sigmaInt} \\
F_m &=  - \dint \frac{1}{\sqrt{2\pi(h(m_1)^2+h(m_2)^2)}}\frac{\Sigma_1(t,m_1)}{m_1}\frac{\Sigma_2(t,m_2)}{m_2} \nonumber \\
&  \qquad \tilde \sigma(m_1,m_2)v_{\mathrm{rel}}M_{\mathrm{red}}(m,m_1,m_2)
dm_1 dm_2 \label{Fm_Sigma}
\end{align}
$\Sigma(m)$ is a short-hand notation for the differential surface density $\frac{d\Sigma}{dm}$.
Further integration over all masses gives the total mass balance:
\begin{align}
\Sigma_{\mathrm{tot}} &=  \int_{m_{\min}}^{m_{\max}} \Sigma(t,m) dm \nonumber \\
\difft \Sigma_{\mathrm{tot}} &=  F_m(m_{\min})- F_m(m_{\max})
\end{align}
The calculation of collisional cross sections is closely related to the underlying
encounter dynamics. A homogenous system introduces no systematic perturbation,
hence an encounter is a pure two-body problem which is analytically solvable.
Thus it is possible to derive the cross section without any approximation.
Since encounters in the field of a central star deviate noticeably from
the pure Kepler solution, the cross sections are also modified.
While the cross section in the high velocity regime reduces to the two-body formula  (except minor corrections),
the low velocity regime is explored best by numerical calculations.
It is not appropriate
to disentangle the different contributions in Eq.~\ref{Fm_Sigma}, but to combine the
various terms to the {collisional probability}
\begin{equation}
P_{\mathrm{coll}}  =  \frac{ \tilde \sigma(m_1,m_2)v_{\mathrm{rel}} }{\sqrt{2\pi(h(m_1)^2+h(m_2)^2)}}
\end{equation}
which can be easily deduced from the fraction of colliding orbits in Monte--Carlo simulations.
An accurate expression for the collisional probability should include the two-body
cross section in the limit of high velocities and the numerical data for the low velocity regime as well.
We use numerical calculations from \cite{Greenberg1991,Greenz1992}
\footnote{Their work includes an averaging over the Rayleigh distributed inclinations 
and eccentricities of the colliding planetesimals.} as a basis for a unified fitting formula
\begin{align}
 \tilde \sigma &=  \sigma \times 0.572 \left(1+3.67v_{\mathrm{Hill}}/v_{\mathrm{rel}}\right)\left(1+1.0\frac{\sigma\Omega^2}{T_z}\right)^{-1/2} \label{fullSig} \\
 \sigma &=   \sigma_{\mathrm{geom} } \left(1+\frac{v_{\infty}^2}{v_{\mathrm{rel}}^2+1.8v_{\mathrm{Hill}}^2} \right)  \label{onlyGrav} \\
 v_{\mathrm{rel}}^2  &=  \frac{1}{2}(T_r+T_{\phi}+T_z) \qquad v_{\mathrm{Hill}}=\Omega r_{\mathrm{Hill}}
\end{align}
which gives an effective cross section $\tilde \sigma$ for planetesimal--planetesimal encounters.
Eq.~\ref{fullSig} reduces to the well-known gravitational focusing formula in the limit of
high velocities:
\begin{align}
 \tilde \sigma & \propto    \sigma_{\mathrm{geom}} \left(1+\frac{v_{\infty}^2}{v_{\mathrm{rel}}^2}\right) \label{sigClas} \\
 v_{\infty}^2 &=  \frac{2G(m_1+m_2)}{R_1+R_2}
\end{align}
If the vertical velocity dispersion is small, the disc becomes two-dimensional and the cross section
is proportional to $R$. The main differences to the two-body cross section~\ref{sigClas} is a
finite gravitational focusing factor, since the Keplerian shear inhibits a zero relative velocity,
and a finite collisional probability for very small velocities, again due to the shear which provides
a finite influx of particles.

The precise calculation of the coagulation kernel should include an integration over all semimajor
axes with a proper weighting kernel. As collisions among particles in the statistical model play only a major
role when the system is still homogenous, we omitted this contribution. In addition,
this helps saving computational time, since the solution of the coagulation equation is very costly.
However, interactions between $N$--body particles and the statistical model include  spatial
inhomogeneities properly (see section~\ref{HybridCode}).

\subsection{Collisional Damping}

Collisions are a dissipative process that removes kinetic energy from the planetesimal system and damps the eccentricity.
Low speed encounters leave the colliding bodies intact and damp the relative velocities
through inelastic collisions.
In contrast, high velocity encounters disrupt the colliding bodies and turn
them into an expanding cloud of fragments. As a major part of the initial
kinetic energy is converted into heat, the fragments disperse with
rather low velocities thus reducing the overall velocity dispersion.
We formulate the dissipation due to collisions analogue to Eq.~\ref{sigmaInt}:
\begin{align}
0 &=  \difft T_k \Sigma(t,m)+  \diffm F_{Q,k}(t,m) \nonumber \\
F_{Q,k} &=  - \dint \frac{1}{\sqrt{2\pi(h(m_1)^2+h(m_2)^2)}}\frac{\Sigma_1(t,m_1)}{m_1}\frac{\Sigma_2(t,m_2)}{m_2} \nonumber \\
&  \qquad \tilde \sigma(m_1,m_2)v_{\mathrm{rel}}Q_{\mathrm{red},k}(m,m_1,m_2)
dm_1 dm_2  \nonumber \\
 k &\in   \{r,z\}
\end{align}
$ Q_{\mathrm{red},k}$ is the {kinetic energy redistribution function} and $F_{Q,k}$ is the
associated flux across the mass distribution. $k$ indicates the two velocity dispersions.
$Q_{\mathrm{red},k}$  is a complex function, since the disruption of a planetesimal produces
fragments with a large scatter in velocities and a complicated velocity field.
The velocity of a fragment consists of two contributions: The ejection velocity  relative to the target and the
velocity $\bar v$ of the target within the corotating coordinate system.
Owing to the strong dissipation, fragment velocities are dominated by
the motion of the centre-of-mass of the two colliding bodies. Thus we neglect the
ejection velocities and estimate the centre-of-mass motion.
The initial kinetic energy of two colliding bodies is:
\begin{align}
 E_{\rm ini} &=  \frac{1}{2}m_1 v_1^2+\frac{1}{2}m_2 v_2^2
\end{align}
We separate $E_{\rm ini}$ into centre-of-mass motion and relative motion and average over ${\bf v}_1,{\bf v}_2$:
\begin{align}
 E_{\rm ini} &=  \frac{1}{2}M\left(\frac{m_1 {\bf v}_1+m_2{\bf v}_2}{M} \right)^2+ \frac{1}{2}\mu(v_1^2+v_2^2)  \nonumber \\
\langle E_{\rm ini} \rangle &=  \frac{1}{2M} \left( m_1^2 T_1+m_2^2 T_2 \right)+\frac{1}{2}\mu(T_1+T_2)
\end{align}
Most of the relative kinetic energy is dissipated during the collision, so we neglect the relative motion after the
incident. The final energy is
\begin{equation}
 \frac{ \langle E_{\rm final} \rangle}{ \langle E_{\rm ini} \rangle}  =  \frac{m_1^2 T_1 + m_2^ 2T_2}{M(m_1 T_1+m_2 T_2)}
\end{equation}
which gives the drift motion ${\bar v}$ of the expanding fragment cloud:
\begin{equation}
{\bar v}^2   =  \frac{2}{M}   \langle E_{\rm final}  \rangle
\end{equation}
$Q_{\mathrm{red}}$ is therefore coupled to the fragment redistribution function $M_{\mathrm{red}}$
\begin{equation}
 Q_{\mathrm{red}}  =  {\bar v}^2 M_{\mathrm{red}}+Q_{\mathrm{diss}}
\end{equation}
where the additional function $Q_{\mathrm{diss}}$ removes the dissipated energy.

\subsection{Correlation}

The statistical model of a planetesimal disc does not only require a large
particle number to assure a proper description of the system by a distribution
function, but also the {uncorrelated motion} of the planetesimals. Each of
the formul{\ae} derived before involves the averaging over different impact
parameters to some extend, in combination with the vital assumption that all
distances are equally probable. As long as all particles are subjected to
perturbations by surrounding bodies, strong correlations are suppressed. This
applies to the early stages, but the formation of protoplanets introduces a few
dominant bodies that are not susceptible to the perturbation of the field
planetesimals. Orbit repulsion gives rise to a regular spacing of the
protoplanets, which prevents mutual collisions. Therefore not all impact
parameters are equally probable due to this strong correlation. Hence a
statistical model is inherently not applicable to the late stages of
protoplanetary growth.

Since statistical models are superior to $N$--body calculations
with respect to speed and (effective) particle number, modifications
have been proposed to remedy this problem.

The statistical model by \cite{Wetherill1993}
uses the following solution: A {gravitational range}  $\Delta a$ (or buffer zone) is attached to each planetesimal,
which represents the minimal spacing that allows for stable orbits.
They propose the expression
\begin{equation}
\Delta a  =  f_{\Delta} R_{\mathrm{Hill}}+\sqrt{2T_r/\Omega^2}
\end{equation}
where $f_{\Delta}$ is the minimal spacing in terms of the Hill radius. The value
of $f_{\Delta}$ is adopted from \cite{Birn1973}, who derived the minimal
spacing that allows for stable orbits:
\begin{equation}
  f_{\Delta}  =  2\sqrt{3}
\end{equation}
Thus it is possible to define a minimum mass $m_{\mathrm{sep}}$ by the assumption that all bodies
larger than this critical mass maintain a clear buffer zone:
\begin{align}
  f &=  \int_{m_{\mathrm{sep}}}^{\infty} \frac{d\Sigma}{dm}\frac{2\pi a\Delta a}{m}  dm \nonumber \\
  f &=  1
\end{align}
$f$ is the area covered by the buffer zones (overlapping is not taken into account,
therefore $f>1$ is possible), normalised to the ring area.
Planetesimals smaller than $m_{\mathrm{sep}}$ can not enforce a minimum distance
to their neighbours, as the whole disc surface is already covered by the buffer
zones of the largest bodies. Owing to the regular spacing introduced by the
buffer zones, planetesimals larger than the critical mass are not allowed to
collide with each other.
This approach has also been employed by \cite{Ina2001}, who adopted
$f_{\Delta}= 10 $, which is the mean distance of protoplanets according to
the orbit repulsion mechanism.

To shed light on the proper gravitational range and the validity
of this approach, we defined an additional quantity $f_C$ which
is the true area (i.\,e. overlapping is handled properly) covered by the buffer  zones
in terms of the total area:
\begin{align}
  f_C &=  \int_{m_{\mathrm{sep}}}^{\infty} \frac{d\Sigma}{dm}\frac{2\pi  a (\Delta a)_C}{m}  dm \nonumber \\
  f_C &\leq   1
\end{align}

\begin{figure}
\resizebox{\hsize}{!}
          {\includegraphics[scale=1,clip]{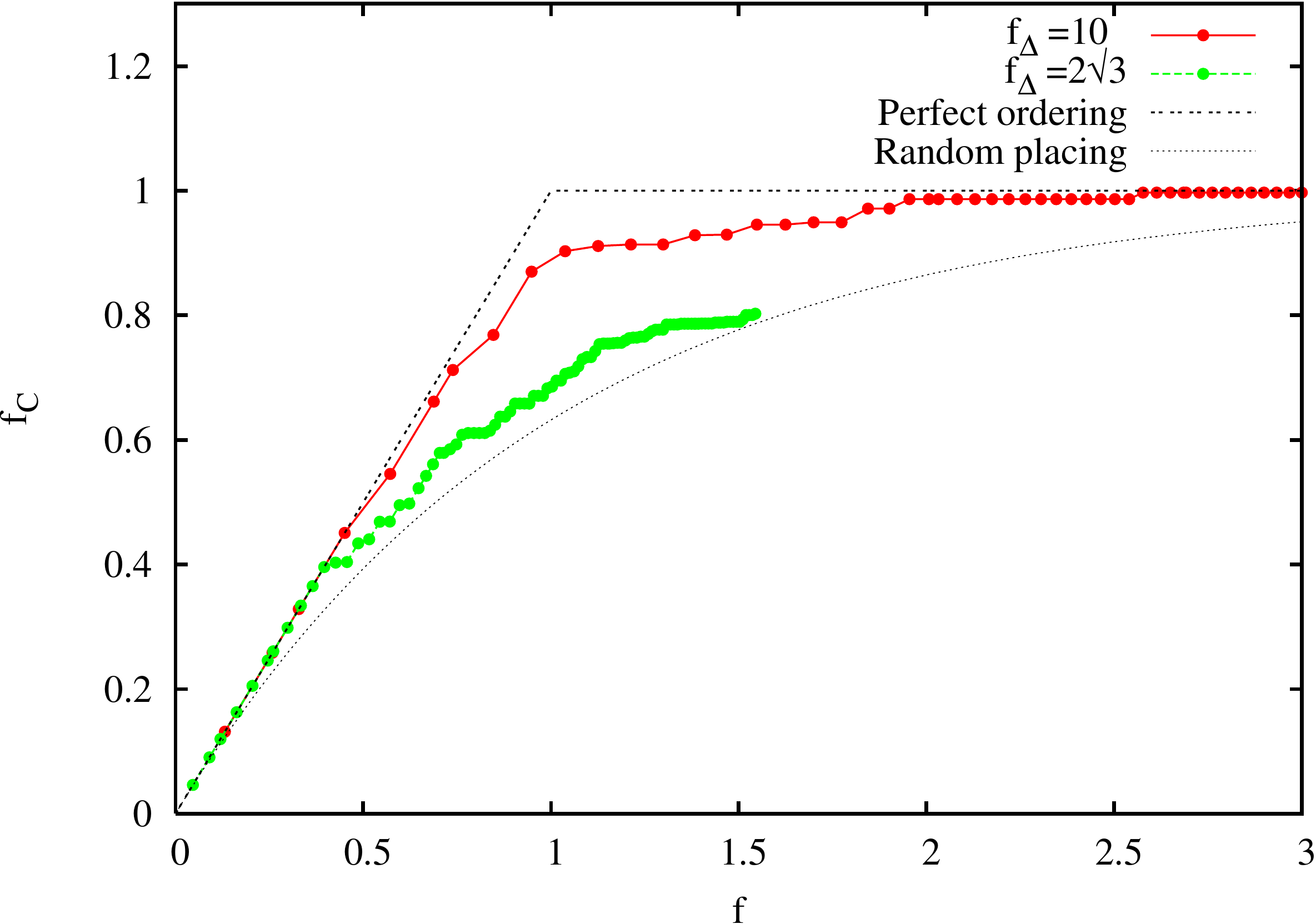}}
\caption
   {  
 Covered fraction $f_C$ as a function of $f$ for simulation  S1FB at $T=10^5$ years.
The protoplanets are already formed and grow oligarchically.  
   }
\label{GravRange}
\end{figure}

Fig.~\ref{GravRange} shows the covered fraction $f_C(m)$ as a function of the integrated buffer zones  $f(m)$ for
one of our hybrid calculations.
Both values for $f_{\Delta}$ are included as well as the two limiting cases {random placing}
and {perfect ordering}. Though we tested also other values of $f_{\Delta}$, a spacing
of ten Hill radii proved to be the best choice.

Our own experience with this method indicates that
it works reasonably well and agrees with \cite{Ina2001} who
used the same technique.
However, this modification includes the regular spacing of the protoplanets
in a prescribed way, so any exploration of later stages, like the
initiation of orbit crossing, is not accessible through this approach.
Therefore we use it for comparison purposes only, since the hybrid code (see section~\ref{HybridCode})
provides a much more general framework.

\subsection{Discretisation}

\begin{figure}
\resizebox{\hsize}{!}
          {\includegraphics[scale=1,clip]{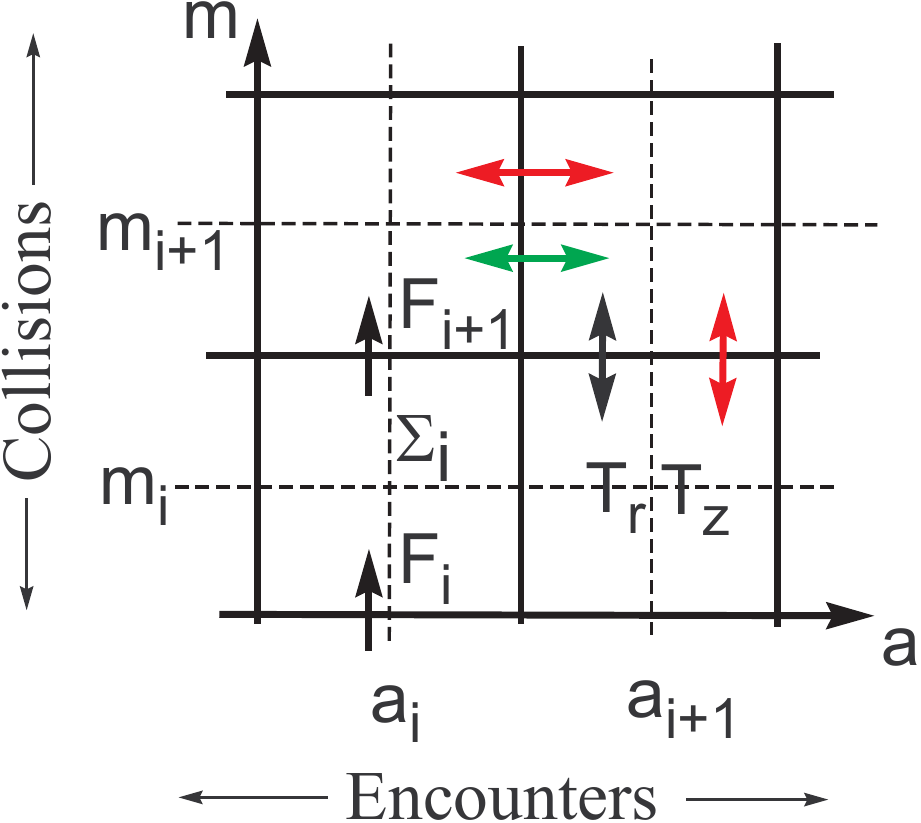}}
\caption
   {  
 Numerical grid. The arrows indicate transport of kinetic energy (red),
spatial transport of mass(green) and accretion (black). Non-neighbouring cells are coupled
 by the coagulation kernel and the radial interpolation kernel.  
   }
\label{ScetchGrid}
\end{figure}

All involved quantities are only functions of $a$ and $m$. Therefore
we introduce a two dimensional grid, where $\Sigma$, $T_r$ and $T_z$ are cell
centred quantities. Fig.~\ref{ScetchGrid} summarises the definition of the two-dimensional grid.
Since the full planetesimal size range covers several
orders of magnitude in mass, we chose a logarithmically equidistant discretisation in mass
to cover the necessary mass range in a reliable way. The radial spacing of the
grid cells is equidistant.
Thus the grid setup for the mass discretisation reads ($N$ grid cells from $m_{\min}\dots m_{\max}$):
\begin{align}
  m_i &=   m_{\min} \delta^{-i} (1/2+\delta/2) \qquad i=1, \dots, N \nonumber \\
 \Delta m_i &=   m_{\min} \delta^{-i}(1-\delta)  \nonumber \\
 \Sigma_i &=   \frac{d\Sigma}{dm} \Delta m_i \nonumber \\
  \delta &=  \left(\frac{m_{\min}}{m_{\max}}\right)^{1/N}
\end{align}
The grid spacing $\delta$ controls the number of cells which are necessary to cover a specified
mass range. As the evaluation of the coagulation equation scales with the third power
of the number of grid cells, $\delta$ should be as large as possible. If the
flux integral (see Eq.~\ref{Eqn1}) is approximated in a standard way
\begin{align}
 F_i &= - \sum_{j=1}^N \sum_{k=1}^N  F_i^{(jk)} \nonumber \\
F_i^{(jk)}  &=   \frac{1}{\sqrt{2\pi(h_j^2+h_k^2)}}\frac{\Sigma_j}{m_j}\frac{\Sigma_k}{m_k}\times
\sigma(m_j,m_k) \nonumber \\
& v_{\mathrm{rel}}M_{\mathrm{red}}(m_i-\Delta m_i/2,m_j,m_k) \label{PartFlux}
\end{align}
a spacing $\delta$ much smaller than 2 is required to guarantee a sufficient
accuracy\footnote{\cite{Ohtsuki1990} give a thorough analysis of the importance of the resolution.}.
However, it is possible to use a spacing of 2 if special precaution is taken.
\cite{Spaute1991} approximated the surface density with a power law,
thus taking the gradient with respect to mass into account. While they reached only a
sufficient accuracy with further special adaptations, we use a more
rigorous approach. A large spacing $\delta$ reduces the accuracy, since the partial
flux (Eq.~\ref{PartFlux}) is strongly varying even inside one grid cell. Thus
we rearrange this expression to identify the most important terms, as we can see in
Eq.~\ref{PartFlux2}.

\begin{figure*}
\begin{align}
F_i^{(jk)}  &=  \frac{v_{\mathrm{rel}}}{\sqrt{2\pi(h_j^2+h_k^2)}}\frac{\Sigma_k}{m_k} \underbrace{ \frac{\Sigma_j}{m_j}\times
\sigma(m_j,m_k)m_j}_{F_V(m_j)} \frac{1}{m_j} M_{\mathrm{red}}(m_i-\Delta m_i/2,m_j,m_k) \label{PartFlux2} \\
& j \geq k \nonumber
\end{align}
\end{figure*}
The strongest varying contribution $F_V$ is now approximated by a power law with
respect to $m_j$:
\begin{equation}
  F_V(m)   \approx   F_V(m_j)\left(\frac{m}{m_j}\right)^q \label{AprFV}
\end{equation}
Thus Eq.~\ref{AprFV} is used to provide improved partial fluxes $F_i^{(jk)}$ and thus we come to Eq.~\ref{PartFlux3}.

\begin{figure*}
\begin{align}
F_i^{(jk)}  &=  \frac{v_{\mathrm{rel}}}{\sqrt{2\pi(h_j^2+h_k^2)}}\frac{\Sigma_k}{m_k}
\int_{m_j-\Delta m_j/2}^{m_j+\Delta m_j/2}  \frac{F_V(m)}{m\Delta m_j} M_{\mathrm{red}}(m_i-\Delta m_i/2,m,m_k) dm \label{PartFlux3} \\
& j \geq k \nonumber
\end{align}
\end{figure*}
Since the fragment redistribution function is a piecewise power law, an analytical solution of the integral
is possible. Eq.~\ref{PartFlux3} gives reliable results even with a spacing $\delta=2$.
The time derivative of the surface density reads
\begin{equation}
\dot \Sigma_m^i  =  - F_{i+1}+F_i
\end{equation}
which assures the conservation of mass within numerical accuracy.

\subsection{Integrator}

All contributions to the evolution of the surface density $\Sigma$ and the velocity
dispersions $T_r$ and $T_z$ are summarised by the following set of differential equations:
\begin{align}
 D & =  \frac{4}{3\Omega^2}\left(\frac{5}{4}\dot T_{r,\mathrm{enc}} +\dot T_{z,\mathrm{enc}} \right) \nonumber \\
\frac{d\Sigma}{dt} &=  \Delta_a (D\Sigma)+\dot \Sigma_{\mathrm{coll}} \nonumber \\
\frac{d \Sigma T_r }{d t} &=   \Delta_a (D\Sigma T_r) +\Sigma \dot T_{r}+\frac{d}{dt}(\Sigma T_r)_{\mathrm{coll}}\nonumber \\
\frac{d \Sigma T_z }{d t} &=  \Delta_a (D\Sigma T_z) +\Sigma \dot T_{z}+\frac{d}{dt}(\Sigma T_z)_{\mathrm{coll}}
\end{align}
The Laplace operator is approximated in accordance with the equidistant radial grid:
\begin{align}
 \Delta_a f  &=  \frac{1}{a}\frac{\partial}{\partial a}\left(a\frac{\partial}{\partial a}f \right) \nonumber \\
 \Delta_a f &\approx   \frac{f_{i+1}(1+\Delta a/(2a_i))-2f_i+(1-\Delta a/(2a_i))f_{i-1}}{(\Delta a)^2} \label{DiscLapl}
\end{align}
We chose the {Heun method}\,\footnote{The name of this method is not unique. Some texts
denote it as the {modified Euler method}. The Heun method is a special case of the Runge--Kutta methods.}
as the basic integrator for the statistical model.
It is a second order accurate predictor--corrector scheme ($X$ is a vector containing all the above quantities):
\begin{align}
  \frac{dX}{dt} &=  f(X) \nonumber \\
      X^p &=  X_n+\Delta t f(X_n) \nonumber \\
      X_{n+1} &=  X^p+\frac{1}{2}\Delta t(f(X^p)-f(X_n))+\mathcal{O}(\Delta t^3)
\end{align}
The Heun method is readily extended to an iterate scheme, which is equivalent to the implicit expression:
\begin{equation}
      X_{n+1}  =  X_n+\frac{1}{2}\Delta t(f(X_{n+1})+f(X_n))
\end{equation}
This adds stability to the method and allows the secure integration of stiff configurations that
may appear during the runaway accretion phase. In practice, three iterations are sufficient to guarantee
a stable integration.
As the diffusive part is discretised with a first order accurate formula (see Eq.~\ref{DiscLapl}),
the whole iterated scheme is equivalent to the {Crank--Nicolsen method}.
We choose a global time step for the statistical model according to the expression
\begin{equation}
      \Delta t  =  \min\left( \eta_{\mathrm{Disc}}\frac{X}{\dot X},X\in \{\Sigma,T_r,T_z\}  \right)
\end{equation}
where the hybrid code (see next section) applies an additional discretisation in powers of
two to achieve a better synchronisation with the $N$--body code component.

\section{Bringing the two schemes together: the Hybrid Code}
\label{HybridCode}

\begin{figure}
\resizebox{\hsize}{!}
          {\includegraphics[scale=1,clip]{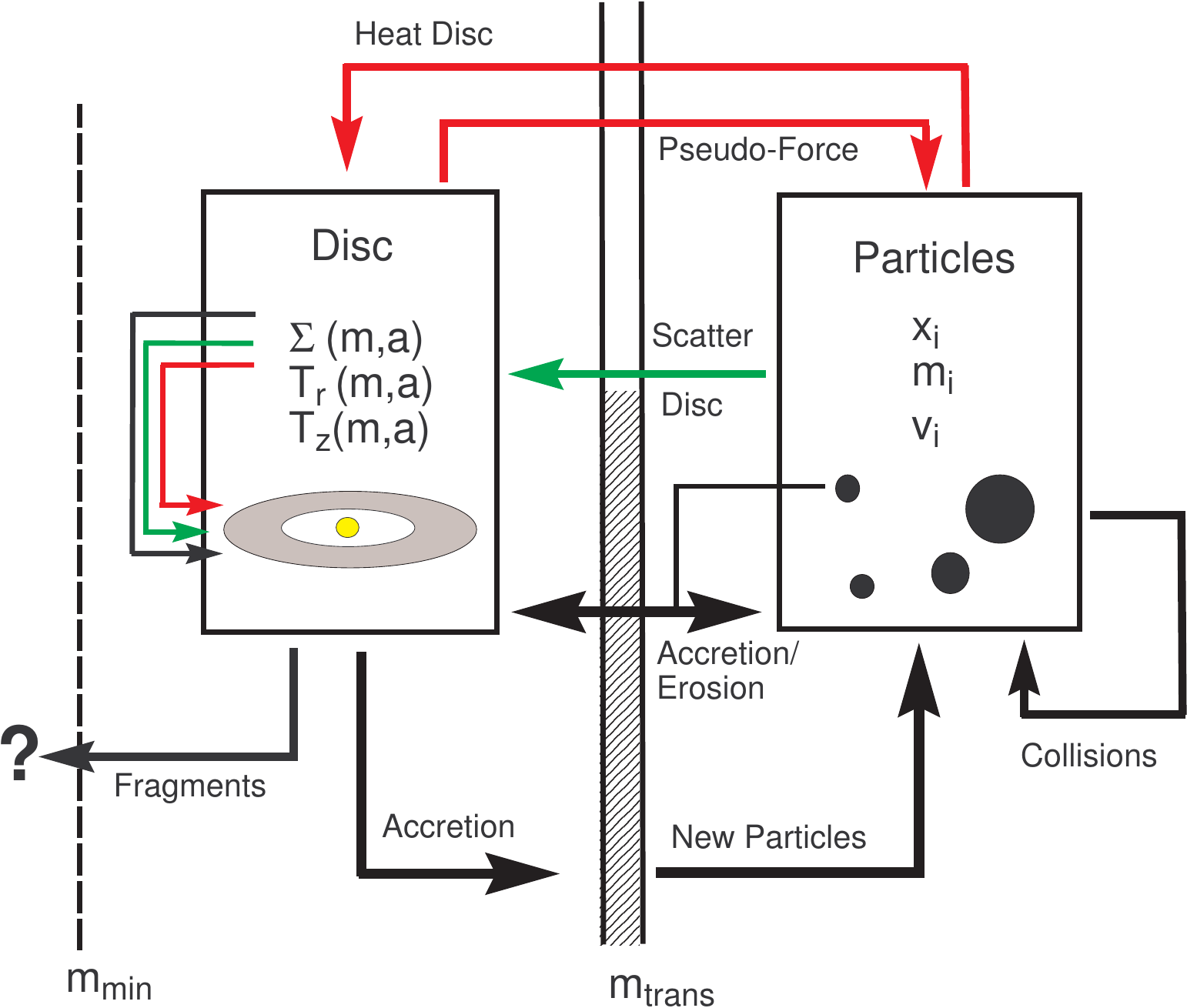}}
\caption
   {  
 Interplay between the $N$--body component and the statistical component of the hybrid
code. Black arrows indicate mass transfer, red arrows exchange of kinetic energy and
green arrows indicate spatial structuring, respectively.  
   }
\label{FlowC}
\end{figure}

We introduced two different methods to solve the planetesimal growth problem.
On the one hand, we modified {\sc Nbody6++}, which has been used so far mainly
for the simulation of stellar clusters, to adapt it to the special
requirements of a long-term integration of planetesimal orbits.
On the other hand, we developed a new statistical code with a consistent
evolution of the velocity dispersion, the capability to treat spatial
inhomogeneities   and a thoroughly constructed collision treatment.
Neither of the two approaches is powerful enough to provide
a complete and accurate description of the planetesimal problem,
since each method is confined to a certain range of the particle
number. However, these restrictions are complementary in the sense
that each method covers a regime where the other method fails.
This intriguing relation stimulated the construction of
a {hybrid code} which combines the benefits of both methods.

The basic idea is to introduce a transition mass $m_{\mathrm{trans}}$, which
separates the two mass regimes. Particles with a lower mass are treated by
the statistical model, whereas larger particles belong to the $N$--body
model. Though both parts are clearly divided in different mass ranges,
they are connected by various interdependencies:
\begin{enumerate}
\item
  Direct collisions between particles lead to a mass exchange.
  One process is the accretion of small particles by $N$--body particles,
  but agglomeration within the statistical model can also produce
  particles larger than the transition mass. This requires the
  generation of new $N$--body particles. Energetic impacts may
  erode larger particles, so a corresponding particle removal is
  also required for consistency.
\item
  Mutual scatterings among $N$--body particles and smaller
  planetesimals transfer kinetic energy. While energy equipartition
  leads to a systematic heating of the smaller field planetesimals,
  a consistent treatment has to include both transfer directions.
\item
  Accretion and scattering by the $N$--body particles
  induce spatial inhomogeneities or even gaps in the planetesimal component,
  if the particles
  have grown massive enough. Likewise, the small particles could induce
  some structure in the distribution of the $N$--body particles.
  Since the spatial structure is dominated by the stirring from few
  protoplanets, we neglect the latter process.
\end{enumerate}
Fig.~\ref{FlowC} summarises this brief overview of the interactions
between the two code components in a schematic diagram.
The following sections explain the implementation  of each interaction
term in more detail.

\subsection{Mass Transfer}

An $N$--body particle accretes smaller particles in its vicinity.
We already derived expressions which describe agglomeration within
the statistical model, so it is manifest to apply these formul{\ae}
to derive the accretion rate of an $N$--body particle.

Most of the material is accreted within the cross-sectional area
$\sigma$ (see Eq.~\ref{onlyGrav}), but the  finite eccentricity
of an orbit extends the accessible radial {feeding zone}. Thus we
assign the following surface density to each particle
\begin{align}
 \Sigma(a) &=  \frac{M}{2\pi a \sqrt{2\pi}l}\exp\left(-\frac{(a-a_0)^2}{2l^2}\right) \label{SigmaNb} \\
  l^2 &=  \sigma/\pi + \frac{1}{2}a^2 e^2 +T_r/\Omega^2
\end{align}
by smearing it out over its feeding zone. $T_r$ is the radial velocity dispersion
of particles in the statistical model with semimajor axis $a$. This density distribution
is projected onto the radial grid to calculate the accretion rate. As the time step
of the statistical model is much larger than the regular step of an $N$--body particle,
the particle mass update is synchronised with the statistical integration.
The projection technique allows the calculation of the accretion rates in a simple way,
which gives the right size of the feeding zone and the proper total accretion rate.

Particle generation is included in the following way:
A ``virtual'' mass bin is introduced as the boundary between the
statistical grid (denoted by the dashed area in Fig.~\ref{FlowC}) and the $N$--body component.
Its sole task is to store mass and kinetic energy that drives
the statistical model towards higher masses. If the mass content
exceeds one $m_{\mathrm{trans}}$, a new particle is created with
inclination and eccentricity according to the stored velocity
dispersions.

The masses of the $N$--body particles are regularly checked to detect
any particle which dropped below the transition mass. While
this procedure would remove the particle and transfer the associated
quantities back to the grid, we never observed such a particle erosion.

\subsection{Disc Excitation}

The projection of an $N$--body particle onto the grid with the help of a proper weight
function is also useful for the calculation of the disc excitation due to stirring by the larger
particles. Since the Hill radius sets the proper length scale for planetesimal encounters,
the weight function is modified to
\begin{align}
 \Sigma(a) &=  \frac{M}{2\pi a \sqrt{2\pi}l}\exp\left(-\frac{(a-a_0)^2}{2l^2}\right)  \nonumber \\
  l^2 &=  R_{\mathrm{Hill}}^2 + \frac{1}{2}a^2 e^2 +T_r/\Omega^2
\end{align}
where $T_r$ is the radial velocity dispersion of the heated planetesimal
component. The velocity dispersion of the stirring $N$--body particle
is (in accordance with Eq.~\ref{Approx_e} and Eq.~\ref{Approx_i}):
\begin{equation}
  T_{r,0} = \frac{1}{2}(\Omega a_0)^2 e^2 \\
  T_{z,0} = \frac{1}{2}(\Omega a_0)^2 i^2
\end{equation}
We employ the orbital elements as  mediators between the fast varying instantaneous position
and velocity of a particle and the slow evolution of the statistical model, which operates on
a longer relaxation timescale. In virtue of the projection of the particle, we readily apply
the standard interaction terms (see section~\ref{StatModel}) to evaluate the additional
heating due to the presence of \mbox{$N$--body} particles.

\subsection{Pseudo--Force}

While an $N$--body particle is moving through the disc, it also interacts gravitationally with the
particles in the statistical model. The collective effect
of all these encounters leads to a change in the orbital elements
of the $N$--body particle. Again, we project the $N$--body particle onto the grid and evaluate
the stirring rates $\dot T_r$ and $\dot T_z$, which correspond to a change
in the orbital elements:
\begin{align}
 \frac{d}{dt} e^2 &=   \frac{2 \dot T_{r}}{(\Omega a_0)^2} \nonumber \\
 \frac{d}{dt} i^2 &=   \frac{2 \dot T_{z}}{(\Omega a_0)^2}
\end{align}
These time derivatives of eccentricity and inclination are translated to
a pseudo-force that effects the desired change of the orbital
elements. We chose the ansatz
\begin{align}
 {\bf F}_{x,y} &=  C_r( {\bf v}_{x,y}- ({\bf v}_K)_{x,y} ) \nonumber \\
 {\bf F}_{z} &=  C_z {\bf v}_{z}
\end{align}
where $ {\bf v}_K$ is the local Keplerian velocity. In addition, we tried a simpler expression
\begin{align}
 {\bf F}_{x,y} &=  2C_r{\bf r}_{x,y} \frac{{\bf r}\cdot{\bf v}}{r^2} \nonumber \\
 {\bf F}_{z}   &=  C_z {\bf v}_{z}
\end{align}
without any significant differences in the  accuracy or the simulation outcome.
The proper friction coefficients are:
\begin{align}
  C_r &=  \frac{\dot T_r}{2T_r} \nonumber \\
  C_z &=  \frac{\dot T_z}{2T_z}
\end{align}
Since the relevant quantities are the time derivatives of the orbital
elements, any other pseudo-force is also applicable. Though this
approach yields the right mean change of the orbital elements, it lacks
the statistical fluctuations from the particle disc. Hence the distribution
of the orbital elements of the $N$--body particles is artificially narrowed,
which is especially important when the $N$--body particles and
the statistical particles have a comparable mass. As the mass contrast
between the two code parts is quite significant in planet formation
simulations, it is safe to neglect the fluctuating part without major restrictions on
the realism of the simulations.

The friction coefficients $C_i$ are kept constant between two integration steps
of the statistical model. While a more frequent update of the coefficients
would be easily possible, a regular update on the basis of the statistical time
step is accurate enough. Moreover, each update poses a considerable computational effort
(roughly equivalent to 1000 force evaluations), so our approach also
saves valuable computational time.

\subsection{Spatial Structure}

\begin{figure}
\resizebox{\hsize}{!}
          {\includegraphics[scale=1,clip]{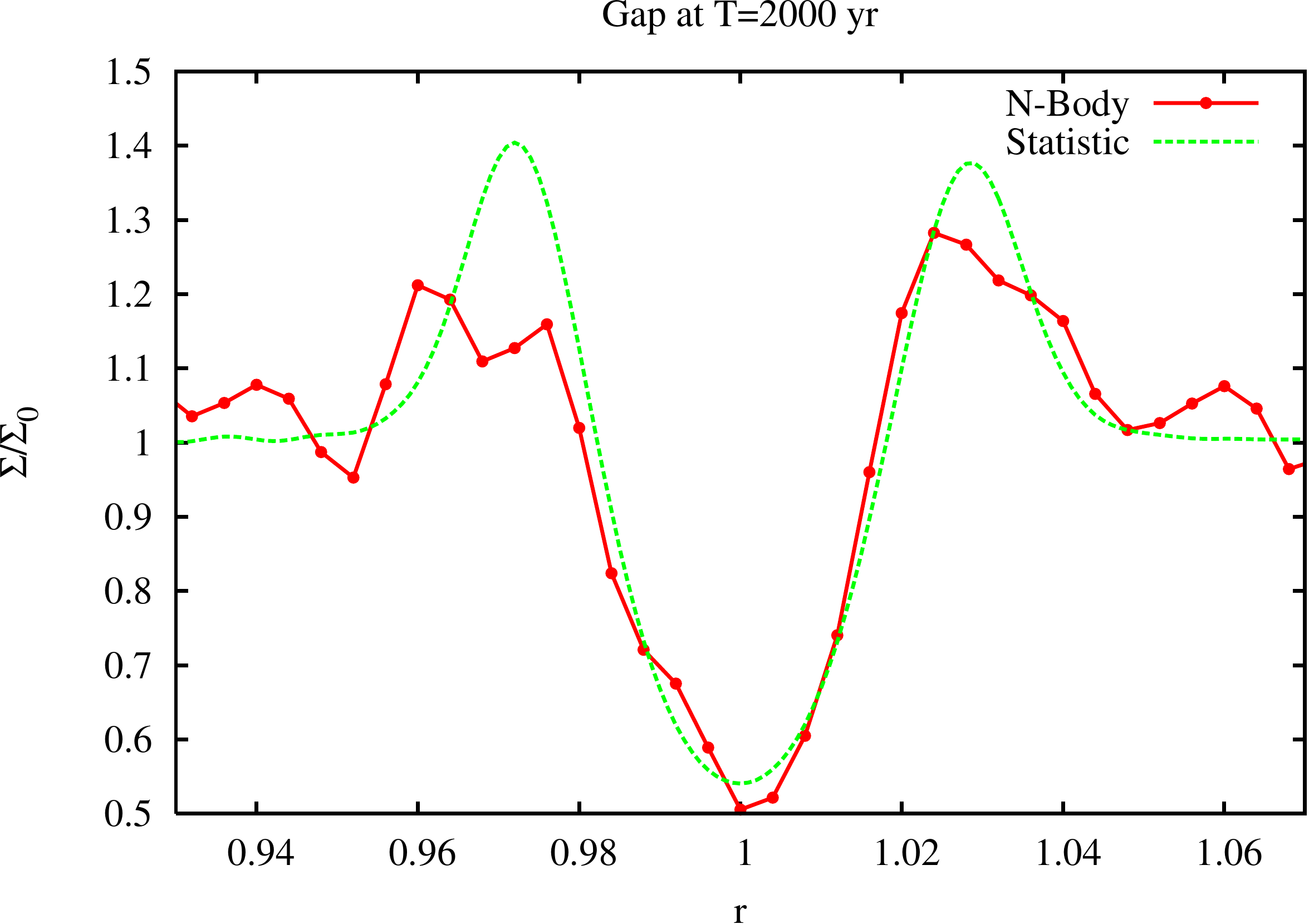}}
\caption
   {  
 Gap opening in a planetesimal disc. The gap is fully developed after 2000 years. Table~\ref{TabGap}
summarises the initial conditions for the comparative runs.  
   }
\label{PlotGap}
\end{figure}

\begin{table*}
{\small
\begin{tabular}{|l|l|l|l|l|l|l|l|l|}  \hline
  No  & $\Sigma$ & $\Delta a$ & $N$ & $N_{\mathrm{rad}}$ & $e^2/h^2$ & $i^2/h^2$ & $m$ & Type \\ \hline
  G1  & $1.1251\times 10^{-6}$ & 0.2 & 1406 & -- & 0.00135 & 0.00135 & $1\times 10^{-9}$  & $N$--body \\ \hline
      & Perturber & --- &  1 & -- & $e=6.1\times 10^{-5}$ & $i=3.2\times 10^{-5}$ & $1\times 10^{-7}$  &  \\ \hline
  G2  & $1.1251\times 10^{-6}$ & 0.2 &  --  & 201 & 0.00135 & 0.00135 & $1\times 10^{-9}$  & Statistic \\ \hline
      & Perturber & --- &  1 & -- & $e=6.1\times 10^{-5}$ & $i=3.2\times 10^{-5}$ & $1\times 10^{-7}$  &  \\ \hline
\end{tabular} }
\caption[Parameters of the gap simulations]
{Parameters of the statistical and the $N$--body gap simulation. The perturber is placed at the centre of the ring.\label{TabGap}}
\end{table*}

The first insight into planetesimal formation was obtained by the
particle--in--a--box method, which invokes the underlying assumption that the planetesimal disc
stays homogeneous throughout the protoplanet growth \citep[see e.g.]{Greenberg1978}. While few large bodies
introduce some coarse-graininess of the surface density, all smaller bodies
are assumed to be evenly spread in the disc. Research on the interaction of
protoplanets showed that this is an oversimplification, as bodies that are massive
enough could open gaps in their vicinity \citep[see e.g.]{Lin1979,Rafikov2001}.
Gap formation
does not only change the overall surface density, but also controls the
accretion onto the protoplanet through the amount of planetesimals in the feeding
zone. If gap formation is too effective, the growth of the protoplanet may well stop
before the isolation mass is reached. Hence any hybrid code should
provide a framework that allows this mechanism to operate. A necessary
condition is a radial density grid with a sufficient resolution to describe possibly
emerging gaps. A too low resolution suppresses local perturbations from the
protoplanets by a simple averaging, thus inhibiting the formation of
any spatial inhomogeneities. A second requirement is that
the interaction terms relating statistical model and $N$--body model
include the local interaction between particles and the statistical
component in a proper way.

\begin{figure}
\resizebox{\hsize}{!}
          {\includegraphics[scale=1,clip]{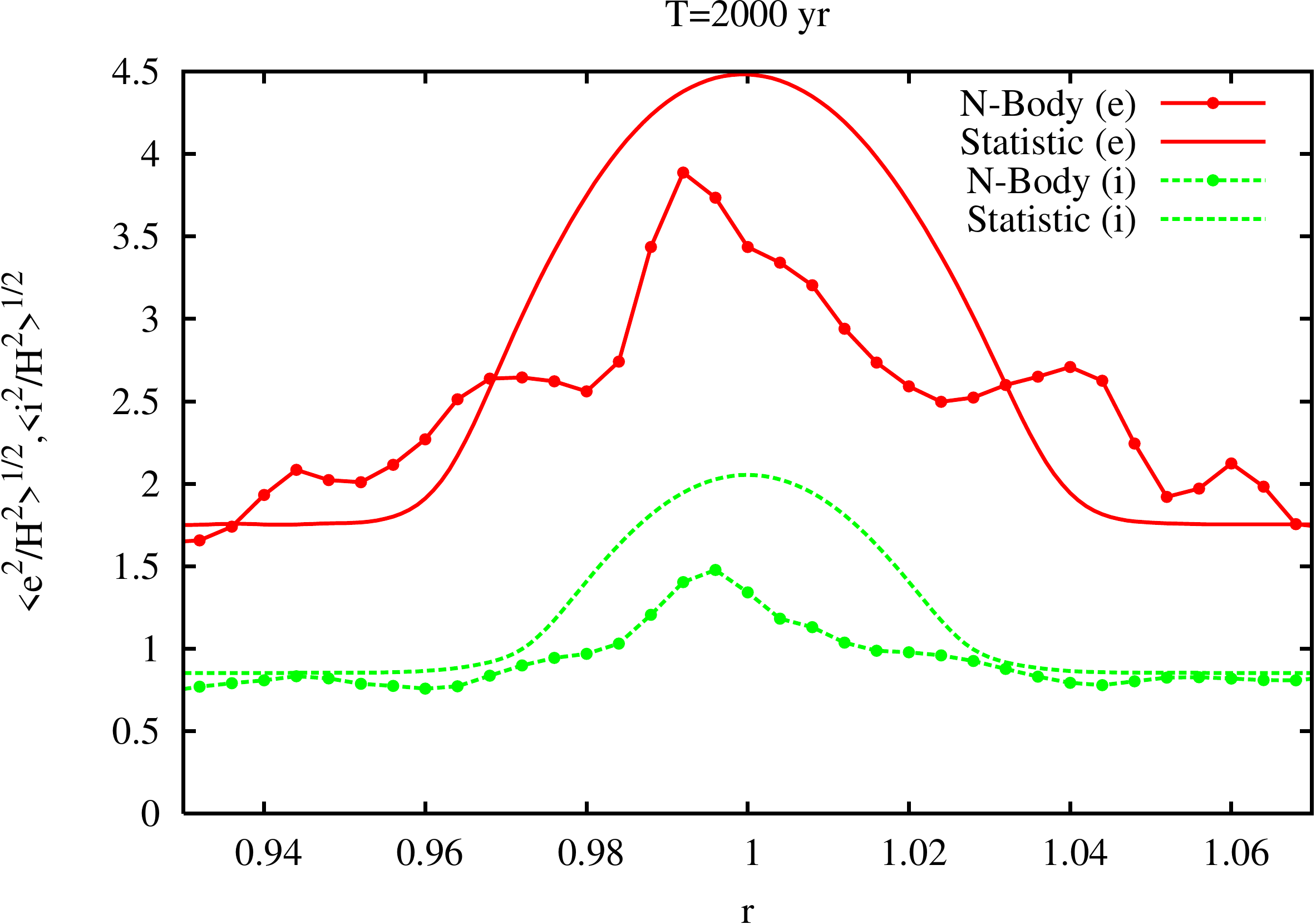}}
\caption
   {  
Mean square eccentricity and inclination of the smaller planetesimals in terms of the reduced Hill radius $H$ of
the protoplanet according to simulation G1 ($N$--body) and G2 (Statistic).   
   }
\label{PlotGap_ei}
\end{figure}

Our hybrid approach includes gap formation implicitly through the diffusive terms.
A protoplanet heats only the planetesimals in its vicinity (defined
by the heating kernel), thus also increasing locally the diffusion
coefficient. Hence the surface density drops due to outward diffusion
of the planetesimals, given that the protoplanet is massive enough.
The minimum gap opening mass is set by the condition that the protoplanet
controls the random velocities of the field planetesimals in its heating zone (see Eq.~\ref{Mheat}),
which is equivalent to the independently derived gap formation criterion (compare Eq.~\ref{Mgap}).

Although our algorithm invokes a simplified picture of the
protoplanet--planetesimal interaction, it is surprisingly accurate with respect
to the width of the forming gap and the opening criterion.
Fig.~\ref{PlotGap} shows a simulation which examines the accuracy
of our approach. The overall performance of the statistical code
is quite remarkable, except a significant overestimation of
the surface density at the gap boundary compared to the $N$--body model.
This deviation is due to the improper treatment of strong planetesimal--protoplanet
encounters, which exceed the diffusive approximation. Moreover,
the higher concentration of
planetesimals near the gap boundary leads to an additional overestimation of the velocity
dispersion of the smaller planetesimals in the statistical calculation (see Fig.~\ref{PlotGap_ei}).
While the comparison with the $N$--body calculation clearly indicates a necessary improvement
of the treatment of spatial inhomogeneities, our approach  catches the main features of gap formation.

\subsection{Transition Mass}

Since the inventory of the new hybrid code is now completed, we turn
to the specification of the transition mass $m_{\mathrm{trans}}$.
The mass boundary between statistical and $N$--body part has a
major influence on the realism and the speed of the simulation.
On the one hand, optimisation with respect to speed favours
a large transition mass, whereas a reasonable resolution of
the transition between the two components introduces some upper limit.

Hence we identify  first the set of large masses, which controls
the velocity dispersion of the disc, since these objects are also
possible candidates for gap opening. The inspection of
all involved stirring terms gives approximately the inequality:
\begin{equation}
 \int_0^{m_{\mathrm{trans}}} \frac{d\Sigma}{dm}m dm  <  \int_{m_{\mathrm{trans}}}^{\infty} \frac{d\Sigma}{dm}m dm \label{Crit_1}
\end{equation}
While this is a necessary condition to select all potential major perturbers, criterion
\ref{Crit_1} does not imply that all particles in the selected mass range exert
indeed a strong influence on the disc. The number of possible gaps
-- and therefore the number of perturbers associated with them -- is ultimately
limited by the available space. Thus we integrate the area of all potential gaps (width $\approx f_{\Delta} R_{\mathrm{Hill}}$)
and normalise it to the total disc area:
\begin{equation}
  f_C   \approx   \int_{m_{\mathrm{trans}}}^{\infty} f_{\Delta} \frac{d\Sigma}{dm}\frac{2\pi a R_{\mathrm{Hill}} }{m} dm
\end{equation}
If the covered fraction $f_C$ is much larger than one, it is possible to increase
the transition mass until the condition
\begin{equation}
  f_C   \lessapprox   1 \label{Crit_2}
\end{equation}
is fulfilled. Of course, condition~\ref{Crit_1} and~\ref{Crit_2} defined only an upper limit
of the transition mass, so the adaptation of a lower value is also possible.  Though
there are two reliable conditions at hand, the transition mass is still a function of
time owing to the time evolution of the density $\Sigma(m)$. Therefore we chose
{a priori} a fiducial value of the transition mass, run the simulation and conduct
an {a posteriori} check, whether the initial choice matches our requirements at any
evolutionary stage of the disc. A reliable value for a solar system analogue
at 1 AU is
\begin{equation}
 m_{\mathrm{trans}}   \approx   3\times 10^{-11} M_{\odot}
\end{equation}
which restricts the number of $N$--body particles to a tractable amount. Later stages
would allow an even larger transition mass, but the current hybrid code does not
include any dynamical adjustment of the transition mass at runtime.

\subsection{Boundary Conditions}

Any numerical simulation is limited to a finite simulation volume and a
finite time interval. Therefore it is mandatory to introduce proper {boundary
conditions} which provide a reasonable closure of the simulation volume.

While boundary conditions with respect to time are the familiar {initial conditions},
the choice of the spatial boundary conditions for the various involved quantities
depends on the problem at hand and the type of the boundary. A simulation boundary
can be due to physical reasons (like walls of a concert hall, surface of a terrestrial
planet) or simply due to  a limitation in computational power that inhibits the
complete numerical coverage of the problem.

The current capability of the hybrid code sets limits on the radial range as well as on
the covered mass range, which a simulation can handle in a reasonable time.
Hence we have to introduce artificial boundaries in radius, and a
lower limit for the mass grid.

Any migration process couples the evolution of a local ring area in the planetesimal disc to the evolution
of the whole disc. Inward (or outward) migrating material also transports
information on the radial zone where the material originated from. As this
information is not available within the frame of a local simulation, any choice
of the boundary condition alters the evolution to some extend.

However, we focus on a formation stage where migration is not a dominant
process, but  provides only removal of the smaller collisional fragments.
Thus we apply closed boundary conditions for the outer and inner
radius of the ring area (i.\,e. all fluxes vanish at the boundary), and an
open boundary for the lower end of the mass range.
While these conditions exclude the study of migrational processes, we
gain clearer insight into the protoplanet growth.

\begin{table*}
\begin{center}
\begin{tabular}{|l|l|l|l|l|l|l|l|l|}  \hline
  No. & $\Sigma$ & $\Delta a$ & $N$ & $N_{\mathrm{rad}}$ & $e^2/h^2$ & $i^2/h^2$ & $m$ & Type \\ \hline
  T1a & $1.1251\times 10^{-6}$ & 0.02 & 1000 & -- & 0.04 & 0.01 & $1.41\times 10^{-10}$  & $N$--body \\ \hline
  T1b & $1.1251\times 10^{-6}$ & 0.02 &  500 & 10 & 0.04 & 0.01 & $1.41\times 10^{-10}$  & Hybrid    \\ \hline
  T1c & $1.1251\times 10^{-6}$ & 0.02 &  --  & 10 & 0.04 & 0.01 & $1.41\times 10^{-10}$  & Statistic \\ \hline \hline
  T2a & $0.5626\times 10^{-6}$ & 0.08 &  800 & -- & 4    & 1 & $5\times 10^{-10}$ & $N$--body  \\ \hline
      & $0.5626\times 10^{-6}$ &      &  200 & -- & 4    & 1 & $2\times 10^{-9}$ & \\ \hline
  T2b & $0.5626\times 10^{-6}$ & 0.08 &  --  & 10 & 4    & 1 & $5\times 10^{-10}$ & Hybrid  \\ \hline
      & $0.5626\times 10^{-6}$ &      &  200 & -- & 4    & 1 & $2\times 10^{-9}$ & \\ \hline \hline
  T3  & Safronov               & --   &  --  & -- & --   & --&  --               & Statistic  \\ \hline \hline
  T4a & $1.1251\times 10^{-6}$ & 0.02  & 10.000   & -- &  4  &  1  & $1.41\times 10^{-11}$  & $N$--body \\ \hline
  T4b & $1.1251\times 10^{-6}$ & 0.02  & --  & 10 & 4  &  1  & $1.41\times 10^{-11}$  & Hybrid  \\ \hline
  T4c & $1.1251\times 10^{-6}$ & 0.02  & --  & 10 &   4  &  1  & $1.41\times 10^{-11}$  & Statistic \\ \hline \hline
  T5  & $1.8789\times 10^{-6}$ & --    & --  & -- &  620 &  155  & $2.4 \times 10^{-15}$  & Statistic \\ \hline
\end{tabular}
\end{center}
\caption[Parameters of all test simulations]
{Parameters of all test simulations. The transition mass in T4b is $m_{\mathrm{trans}}=3.1\times 10^{-10}$ \label{SimData}
Only simulations T3, T4a--T4c and T5 include collisions. All values are scaled to $M_c=G=r_0=1$.}
\end{table*}

\section{Discussion and conclusions}
\label{discussion}

The formation of planetary systems represents a challenge from a numerical
standpoint.  The dynamical problem spans over many orders of magnitudes in
length and demands the combination of different techniques. We have presented a
composite algorithm that brings together the advantages of direct-summation
tools and statistics for the description of the planetesimal disc.
Direct-summation $N-$body techniques have been around for some decades and have
proven their accuracy in a very large number of studies of stellar clusters
such as galactic nuclei and globular and open clusters. We deem it to be the
numerical tool to integrate the motion of the bodies for the very precise
integration of the orbits and treatment of close encounters. Typically, in a
simulation of a stellar system, the energy is conserved in each timestep by
$E/\Delta E \sim 10^{-11}$ (where $E$ is the total energy and $\Delta E$ the
difference between the former and current total energy for a specific time), so
that even if we integrate for a long time the cluster, the accumulated energy
error is negligible. Nevertheless, porting the numerical tool to the problem of
planetary dynamics is not straighforward and requires important modifications
and additions. In this work we present them in detail: the neighbour radius
selection for the protoplanets, the Hermite iteration and we introduce for the
very first time the new extended Hermite scheme, since the usual Hermite scheme
is not sufficient to integrate planetesimal orbits accurately enough. Then we
bring in new forces to the problem, namely the introduction of the central
potential of the star, as well as the drag forces, which depend on the gas
density and size of the planetesimals. Hence, the regularisation scheme,
crucial to exactly integrate the close encounters, has to be accordingly
modified. We then introduce the disc geometry and discuss the required changes
to the neighbour scheme and prediction, as well as the communication algorithm
and block size distribution.

For the statistical description of the planetesimal disc we employ a
Fokker-Planck approach. We include dynamical friction, high- and low-speed
encounters, the role of distant encounters as well as gas and collisional
damping and then generalise the model to inhomogenous discs.  We then describe
the combination of the two techniques to address the wole problem of
planetesimal dynamics in a realistic way via a transition mass to integrate the
evolution of the particles according to their masses.

In particular, we introduce and describe the extended Hermite scheme, which
reduces the the energy error by three orders of magnitude with the same number
of force evaluations, compared to the standard version of {\sc Nbody6++}.

While the implementation and some code details are newly introduced to the
field of planet formation simulations, the first hybrid approach was developed
in the early 90's. \cite{Spaute1991} \citep[further improved in][]{Weiden1997}
constructed a hybrid code with a statistical component to treat the smaller
particles and a special treatment for the larger particles. A statistical model
covers the field planetesimals with the help of a distribution function
(similar to \cite{Wetherill1989}), whereas the larger particles are
individually stored and characterised by mass, semimajor axis, eccentricity and
inclination. While the interaction between these single particles and the
statistical component is expressed by standard viscous stirring and dynamical
friction terms, perturbations among the single particles are equated in a
different way. First, the probability of an encounter of two neighbouring
particles is calculated. This probability is used in a second step to decide
whether a (numerically integrated) two-body encounter of the neighbouring
particles is carried out to derive the change in the orbital elements.  Though
these two well-defined code components justify to speak about a hybrid
approach, the Monte--Carlo like integration of the largest particles is still
closely related to a statistical treatment. 

A modified $N$--body approach is used in the work of
\cite{LevisonMorbidelli2007}. Their method covers the largest particles by a
direct $N$--body code, which includes the smaller particles as ``tracer''
particles.  The term ``tracer'' indicates that each particle represents a whole
ensemble of planetesimals. In a similar line of approach and inspired by this
idea, \cite{LevisonThommesDuncan2010} modified a symplectic algorithm, {\sc
Symba}, to study the formation of giant planet cores.  However, they made some
assumptions in order to calculate the gravitational interaction between the
planetesimals.  In particular, they ignored totally close encounters between
planetesimals.

\cite{OrmelSpaans2008} present in their work a scheme based on Monte Carlo
techniques to cover the vast range of sizes. For this, they assign more
resolution to those particles that are more relevant to the interactions,
typically the largest bodies. Smaller particles are grouped and treated
collectively, which means that they all share the same mass and structural
parameters. This classification is done in accordance to the ``zoom factor'', a
free paramenter. Later, \cite{OrmelEtAl2010a} presented an detailed comparison
of their Monte Carlo code with other techniques, in particular with pure
direct-summation $N-$body results and other statistical studies and found that
system leaves the runaway at a larger radius, in particular at the outer disc.
With their simulations, the authors propose a new criterion for the runaway
growth-oligarchy transition: from several hundreds of km in the inner disk
regions up to a thousand km for the outer disc \citep{OrmelEtAl2010b}.

\cite{BromleyKenyon06} published a description of a hybrid method with a basic
approach similar to our work. They employ two velocity dispersions and the
surface density of the planetesimals to describe the planetesimal system.  The
statistical component includes migration of the planetesimals and dust
particles due to gas drag and Pointing--Robertson drag.  In contrast to our
approach, they did not include mass transport due to the diffusion of the
planetesimals, which precludes the study of spatial structures induced by the
protoplanets. One must note also that their method uses the standard
discretisation of the collisional flux (see Eq.~\ref{PartFlux}) and thus
restrict the spacing factor to $\delta \lesssim 1.25$ \citep{Kenyon1998}.
\cite{BromleyKenyon06} chose a set of test calculations which focused less on
the technical aspects of their method, but on an overall comparison with a
selected set of standard works on planet formation. Their test simulations are
in good agreement with the references simulations, thus indicating a comparable
quality of the method.  Four years later, the authors presented an updated
version of their code for planet formation. The new characteristics of the code
included 1D evolution of the viscous disc, gas accretion on to massive cores,
as well as accretion of small particles in planetary atmospheres
\citep{BromleyKenyon10}.

While a variety of hybrid approaches emerged over the past years, this
technique is still far from a routinely application and is still challenged by
many open issues. Hybrid codes bear the potential to address the dynamical
evolution of a whole planetary system, the later stages of protoplanet
formation initiate a strong interaction with the gaseous disc, which may
require more diligence than the inclusion of a few additional interaction terms.
However, the development is picking up speed, which places our work in a good
position for further research.

\begin{appendix}

\section{Central Force -- Derivatives} 
\label{Force23Ana}

Central force $ {\bf F}$ per mass (i.\,e. acceleration) and its time derivatives are:
\begin{align}
 {\bf F}        &=  - \frac{ {\bf x} M}{x^3} \nonumber \\
 {\bf F}^{(1)}  &=  -\frac{ {\bf v} M}{x^3} -3A{\bf F} \nonumber \\
 {\bf F}^{(2)}  &=  -\frac{ {\bf a} M}{x^3} -6A{\bf\dot F}-3B{\bf F} \nonumber \\
 {\bf F}^{(3)}  &=  -\frac{ {\bf\dot a} M}{x^3} - 9A {\bf F}^{(2)}-9B{\bf F}^{(1)}-3C{\bf F} \nonumber \\
  {\bf a}       &=  {\bf \dot v} \nonumber \\
   A            &= \frac{ {\bf x} \cdot {\bf v} }{x^2} \nonumber \\
   B            &= \frac{v^2}{x^2}+ \frac{{\bf x}\cdot {\bf a}}{x^2} +A^2 = \dot A +3A^2 \nonumber \\
   C            &= \frac{3{\bf v}\cdot {\bf a}}{x^2} +\frac{{\bf x}\cdot {\bf \dot a}}{x^2} +A(3B-4A^2)
\end{align}

The ${\bf F}^{(i)}$ denote the central force and its time derivatives, whereas ${\bf a}$ and ${\bf\dot a}$ refer to the
total acceleration of the particle. The assumption that ${\bf x},{\bf v},{\bf a}$ and ${\bf\dot a}$
are independent of each other allows the derivation of averaged expressions for particle--particle interactions:
\begin{align}
  \langle({\bf F})^2\rangle &=  m^2 \frac{1}{x^4} \nonumber \\
  \langle({\bf F}^{(1)})^2\rangle &=  m^2  \frac{2v^2}{x^6} \nonumber \\
  \langle({\bf F}^{(2)})^2\rangle &=  m^2 \left( 12\frac{v^4}{x^8}+2\frac{a^2}{x^6} \right) \nonumber \\
  \langle({\bf F}^{(3)})^2\rangle &=  m^2 \left( 144\frac{v^6}{x^{10}}+126\frac{a^2v^2}{x^8}+2\frac{\dot a^2}{x^6} \right)
\end{align}
We combine these expressions with Aarseth's time step formula to derive the
regular time step as a function of the neighbour sphere radius $R_s$:
\begin{align}
 \Delta t_{\mathrm{reg}} & \approx   \sqrt{\eta_{reg}}\frac{R_s}{\bar v}
    \frac{1}{1+\sqrt{ R_s/ R_0} }  \nonumber \\
    R_0 &=  4\frac{ \bar v^2}{ a} \approx \frac{4 \bar v^2\bar r^2}{Gm}  =4\frac{\bar r^2}{r_{\mathrm{close}}}
\end{align}
$\bar r$ is the average particle distance and $r_{\mathrm{close}}$ is the impact parameter for a 90--degree deflection.

\section{Scalable Collisions Flux} 
\label{ScalFlux}

The mass flux according to the perturbation equation~\ref{EqDeltan} is:
\begin{align}
 F_p &=  -\dint (n(m_2)\Delta n(m_1)+n(m_1)\Delta n(m_2))\nonumber \\
 &\sigma(m_1)v(m_2)m_1 f_m(m_1/m,\epsilon) dm_1 dm_2 \nonumber \\
   &=  F^{(1)} + F^{(2)}
\end{align}
Firstly, we employ the substitution
\begin{align}
  m_1 &=  m x_1 \nonumber \\
  m_2 &=  m_0 \left( \frac{m_1}{m_0}\right)^{\frac{1+\alpha}{1+2w}}\left(\tilde S\right)^{\frac{1}{1+2w}}
    \epsilon^{\frac{1}{1+2w}}
\end{align}
to solve for the partial flux  $F^{(1)}$:
\begin{align}
F^{(1)} &=  -n_0^2 m_0^3 \sigma_0v_0 \int  g(mx_1) F_1(x_1) dx_1 \nonumber \\
F_1(x_1) &=  {\tilde S}^{-k'} \int  \epsilon^{-\frac{w+s+3+\alpha}{2+\alpha+2w}}\frac{f_m(x_1,\epsilon)}{x_1(1+2w)}d\epsilon
\end{align}
The second contribution $F^{(2)}$ requires a slightly different transformation:
\begin{align}
  m_1 &=  m x_1 \epsilon^{-1/(1+\alpha)} \nonumber \\
  m_2 &=  m_0 \left( \frac{m x_1}{m_0}\right)^{\frac{1+\alpha}{1+2w}}\left(\tilde S\right)^{\frac{1}{1+2w}}
\end{align}
Thus the partial flux  $F^{(2)}$ is:
\begin{align}
F^{(2)} &=  -n_0^2 m_0^3 \sigma_0 v_0 \int g(m_2) F_2(x_1) dx_1\nonumber \\
F_2(x_1) &=  {\tilde S}^{-k'}\int
 \epsilon^{-\frac{w+s+3+\alpha}{2+\alpha+2w}}
 \frac{f_m(x_1 \epsilon^{-1/(1+\alpha)},\epsilon )}{ x_1(1+2w) } d\epsilon
\end{align}
We change to a new set of logarithmic coordinates
\begin{equation}
  u  =  \ln(m/m_0) \qquad u_1 = \ln(x_1 ) \qquad \tilde s = \frac{\ln(\tilde S)}{1+\alpha}
\end{equation}
which transforms the total flux $F_m$ to a convolution integral:
\begin{align}
  F_p & =  -n_0^2 m_0^3 \sigma_0 v_0 \int [ g(u+u_1)G_1(u_1)+ \nonumber \\
      & g \left(p(u+u_1+s_1) \right)G_2(u_1) ] du_1 \label{FpSum} \\
  p   & =  \frac{1+\alpha}{1+2w}
\end{align}
$p=1$ refers to the already derived solution for self-similar collisions. Hence we expand Eq.~\ref{FpSum} at
$p=1$ and retain only the zeroth-order moment of the fragmentation kernel:
\begin{align}
F_p &=-n_0^2 m_0^3\sigma_0 v_0 \Big[ g(u)G_{1,0}+(g(u)+\nonumber \\
& u(p-1)\frac{\partial g}{\partial u} )G_{2,0} \Big]
\end{align}
This expression is equivalent to
\begin{align}
F_p  &= -n_0^2 m_0^3\sigma_0 v_0 ( g(u)G_{1,0}+\nonumber \\
& [g(u)+(p-1)(g(u)-g(0))  ]G_{2,0} )
\end{align}
where higher derivatives of $g(u)$ are neglected.
Hence we recover the same functional form of the perturbed mass flux $F_p$ as for self-similar collisions:
\begin{align}
F_p &= -n_0^2 m_0^3\sigma_0 v_0 g(u)\left( G_{1,0}+p \,G_{2,0} \right) + \mbox{const.}\nonumber \\
  & \propto   {\tilde S}^{-k'}
\end{align}

\section{Coagulation Equation} 
\label{AprCoagEq}

While the success of a general approximation of the coagulation equation depends
heavily on the used coagulation kernel, we nevertheless provide a more general
approach to embed section~\ref{CollCasc} in a broader context. The standard
coagulation equation is:
\begin{align}
0 &=  \difft mn(t,m)+  \diffm F_m(t,m)  \nonumber \\
F_m &=  - \dint
n(t,m_1)n(t,m_2)\sigma(m_1,m_2)\nonumber \\
& v_{\mathrm{rel}}M_{\mathrm{red}}(m,m_1,m_2)
dm_1
dm_2
\end{align}
In virtue of our experience drawn from the perturbation expansion, we
transform the coagulation equation to logarithmic coordinates
\begin{equation}
   u  =  \ln(m)
\end{equation}
and employ the size distribution $g(u)$ relative to the steady-state
solution  $n_{\mathrm{eq}}(m)$:
\begin{align}
0 &=  \difft g(u,t)n_{\mathrm{eq}}(u)e^{2u} +  \frac{\partial}{\partial u} F_u(t,m)  \nonumber \\
F_u &=  - \dint g(t,u_1)g(t,u_2)K(u,u_1,u_2) du_1 du_2
\end{align}
$K(u,u_1,u_2)$ is the properly transformed new coagulation kernel.
$g(u)$ is expanded under the integral to arrive at a moment expansion of the flux $F_u$:
\begin{align}
F_u &=  - K_{00}(u)g(u)^2 - (K_{10}(u)+K_{01}(u))g(u)\frac{\partial g}{\partial u} + \dots \nonumber \\
  K_{ij} &=  \dint K(u,u_1,u_2)u_1^i u_2^j du_1 du_2
\end{align}
Retaining only the leading order terms, we recover an approximate coagulation equation which is similar
to the {inviscid Burgers' Equation}\,\footnote{This notion goes back to \cite{Burgers1948},
but the equation was already introduced by \cite{Bateman1915}.}:
\begin{equation}
0  =  \difft g(u,t)n_{\mathrm{eq}}(u)e^{2u} -  \frac{\partial}{\partial u} \left(K_{00}(u)g(u)^2\right)
\end{equation}

\end{appendix}

\section*{Acknowledgments}

It is a pleasure to thank Sverre Aarseth, Cornelis Dullemond and Phil Armitage
for comments on the manuscript.  PAS thanks the National Astronomical
Observatories of China, the Chinese Academy of Sciences and the Kavli Institute
for Astronomy and Astrophysics in Beijing, for an extended visit, as well as
the Aspen Center of Physics and the organizers of the summer meeting, where
this work was finished. PAS expresses his utmost gratitude to Hong Qi, Wenhua
Ju and Xian Chen for their hospitality during his stay in Beijing. He is also
somehow marginally indebted with the 2011 winter strain of German H1N2, which
allowed him to skip all possible duties and focus for an extended period of
time on this work at home.  RS acknowledges support by the Chinese Academy of
Sciences Visiting Professorship for Senior International Scientists, Grant
Number 2009S1-5 (The Silk Road Project). The special supercomputer Laohu at the
High Performance Computing Center at National Astronomical Observatories,
funded by Ministry of Finance under the grant ZDYZ2008-2, has been used.
Simulations were also performed on the GRACE supercomputer (grants I/80 041-043
and I/84 678-680 of the Volkswagen Foundation and 823.219-439/30 and /36 of the
Ministry of Science, Research and the Arts of Baden-urttemberg). Computing time
on the IBM Jump Supercomputer at FZ J{\"u}lich is acknowledged.

\label{lastpage}
\end{document}